# Light interaction with photonic and plasmonic resonances

Philippe Lalanne\*, Wei Yan, Kevin Vynck

LP2N, Institut d'Optique Graduate School, CNRS, Univ. Bordeaux, 33400 Talence, France

Christophe Sauvan, Jean-Paul Hugonin

Laboratoire Charles Fabry, Institut d'Optique Graduate School, CNRS, Université Paris-Saclay, 91127 Palaiseau Cedex, France

\* E-mail: philippe.lalanne@institutoptique.fr

**Abstract**

In this Review, the theory and applications of optical micro- and nano-resonators are presented from the underlying concept of their natural resonances, the so-called quasi-normal modes (QNMs). QNMs are the basic constituents governing the response of resonators. Characterized by complex frequencies, QNMs are initially loaded by a driving field and then decay exponentially in time due to power leakage or absorption. Here, the use of QNM-expansion formalisms to model these basic effects is explored. Such modal expansions that operate at complex frequencies distinguish from the current user habits in electromagnetic modeling, which rely on classical Maxwell's equation solvers operating at real frequencies or in the time domain; they also bring much deeper physical insight into the analysis. An extensive overview of the historical background on QNMs in electromagnetism and a detailed discussion of recent relevant theoretical and numerical advances are therefore presented. Additionally, a concise description of the role of QNMs on a number of examples involving electromagnetic resonant fields and matter, including the interaction between quantum emitters and resonators (Purcell effect, weak and strong coupling, superradiance, . . . ), Fano interferences, the perturbation of resonance modes, and light transport and localization in disordered media is provided.

## Content

# 1. Introduction

The last decades have witnessed a proliferation of optical micro/nanoresonators providing an unprecedented control over light-matter interaction on wavelength and subwavelength scales. This progress was made possible thanks to the development of bottom-up and top-down nanofabrication technologies, which allowed fabricating and assembling structures of varying shape, composition and size (Fig. 1). Optical resonators play an essential role in current developments in nanophotonics, such as optical metamaterials, integrated photonic circuits, optical sensing, and light tweezers, and find use in many areas of science and technologies, from DNA nanotechnologies to microfluidic devices, to quantum information processing.

Most previous reviews [Vah03,Shi07,Lal08,Bha09,Not10,Gia11,Agi12,Bia12] on micro-nanoresonators were mostly concerned with novel applications offered by light confinements at wavelength and subwavelength scales. In contrast, this Review focuses on basic concepts, essentially in relation with the description of the interaction of light with micro/nanoresonators via their resonant modes, in a similar manner as to what is done in integrated optics with guided and radiation modes. The approach is similar to that previously adopted in a perspective paper [Kr14a]. However, the present review is more comprehensive and detailed, and additionally reflects important progresses achieved in the domain in the past four years. The interaction of light with electromagnetic resonators fundamentally relies on the resonant modes of the structure. Under excitation by a pulse, the modes are initially loaded, and then release their energy by exponentially decaying in time. Consistently with the literature on non-conservative open systems described by non-Hermitian operators [Cha96], we will refer to these natural resonant modes as *quasinormal modes* (QNMs) hereafter. In early works on open optical microcavities, the prefix "quasi" has been used to emphasize that QNMs are the modes of non-conservative systems with complex frequencies, in contrast to normal modes that are the modes of conservative systems. The eigenmodes of open systems are also known in the literature as decaying states [Mor71], resonant states [Mor73,Mul10], and leaky modes [Sny83]. A QNM is a formal solution of differential equations with a complex eigenvalue and as such, are found in many areas of wave physics, e.g., gravitational waves [Kok99]. QNMs are intrinsic quantities (they are solutions of the homogeneous problem without source), which characterize the system independently of the excitation field.

Unfortunately, the complex eigenfrequencies cause mathematical difficulties with QNM normalization. For a long time, the literature has bypassed the difficulty by considering energy dissipation as a perturbation of an ideal closed and nonabsorbing system [Cha96]. In the initial conservative case, the system is Hermitian and admits a complete set of discrete normal modes with real eigenfrequencies. The perturbation just results in a broadening of the eigenstates with a Lorentzian line shape, but not in a change of the resonance frequencies and field distributions. Even though this phenomenological approach might work properly for resonators with small energy dissipation (i.e., a high quality factor), it becomes largely unsubstantiated for less confined resonances characterized by low quality factors, especially plasmonic resonances [Sau13]. Today the normalization issue is largely solved even for three-dimensional (3D) resonators made of dispersive materials and placed in a non-uniform background, and the first solvers to compute and normalize QNMs for the general case start to be available as open-source codes. It is one of the objectives of this Review to examine the progress made recently, so that we may precisely predict how QNMs are excited and focus our attention on the physics.

Each QNM is characterized by two main quantities that will be defined in detail in the following Sections and introduced intuitively for now: the mode volume $V$ and the quality factor $Q$. The former is related to the spatial extent of the electromagnetic mode and will be precisely defined in the review and the latter is proportional to the confinement time in units of the optical period. Thus, $Q$ and $1/V$ can be seen as the spectral and spatial energy density associated to the QNM, respectively. When trapped for sufficiently long times in a small space, photons strongly interact with the host material and can create significant nonlinear, quantum and optomechanical effects, to mention only a few [Gib85,Yok95]. An emblematic phenomenon is the modification of the spontaneous emission rate of atoms placed in resonance with the resonator mode (the Purcell effect) [Pur46].

There are essentially two types of microresonators in optics. In the first type (Fig. 1a), high $Q$'s are achieved with lossless dielectric optical materials, with mode volumes of the order of a cubic wavelength, essentially limited by the diffraction limit [Vah03,Not10]. Microscale volumes ensure that resonant frequencies are more sparsely distributed throughout the spectrum than they are in macroscale resonators, such as a Fabry-Perot cavity formed by two metallic mirrors separated by thousands of wavelengths. However, the physical mechanism of the confinement is essentially identical and the electromagnetic mode is a standing wave pattern composed of progressive waves that are bouncing back and forth between mirrors. Inside the resonator, the energy is then transferred every half period between the electric field energy $u_E = \frac{1}{2}\epsilon E^2$ and the magnetic field energy $u_H = \frac{1}{2}\mu_0 H^2$ (like in progressive waves), similarly to an oscillating mass on a spring, where the energy is transferred back and forth between the potential and the kinetic energies. Because the system size is larger than the diffraction limit, the energy is essentially conserved while transferred from one form to another, and self-sustaining oscillations with ultra-high $Q$'s are possible. Famous examples of high-$Q$ microresonators with $Q \sim 10^6$ are micropillar cavities, microtoroid resonators, photonic-crystal cavities [Vah03].

In the second type of resonators, the characteristic size is well below the diffraction limit, routinely by 2-3 orders of magnitude. In such small volumes, self-sustaining oscillations are no longer possible between the electric-field and magnetic-field energies. A tiny piece of dielectric material essentially behaves as a poor antenna that radiates energy rather than storing it. With metals, however, the energy can also be stored by the free electrons in the form of kinetic energy, $u_K$. Depending on the exact geometry, the energy balance $u_E = u_H + u_K$ can be restored in metallic nanoantennas [Bha09] at the surface-plasmon polariton frequency, typically a fraction of the plasma frequency of the metal. As a result, the diffraction limit can be beaten with the help from free carriers [Khu15]. This success however comes at a price: the energy stored in the form of kinetic energy is lost at a rate that is commensurate with the rate of scattering of electrons in metal, which is on the scale of 10 fs. Thus, as the optical mode becomes deeply sub-wavelength in all three dimensions, independent of its shape, the $Q$-factor is limited to $\sim 10$.

In sum, the field of optical resonances is mainly composed of two completely opposed geometries, dielectric microcavities with $Q \sim 10^6, V \sim \lambda^3$, and plasmonic nano-resonators with $Q \sim 10, V \sim 10^{-4}\lambda^3$, which lead to radically different situations, both with pros and cons. The paradigm of ultrahigh $Q$'s with huge spectral confinements is sometimes impractical because it also implies ultranarrow bandwidths with an incredibly strong accuracy of the tuning of the desired wavelength. Conversely, the opposite paradigm of ultimate confinement by plasmonics means detrimental dissipative loss. To mitigate these two extreme situations, it is also possible to marry dielectric resonators and plasmonics, by inserting plasmon antennas inside dielectric microresonators (Fig. 1c), thereby enabling intermediate $Q$'s and $V$'s [Ang10]. The present Review addresses the modeling of light interaction with these three different types of optical resonators.

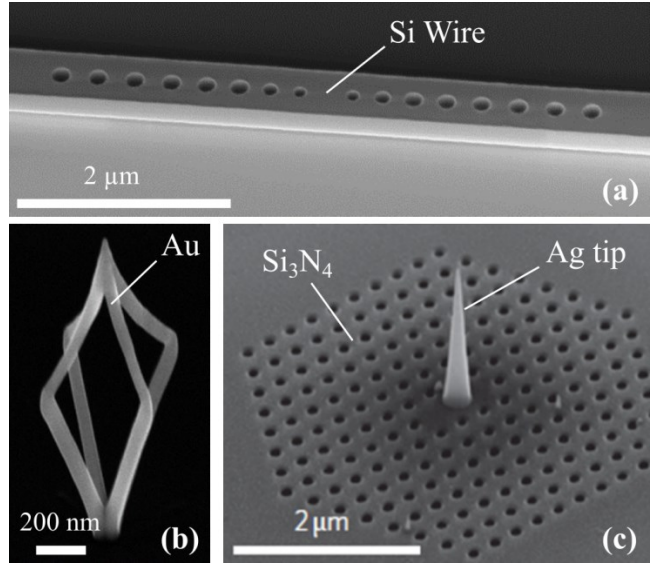

**Fig. 1 A few examples illustrating the diversity of optical micro-nanoresonators**. (**a**) Fabry-Perot microcavity with tapered Bragg mirrors etched in a Si rib-waveguide on SiO$_2$ [Lal08]. (**b**) Plasmonic crown fabricated by 3D focused electron beam induced deposition [Fow16]. (**c**) Hybrid plasmonic-photonic resonator formed with a silver tapered waveguide on a silicon photonic-crystal microcavity [Ang10].

There is an enormous literature on light interaction with resonances of electromagnetic systems and, in the vast majority of cases, the response is described via scattering theory with classical Maxwell's equation solvers operating either in the real-frequency domain, or in the time domain. There exists nowadays a wide number of good solvers, which have considerably progressed over the last decades, to numerically compute the electric field $\mathbf{E}_S(\mathbf{r}, t, \mathbf{e})$, or $\mathbf{E}_S(\mathbf{r}, \omega, \breve{\mathbf{e}})$ in the frequency domain, scattered by the resonator for an incident pulse $\mathbf{e}(\mathbf{r}, t)$ or an harmonic driving field $\breve{\mathbf{e}}(\mathbf{r}, \omega)$. However, it is certainly not a panacea.

For frequency-domain solvers, the entire computation has to be repeated for each individual frequency and for time-domain solvers, one needs to repeat the entire computation to predict the response if varying the excitation field, such as the pulse shape or duration, the polarization, the incidence angle… Even for simple micro-nanoresonators, the analysis is computationally heavy and some important complex geometries of contemporary optics, such as multi-resonator systems or hybrids combining resonances of different nature, are hardly modeled even with multi-core processors. Perhaps more importantly, interpretation and understanding of the underlying physics is not straight ahead with current electromagnetic solvers that rely on a brute-force discretization of Maxwell's differential equations and hide the beauty and the "simplicity" of the physics at hand, i.e., the excitation of a few modes.

The Review focuses on a completely different approach, which explicitly considers the natural modes, trying to reconstruct the optical response (at least in a compact subspace of $\mathbb{R}^3$) with a QNM expansion of the scattered field

$$\begin{bmatrix} \mathbf{E}_S(\mathbf{r}, \omega) \\ \mathbf{H}_S(\mathbf{r}, \omega) \end{bmatrix} = \sum_m \alpha_m(\omega) \begin{bmatrix} \tilde{\mathbf{E}}_m(\mathbf{r}) \\ \tilde{\mathbf{H}}_m(\mathbf{r}) \end{bmatrix}, \quad (1.1)$$

in the frequency domain, or assuming that the excitation pulse can be Fourier transformed,

$$\begin{bmatrix} \boldsymbol{E}_S(\boldsymbol{r},t) \\ \boldsymbol{H}_S(\boldsymbol{r},t) \end{bmatrix} = \text{Re}\left(\sum_m \beta_m(t) \begin{bmatrix} \widetilde{\boldsymbol{E}}_m(\boldsymbol{r}) \\ \widetilde{\boldsymbol{H}}_m(\boldsymbol{r}) \end{bmatrix}\right), \tag{1.2}$$

in the time domain. In Eqs. (1.1) and (1.2), $\widetilde{\boldsymbol{E}}_m(\boldsymbol{r})$ and $\widetilde{\boldsymbol{H}}_m(\boldsymbol{r})$ respectively denote the electric and magnetic field distributions of the $m^{\text{th}}$ QNM and the $\alpha_m$'s and $\beta_m$'s are the modal excitation coefficients, which describe how the QNMs are loaded and release their energy. We will give analytical formulas for the modal excitation coefficients as overlap integrals between the driving field $\check{\boldsymbol{e}}(\boldsymbol{r},\omega)$ or $\boldsymbol{e}(\boldsymbol{r},t)$ and the QNM field. Throughout the manuscript, we will use a tilde to differentiate the QNM fields from other fields, for instance the scattered or driving fields. Consistently, we will also use a tilde to denote the complex frequency $\widetilde{\omega}_m$ of the $m^{\text{th}}$ QNM, in contrast with the excitation frequencies that will be denoted by $\omega$ as it is conventionally. The intrinsic force of modal expansions is that they provide key clues towards understanding the physics of the resonator response. The latter is readily available and unambiguous since the resonance modes are explicitly considered, in sharp contrast with classical scattering theories.

The present period is marked by a deployment of QNM concepts in various applications, quantum plasmonics [Yan15,Ge15,Dez17a,Dez17b,Zha17,Fau17,Zam15], spectral filtering with diffraction gratings [Vi14b], energy loss spectroscopy in plasmonic nanostructures [Ge16], second-harmonic generation in metal nanoparticles [Ber16,Bra07], coupled cavity-waveguide systems [Kr14b,Kr17b,Fa17b], single-photon antennas [Fag15,Fei17], ultrafast-dynamics nanooptics [Fag17], scattering-matrix reconstruction in complex systems [Alp17], spontaneous emission at exceptional points [Pic17], wave transport in disordered media [Wan11,Pen14], spatial coherence in complex media [Sau14], mode hybridization and exceptional points in complex photonic structures [Van09,Bac14], random lasing [Tur06,And11,Goa11], light localization and cooperative phenomena in cold atomic clouds [Ski14,Bel14,Sch16], spatially nonlocal response in plasmonic nanoresonators [Dez17c], and thermal emission [Liu17,Li17b]. We discuss some of these numerous applications of QNM concepts in this Review and Fig. 2 summarizes a few.

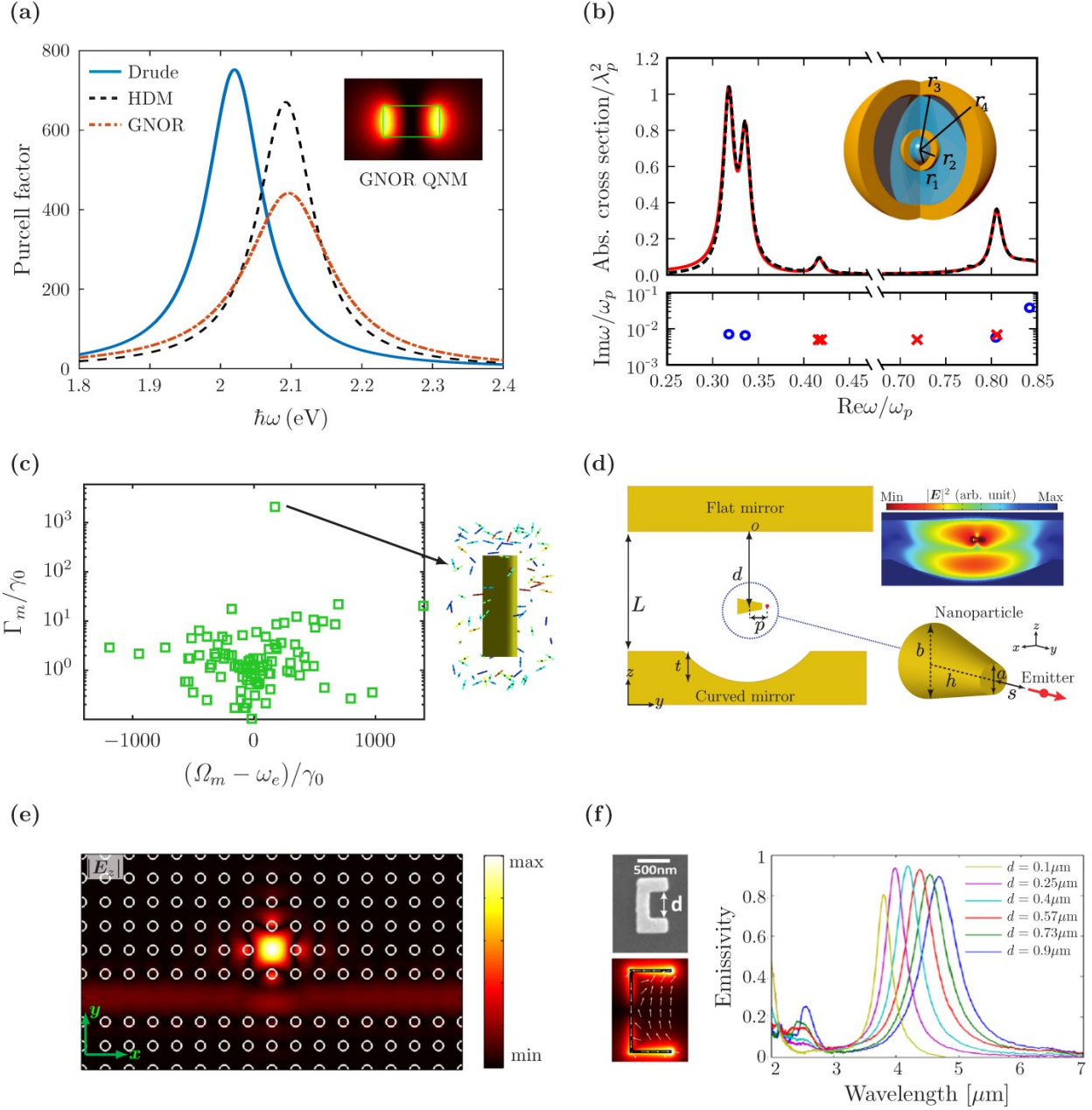

**Fig. 2. Examples of applications whose analysis benefit from QNM-expansion approaches. (a) Nonlocal plasmonics.** QNMs can be used to predict the nonlocal response of free electron gas on the Purcell factor of an emitter placed in the near-field of a gold nanorod. Predictions from two different nonlocal models — the hydrodynamic Drude model (HDM) and the generalized nonlocal optical response (GNOR)Model — are shown and compared to classical predictions obtained with a Drude model. The inset shows the electric field intensity of the nonlocal GNOR QNM. Adapted from [Dez17c]. **(b) QNM expansion of the scattering matrix.** (Upper panel) Absorption cross section of a multi-layered metallic-dielectric sphere in air, demonstrating the good agreement between the results obtained with Mie's scattering theory (dashed black curve) and a QNM-expansion formalism for the scattering matrix (solid red curve) computed with the QNM eigenfrequencies shown in the Lower panel. After

[Alp17]. **(c) Quantum hybrids.** Supperradiant and subradiant decay rates $\Gamma_m$ and energies $\Omega_m$ of a quantum hybrid formed by 100 molecules that are randomly distributed and oriented around a silver nanorod (diameter 30 nm, length 100 nm), predicted with the QNM formalism. $\gamma_0$ denotes the decay rate of every individual molecule in the vacuum. The left inset shows the superradiant state of the hybrid, with a large cooperativity involving more than one half of the molecules. After [Fau17]. **(d) Quantum-electrodynamic analysis.** Strong coupling of a quantum emitter to an hybrid plasmonic-photonic resonators formed by a metallic nanocone embedded in a Fabry-Perot microcavity leads to detuned hybrid-Fano QNMs (intensity distribution shown in the left inset). After [Gur07]. **(e) Coupled waveguide-cavity systems.** QNM theory can be used to study the leakage of photonic-crystal cavities into periodic waveguide mode and cavity resonances. After [Kr14b] and [Fa17b]. **(f) Thermal emission.** QNM theory can be used to design narrow-band and tunable metasurfaces emitting in the mid infrared with a near-unity emissivity. For resonant structures, the thermal emission power density can be formulated in terms of a single, dominant QNM shown in the bottom inset, with semi-analytical expression that can be used to ease the design and the optimization. After [Liu17] and [Li17b].

The Review is divided up into nine additional sections. Sections 2-5 present the important theoretical concepts: from the definition, computation and normalization of QNMs to the calculation of their excitation coefficients. Then, Sections 6-9 describe how QNM formalisms can be used and what they can offer in a few canonical situations of resonant light interaction. These Sections are less technical than the first ones and we intended to write them in such a way that they can be read independently of each other.

We provide in Section 2 a general definition of QNMs. For pedagogical reasons, we then describe in detail the case of a 1D Fabry-Perot resonator for which the QNMs can be computed analytically. The last part of the Section is devoted to the energy definition of the QNM quality factor, an important definition often disregarded by the literature.

Section 3 provides an overview of the different numerical methods used to compute the QNMs of micro/nanoresonators. For dielectric resonators, dispersion can be neglected in most cases and solvers are readily available. For the most general case of resonators with dispersive materials, QNMs are solution of a nonlinear eigenvalue problem and effective QNM solvers are presently under development. These developments are crucial to enable a widespread use of QNM formalisms and to benefit from the in-depth physical analysis that they provide. .

Once QNMs are calculated arises the delicate question of their normalization. This question, which we address in Section 4, has been the subject of major advances and vivid debates in recent years. We attempt to clarify the situation by providing an historical perspective and a summary of the different safe approaches available for normalizing QNMs. Additionally, we provide the derivation of an important formula on the response of resonators for driving frequencies $\omega \approx \widetilde{\omega}_m$, which highlights the key role of normalization in the definition of the modal excitation coefficients.

Section 5 reviews the two main approaches developed in the literature to expand the response of micro-nanoresonators as a sum over QNMs, the orthogonality-decomposition and the residue-decomposition approaches. We provide a clear link between the two approaches, highlight their similarities and differences and summarize the most important formulas for modal excitation coefficients in the frequency domain, the $\alpha_m$'s of Eq. (1.1). Sections 4 and 5 are carried out at a more technical level than the rest of the article. They can be skipped at first reading, but the main theoretical results that they summarize are worth being known.

Section 6 aims at providing a QNM interpretation of the classical experiment of how the fluorescence lifetime of a molecule coupled to an electromagnetic resonator depends on the exact position of the molecule inside the resonator. This leads us to revisit the classical concepts of local density of electromagnetic states and Purcell factor with QNMs and to provide an accurate definition for the mode volume of optical resonances.

We address in Section 7 the perturbation theory of an optical resonator. It allows predicting how the performance of a resonator is affected when a foreign object approaches it – a problem that has many applications in diverse areas, e.g. sensing. Cavity perturbation theory has been elaborated and used over many years, but has received a recent new birth thank to the use of a mathematically-correct QNM normalization which allows for a correct prediction of the resonance shift and broadening for the first time.

Section 8 presents a QNM description of the strong coupling regime – in line with Section 6 that was devoted to the so-called weak-coupling regime - for which new hybrid light/matter states are produced by mixing the initial states of the separated individual oscillators. These mixed states are forming building blocks for quantum information systems and for ultralow-power switches and lasers.

QNMs are also highly relevant to studies on light scattering by resonant particles and Section 9 is devoted to this important topic. Subsection 9.1 deals with light scattering by complex nanoparticles, Subsection 9.2 shows how QNM expansions may be applied in the time-domain to analyze the ultrafast temporal dynamics of optical nanoresonators, and Subsection 9.3 illustrates how QNM concepts may be used to analyze coherent phenomena in complex, disordered media.

Section 10 summarizes the Review and provides some perspectives for future works.

## 2. Quasinormal modes of electromagnetic resonators

Resonances occur with all types of vibrations or waves. They may cause spectacular events, such as the collapse of macroscopic structures, a famous example being the fall of the Broughton suspension bridge near Manchester in 1831. Resonances also play a pivotal role in the operation of musical instruments. Figure 3 shows the oscillogram recorded by a microphone placed inside a spherical Helmholtz resonator, initially knocked by the palm of one hand. For $t > 0.1$ s, one clearly sees the exponential decay of the fundamental resonance mode (the natural mode), which gradually dies away as it loses energy through viscous drag and sound radiation. Closer examination also shows an initial transient regime characterized by rapid oscillations that result from the excitation and decay of high-order resonances. The same wave physics occurs in electromagnetism and in other areas.

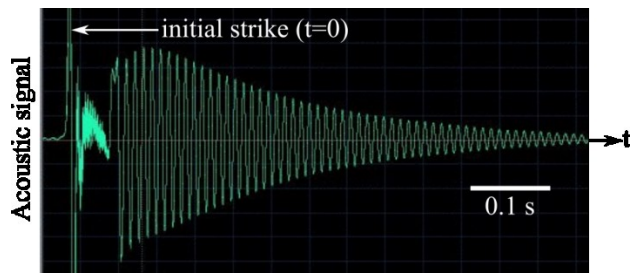

**Fig. 3. Illustration of the concept of QNM** with an acoustic spherical Helmholtz resonator with a volume of 3000 cm$^3$, a cylindrical neck of length 8 cm, and a cross-sectional area of 8.3 cm$^2$. Oscillogram recorded by a microphone placed inside the resonator, initially knocked by the palm of one hand. For $t > 0.1$ s, the acoustic signal oscillates at the resonance frequency and exponentially decays in time because the longest lived resonance, i.e., the fundamental natural leaky mode of the resonator, initially fed by the knock, gradually releases its energy through dissipation. The same wave physics occurs in other areas, e.g. electromagnetic fast pulses scattering with time-domain-type radars, photonics, and nanophotonics. After [html2].

In this Section, we define the electromagnetic modes of open, absorbing and dispersive systems. A system is said to be closed if the fields are confined to a finite region in space, for example in an enclosed microwave cavity, or more generally if the energy density is integrable. In contrast, it is said to be open if the fields are not strictly confined and leak to the whole universe. All optical cavities with some degree of output coupling belong to the latter category and we shall use the terms "open system" and "leaky cavity" interchangeably. When the leakage is small, concepts pertaining to closed systems can often be applied with a high degree of accuracy and a huge physical intuition. However, such approach is an approximation that fails as soon as the leakage cannot be neglected. This approximation of the low-leakage limit will be used in this review at some points for pedagogical reasons, but we start by the most general case that we introduce in Section 2.1. Then, in order to clarify the main issues arising in open systems, we describe in Section 2.2 the simple case of one-dimensional (1D) Fabry-Perot resonators, for which closed-form expressions for the eigenmodes can be derived. Finally, we provide in Section 2.3 an energy interpretation of one of the most important figures of merit of resonators, their quality factor.

**2.1 General definition**

Let us consider a damped mechanical, acoustic or electromagnetic resonator that is initially excited for $t < 0$ and then let free to evolve. Undergraduate physics textbooks tell us that the temporal response of the system will be an oscillation at the resonator eigenfrequency with an exponentially decaying envelope that characterizes the damping rate. This is illustrated in Fig. 3 with the time evolution of an acoustic Helmholtz resonator. In general, such a system however exhibits several resonances, with different oscillation frequencies and damping rates. In the frequency domain, these various resonances typically give rise to different peaks of varying widths, which may be observed in the spectrum of a measured quantity, such as a cross-section, the local density of states, or the absorption, see Fig. 4. As a consequence, one expects that the field in an electromagnetic resonator be written as a sum over the different resonant modes of the system, as $\boldsymbol{E}(\mathbf{r},t) = \mathrm{Re}\{\sum_m A_m(t)\widetilde{\boldsymbol{E}}_m(\mathbf{r})\exp(-i\Omega_m t)\exp(-\Gamma_m t/2)\}$, with $A_m$ the excitation coefficient, $\widetilde{\boldsymbol{E}}_m(\mathbf{r})$ the spatial mode profile, $\Omega_m$ the resonance frequency and $\Gamma_m$ the damping rate of mode $m$. The $\exp(-i\omega t)$ convention for time harmonic fields will be assumed throughout this Review. The modes of a leaky cavity are thus time-harmonic fields with a complex (angular) frequency

$$\widetilde{\omega}_m = \Omega_m - i\Gamma_m/2, \tag{2.1}$$

where the imaginary part gives the leakage rate. The latter can be easily related to the mode lifetime $\tau_m$ through the relations $\Gamma_m = 1/\tau_m$. In a spectrum, small (resp. large) values of $\Gamma_m$ therefore correspond to narrow (resp. broad) resonances.

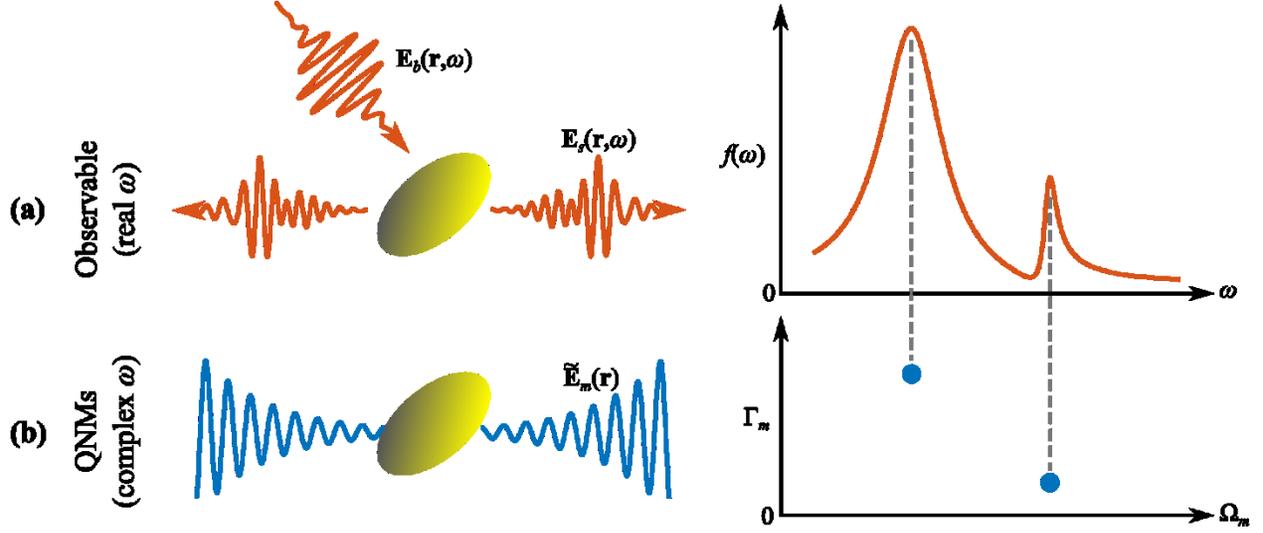

**Fig. 4. Concept of QNM to describe light interaction with resonant systems. (a)** Real frequency domain: An incident field $E_b(r, \omega)$ at real frequency $\omega$ interacts with an electromagnetic resonator, producing a scattered field $E_s(r, \omega)$. The spectrum of the optical response (an observable $f$, which may be the scattering efficiency, the Purcell factor, ...) typically exhibits several peaks or resonances that may overlap. **(b)** Complex frequency domain: The electromagnetic resonator possesses several leaky resonances, or QNMs, each of them being described by an electric field $\tilde{E}_m(r)$ and a complex frequency $\tilde{\omega}_m = \Omega_m - i\Gamma_m/2$ (blue dots), which are formally the eigenmodes and eigenvalues of Eq. (2.2) with outgoing-wave boundary conditions. An important feature of QNMs is the exponential divergence of their field distribution at $|r| \to \infty$. Knowledge of the QNMs field distribution, resonant frequency and decay rate, provides significant insight into the physical mechanisms underlying the optical response of resonant systems.

Formally, the resonances (i.e., the QNMs) of an open electromagnetic system are found by solving the time-harmonic source-free Maxwell's equations

$$\begin{bmatrix} 0 & i\varepsilon^{-1}(r,\tilde{\omega}_m)\nabla \times \\ -i\mu^{-1}(r,\tilde{\omega}_m)\nabla \times & 0 \end{bmatrix} \begin{bmatrix} \tilde{E}_m(r) \\ \tilde{H}_m(r) \end{bmatrix} = \tilde{\omega}_m \begin{bmatrix} \tilde{E}_m(r) \\ \tilde{H}_m(r) \end{bmatrix}, \quad (2.2)$$

where $\varepsilon(r, \tilde{\omega}_m)$ and $\mu(r, \tilde{\omega}_m)$ are the position and frequency-dependent permittivity and permeability tensors of the resonator and its surrounding background. Equation (2.2) takes the form of an eigenproblem with $\tilde{\omega}_m$ and $[\tilde{E}_m(r), \tilde{H}_m(r)]$ being its eigenvalues and eigenvectors, respectively. In open systems, the eigenmodes should also satisfy outgoing-wave boundary conditions to insure that the energy leaks away from the resonator.

Without restriction, the background can be free space or any non-uniform medium. The resonator can, for instance, lie on a substrate, be embedded in a stratified medium or be coupled to a waveguide. The materials composing the resonator and/or the background can be dispersive (in which case the eigenproblem becomes nonlinear). The permittivity and permeability in Eq. (2.2) are defined at the complex frequency $\tilde{\omega}_m$. This implies that, for dispersive materials, an analytic continuation in the complex frequency plane has to be known, although the material constants are measured at real frequencies. There exist various methods to infer the permittivity and permeability values at complex frequencies from data measured at real frequencies over a finite spectral range [Gar17]. Note that all dispersion models deduced from a microscopic description of charges motion (Drude, Lorentz) lead to an analytic form. In general, any dispersive material can be modeled with several Lorentz poles [Zha13].

The open character of the eigenproblem results in an unusual yet critical feature of QNMs, being that the field distributions $[\widetilde{\mathbf{E}}_m(\mathbf{r}), \widetilde{\mathbf{H}}_m(\mathbf{r})]$ should diverge for $|\mathbf{r}| \to \infty$, as illustrated in Fig. 4(b). Indeed, since the energy damping in time imposes a negative imaginary part for $\widetilde{\omega}_m$, see Eq. (2.1), a spherical outgoing wave of the form $\exp[-i\widetilde{\omega}_m(t - r/c)]/r$, as encountered in the far field of the resonator, grows as $\exp[\Gamma_m r/(2c)]/r$ because of the necessary minus sign of the propagation term $(t - r/c)$. One may already guess that this field divergence is problematic to define the QNM norm with some integral of its electromagnetic field over the whole space, as commonly done with the eigenmodes of Hermitian systems.

**2.2 QNMs of 1D Fabry-Perot resonators**

To gain insight into the properties of QNMs, it is instructive to investigate the case of a 1D Fabry-Perot resonator. Let us consider a slab of length $L$ and refractive index $n(\omega)$ embedded in a homogeneous medium of refractive index $n_1$. We will consider a dispersive slab $n(\omega)$ for the general presentation, but neglect the dispersion in the numerical example below for the sake of simplicity. In this system, the eigenmodes are formed by plane waves bouncing back and forth between the two interfaces, see Fig 5(a). We denote by $r$ and $t$ their reflection and transmission coefficients.

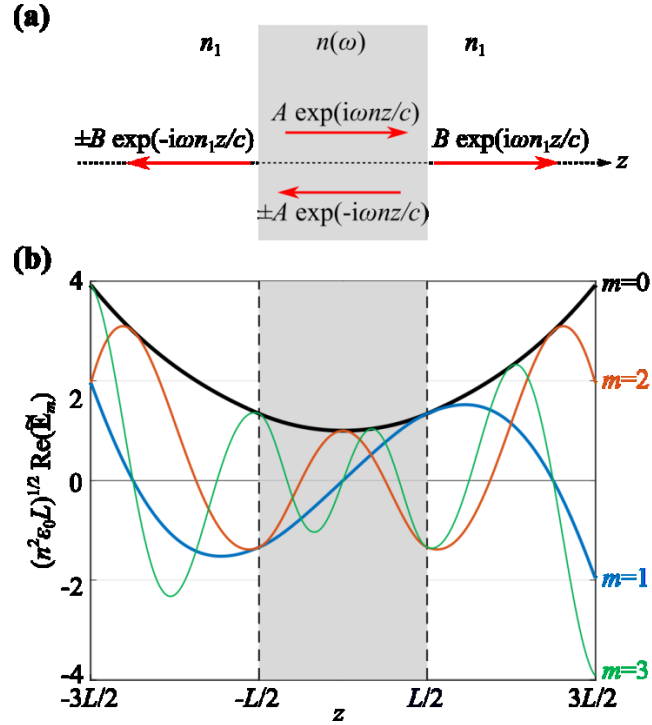

Fig. 5. (a) One-dimensional Fabry-Perot cavity of length $L$. The complex reflection and transmission coefficients of the mirrors are denoted by $r$ and $t$. The refractive index, $n(\omega)$, inside the cavity is assumed to be dispersive, and we denote by $n_g(\omega)$ the group index of the two counter-propagative plane waves that are bouncing back and forth between the two interfaces. The red arrows represent the plane waves that form the QNMs. (b) Field distribution profiles of the four lowest-frequency QNMs for $n = 1.5$ and $n_1 = 1$, inside the dielectric layer (gray area) and in the surrounding medium. Only the real part of the QNM electric field is represented and the QNMs are normalized. Symmetric and anti-symmetric modes correspond to even and odd $m$'s. A special feature of QNMs is that the field is exponentially diverging away from the resonator (here for $|z| > L/2$).

In the absence of an incident wave, the field at frequency $\omega = ck$ can be written as the superposition of two counter-propagating plane waves inside the cavity and two outgoing plane waves outside,

$$E(z) = \begin{cases} A\exp(iknz) \pm A\exp(-iknz), & \text{for } |z| < \frac{L}{2} \\ B\exp(ikn_1 z), & \text{for } z > \frac{L}{2} \\ \pm B\exp(-ikn_1 z), & \text{for } z < -\frac{L}{2} \end{cases}. \qquad (2.3)$$

The pluses (minuses) correspond to symmetric (anti-symmetric) modes. The field amplitude $B$ outside the cavity is related to the amplitude $A$ inside it by $B\exp(ikn_1 L/2) = tA\exp(iknL/2)$. The condition to build up a resonance is that the field amplitude $A$ should be recovered after one round trip in the resonator, $A = Ar^2 \exp(2iknL)$. Since the cavity loses energy via leakage ($|r|^2 < 1$) and/or absorption ($\text{Im}(n) > 0$), the resonance condition can be satisfied only for complex frequencies $\widetilde{\omega}_m = c\widetilde{k}_m = \Omega_m - i\Gamma_m/2$, which verify

$$1 - r^2(\widetilde{\omega}_m) \exp\left(2i \frac{\widetilde{\omega}_m}{c} n(\widetilde{\omega}_m) L\right) = 0. \qquad (2.4)$$

The imaginary part of the complex frequency has a clear physical role here, as it *restores a stationary state in the lossy cavity by amplifying the waves*, similarly to the amplified propagation in lasers, see Eq. (2.4). Inside the cavity, the fields of the symmetric and antisymmetric modes take the form of cosine and sine functions with argument $\frac{\widetilde{\omega}_m}{c} n(\widetilde{\omega}_m) z$ and are therefore complex-valued.

Outside the cavity, as expected, the field diverges exponentially. This divergence yields difficulties for the normalization of QNMs; this issue will be discussed in Section 4. Hopefully for our understanding at this early stage, the particular case of 1D Fabry-Perot resonator does not raise any divergence problem. Indeed, the electric and magnetic fields of the outgoing diverging plane wave outside the resonator are related through the vacuum impedance $\frac{\widetilde{E}_m(z)}{\widetilde{H}_m(z)} = \frac{1}{n_1}\sqrt{\frac{\mu_0}{\varepsilon_0}}$. This implies that $\varepsilon_0 n_1^2 \widetilde{E}_m^2(z) = \mu_0 \widetilde{H}_m^2(z)$ for $|z| > L/2$ such that the contribution of the uniform medium surrounding the resonator to the normalization integral of Eq. (4.5) in Section 4 is simply equal to zero thanks to the minus sign. By contrast with the QNMs of 3D resonators, for which the electric and magnetic contributions in the normalization integral outside the cavity do not compensate, the QNM normalization of 1D resonators only involves the *non-diverging* field in a finite domain between the two mirrors.

In the case of 1D Fabry-Perot resonators with dispersive materials, the $\widetilde{\omega}_m$'s can be found numerically by searching for the complex roots of Eq. (2.4). In the simple case of a dielectric slab with a frequency-independent refractive index $n$, one simply finds $\widetilde{\omega}_m \frac{nL}{c} = m\pi - \frac{i}{2}\log\left(\frac{1}{r^2}\right)$ with $r = \frac{n-n_1}{n+n_1}$ and $m \in \mathbb{Z}$. Note that an increase of $r$ (i.e., of the index contrast) yields a decrease of the imaginary part of the complex frequency, consistently with the fact that the light remains trapped in the cavity for longer times. Upon proper QNM normalization, the complex amplitude reads $A^s = \frac{\pm 1}{2n\sqrt{\varepsilon_0 L}}$ for the symmetric modes ($m$ even) and $A^a = \frac{\pm i}{2n\sqrt{\varepsilon_0 L}}$ for the antisymmetric modes ($m$ odd). Note that the fields of normalized QNMs are defined up to a plus/minus sign. Details on the QNM formalism for 1D non-dispersive Fabry-Perot resonators are given in Annex 1 for the interested reader.

Figure 5(b) shows the electric field distributions (real part only) of the four lowest-frequency QNMs ($m = 0, ..., 3$).

It is worth noting that (i) the $m = 0$ mode with $\Omega_0 = 0$ is not constant in space due to the non-zero imaginary part of the frequency (the field is a hyperbolic cosine function), and (ii) the exponential divergence is the same for all modes because the imaginary part of the frequency does not depend on $m$. In Section 5.4, we will show numerically that the QNM set is complete (inside the cavity) as it allows a perfect reconstruction of the field distribution in the cavity under plane-wave excitation.

## 2.3 Quality factor and energy balance

An important figure of merit of resonators is the quality factor or $Q$-factor. The latter is a dimensionless parameter that describes how damped a resonance is, or equivalently, characterizes a resonance bandwidth relative to its center frequency. A higher $Q$ indicates a lower rate of energy loss relative to the energy stored in the resonator; the oscillations of the temporal response die out more slowly and the resonator rings longer. A pendulum suspended from a high-quality bearing, oscillating in air, has a high $Q$, while a pendulum immersed in oil has a low one.

The $Q$-factor of a resonance is often defined as $2\pi$ times the ratio of the time-averaged energy stored in the cavity to the energy loss per cycle (or equivalently, the energy supplied by a generator, per cycle, to keep the signal amplitude constant) [Jac99],

$$Q_m = \Omega_m \frac{\text{Stored energy}}{\text{Power loss}}. \tag{2.5}$$

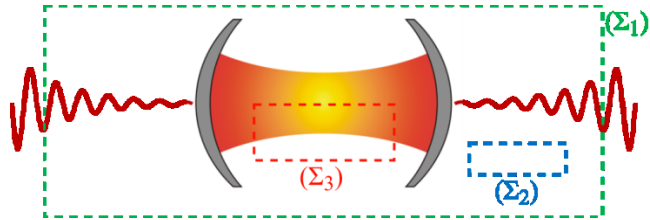

**Fig. 6.** Sketch of an electromagnetic resonator with its quasinormal mode. Because the QNM is a solution of Maxwell's equations for a complex frequency, its field distribution is a stationary wave inside the cavity and a diverging outgoing wave outside. The $Q$-factor of the mode can be derived from the field distribution in any closed surface that either surrounds the cavity (surface $\Sigma_1$), or is located entirely outside the resonator in the clad (surface $\Sigma_2$), or encloses part of the mode (surface $\Sigma_3$). However, because of the field divergence, the closed surface should enclose a finite volume.

The definition of Eq. (2.5) is based on an energy balance and it is instructive to relate it to the complex frequency using Poynting's theorem. Let us then consider one QNM $(\widetilde{\mathbf{E}}_m, \widetilde{\mathbf{H}}_m, \widetilde{\omega}_m)$ of an arbitrary resonator, which, by definition, is one solution to the source-free Maxwell's equations that satisfies outgoing wave boundary conditions. Conjugating Eq. (2.2), one shows that $(\widetilde{\mathbf{E}}_m^*, \widetilde{\mathbf{H}}_m^*, -\widetilde{\omega}_m^*)$ is another solution with the same permittivity and permeability distribution, since $\varepsilon^*(\mathbf{r}, \omega) = \varepsilon(\mathbf{r}, -\omega^*)$ and $\boldsymbol{\mu}^*(\mathbf{r}, \omega) = \boldsymbol{\mu}(\mathbf{r}, -\omega^*)$ due to the Hermitian symmetry of real Fourier transforms. This second solution also satisfies outgoing wave conditions, since its Poynting vector is identical to the first one, $\frac{1}{2}\text{Re}(\widetilde{\mathbf{E}}_m \times \widetilde{\mathbf{H}}_m^*)$ in both cases. For the sake of simplicity, we assume here that the resonator is made of isotropic materials but the following derivation can easily be extended to anisotropic materials.

Applying the divergence theorem to the vector $\widetilde{\mathbf{E}}_m^* \times \widetilde{\mathbf{H}}_m + \widetilde{\mathbf{E}}_m \times \widetilde{\mathbf{H}}_m^*$ on an arbitrary closed surface $\Sigma$ that defines a finite volume $V$ leads to

$$\iint_\Sigma \text{Re}(\tilde{\mathbf{E}}_m \times \tilde{\mathbf{H}}_m^*) \cdot d\mathbf{S} = - \iiint_V \left[\text{Im}[\tilde{\omega}_m \varepsilon(\tilde{\omega}_m)]\tilde{\mathbf{E}}_m \cdot \tilde{\mathbf{E}}_m^* + \text{Im}[\tilde{\omega}_m \mu(\tilde{\omega}_m)]\tilde{\mathbf{H}}_m \cdot \tilde{\mathbf{H}}_m^*\right] dV, \tag{2.6}$$

with the permittivities and permeabilities taken at the complex frequency $\tilde{\omega}_m$. This expression can be conveniently rewritten as

$$\tfrac{1}{2}\iint_\Sigma \text{Re}(\tilde{\mathbf{E}}_m \times \tilde{\mathbf{H}}_m^*) \cdot d\mathbf{S} + \tfrac{\Omega_m}{2} \iiint_V \left[\text{Im}[\varepsilon(\tilde{\omega}_m)]|\tilde{\mathbf{E}}_m|^2 + \text{Im}[\mu(\tilde{\omega}_m)]|\tilde{\mathbf{H}}_m|^2\right] dV =$$

$$\tfrac{\Gamma_m}{4} \iiint_V \left[\text{Re}[\varepsilon(\tilde{\omega}_m)]|\tilde{\mathbf{E}}_m|^2 + \text{Re}[\mu(\tilde{\omega}_m)]|\tilde{\mathbf{H}}_m|^2\right] dV. \tag{2.7}$$

Equation (2.7) is valid for any surface enclosing a finite volume. It may surround the whole cavity (surface $\Sigma_1$ in Fig. 6) or only part of it (surface $\Sigma_3$). It may even be located outside the physical resonator in the clad (surface $\Sigma_2$). The surface may potentially enclose absorbing dielectric materials or metals. However, if the surface $\Sigma$ encloses only non-dispersive materials ($\varepsilon$ and $\mu$ are frequency-independent), one readily sees that the left-hand side of Eq. (2.7) corresponds to the sum of the power dissipated by radiation (Poynting vector flux) and the power dissipated by absorption, while the volume integral in the right-hand side can be interpreted as the time-averaged electromagnetic energy stored in the volume $V$. Following Eq. (2.5), this interpretation of Eq. (2.7) directly leads to $Q_m = \Omega_m/\Gamma_m$, or

$$Q_m = -\frac{\text{Re}(\tilde{\omega}_m)}{2\,\text{Im}(\tilde{\omega}_m)}. \tag{2.8}$$

This reasoning is more delicate when the surface $\Sigma$ encloses dispersive materials but it can be made in the case when the variation in the permittivity and permeability of the materials is small over the bandwidth of the resonance. This assumption is valid with either high-Q resonances or weakly-dispersive materials. A first-order Taylor expansion of $\varepsilon$ and $\mu$ around the resonance frequency $\Omega_m$ yields $\varepsilon(\tilde{\omega}_m) = \varepsilon(\Omega_m) - i\frac{\Gamma_m}{2}\frac{\partial \varepsilon}{\partial \omega}\Big|_{\Omega_m}$ and similarly for $\mu$. Inserting these expressions into Eq. (2.7) and noting that $\text{Re}(\varepsilon) + \text{Re}(\omega)\text{Re}\left(\frac{\partial \varepsilon}{\partial \omega}\right) - \text{Im}(\omega)\text{Im}\left(\frac{\partial \varepsilon}{\partial \omega}\right) = \text{Re}\left(\frac{\partial \omega \varepsilon}{\partial \omega}\right)$, we obtain

$$\tfrac{1}{2}\iint_\Sigma \text{Re}(\tilde{\mathbf{E}}_m \times \tilde{\mathbf{H}}_m^*) \cdot d\mathbf{S} + \tfrac{\Omega_m}{2} \iiint_V \left[\text{Im}[\varepsilon(\Omega_m)]|\tilde{\mathbf{E}}_m|^2 + \text{Im}[\mu(\Omega_m)]|\tilde{\mathbf{H}}_m|^2\right] dV =$$

$$\tfrac{\Gamma_m}{4} \iiint_V \left[\text{Re}\left(\tfrac{\partial \omega \varepsilon}{\partial \omega}\right)\Big|_{\Omega_m} |\tilde{\mathbf{E}}_m|^2 + \text{Re}\left(\tfrac{\partial \omega \mu}{\partial \omega}\right)\Big|_{\Omega_m} |\tilde{\mathbf{H}}_m|^2\right] dV. \tag{2.9}$$

Compared to Eq. (2.7), we note that the material parameter values are taken at the resonance frequency $\Omega_m$ and the volume integral in the right-hand side corresponds to the time-averaged electromagnetic energy in absorbing and weakly-dispersive materials [Jac99]. This shows that the complex-frequency definition of the $Q$-factor, Eq. (2.8), derives from an energy balance also for weakly dispersive materials. In light of this first-order perturbative derivation, we remark that the second term in the left-hand side of the general expression, Eq. (2.7), also contains a fraction of the stored energy, and thus, cannot be identified as the power dissipated by absorption in dispersive materials. In sum, while Eq. (2.7) allows deriving the $Q$-factor from the field distribution in any closed surface, it does not strictly correspond to an energy balance.

To conclude this section, it is instructive to link Eq. (2.8) with the usual formula for the $Q$-factor of Fabry-Perot cavities. For that purpose, we solve the resonance condition given by Eq. (2.4) in the case of a symmetric 1D resonator with a dispersive material and dispersive mirrors. Since usual Fabry-Perot formulas are generally valid for large $Q$'s, it is convenient to neglect the absorption ($n$ is real) and assume that the mirror reflection coefficient $r$ is frequency-dependent via its argument only, such that one may write $r(\omega) = \sqrt{R}\exp[i\phi_r(\omega)]$. We further define the group index $n_g(\omega) = n(\omega) + \omega \frac{\partial n(\omega)}{\partial \omega}$ of the plane wave bouncing inside the cavity and the mirror penetration length $L_p(\omega) = \frac{c}{2n_g}\frac{\partial \phi_r(\omega)}{\partial \omega}$, which accounts for the dispersive nature of the phase of the reflection coefficient. A first-order Taylor expansion of Eq. (2.4) near the resonance frequency $\Omega_m$ straightforwardly leads to

$$\Omega_m = [m\pi - \phi_r(\Omega_m)]\frac{c}{n(\Omega_m)L}, \tag{2.10a}$$

$$\Gamma_m = (1-R)\frac{c}{n_g(\Omega_m)\bigl(L+2L_p(\Omega_m)\bigr)}, \tag{2.10b}$$

where we have assumed that $R$ approaches unity so that $\ln(1/R) \sim 1-R$. Equation (2.10a) sets the resonance wavelength and Eq. (2.10b) allows calculating the quality factor $Q_m$ associated to the $m$-th Fabry-Perot resonance

$$Q_m = \frac{\Omega_m}{c}\frac{n_g(\Omega_m)\bigl(L+2L_p(\Omega_m)\bigr)}{(1-R)}. \tag{2.11}$$

The expression reveals the three degrees of freedom that can be used to boost the $Q$'s of photonic crystal cavities [Lal08]: increasing the reflectivity $R$ of the individual mirrors, increasing the effective cavity length $(L+2L_p)$, and slowing down the speed $c/n_g$ of the waves bouncing back and forth between the mirrors. Let us note that the textbook expression $Q_m = \frac{\Omega_m}{c}\frac{nL}{(1-R)}$ [Yeh88] is recovered in the limits where $n$ and $\phi_r$ do not vary with the frequency. In the presence of absorption (complex refractive index $n + i\kappa$), the reflectivity $R$ has to be replaced by an effective reflectivity that takes into account the energy absorbed during to one half round trip inside the cavity, $R_{\text{eff}} = R\exp\left(-2\frac{\omega}{c}\kappa L\right)$.

## 3. Overview of QNM solvers

This Section aims at a simplified presentation of the different methods and software that have been developed in the literature. Our goal is not to dress an exhaustive list but rather to categorize the main methods in three groups. We skip the details and complexity that make computational tools so effective.

QNMs are solutions of the source-free Maxwell's equations, Eq. (2.2), with eigenfrequencies $\widetilde{\omega}_m$ and eigenvectors $\bigl[\widetilde{\boldsymbol{E}}_m(\boldsymbol{r}), \widetilde{\boldsymbol{H}}_m(\boldsymbol{r})\bigr]$, which satisfy outgoing-wave boundary conditions, i.e., the energy leaks away far from the resonator. For nondispersive materials with frequency-independent permittivities and permeabilities, Eq. (2.2) defines a linear eigenvalue problem and various mode solvers, including commercial ones such as COMSOL Multiphysics [COMSOL], are available to compute QNMs very efficiently with iterative methods, e.g. Arnoldi-Lanczos iterative algorithm. We focus hereafter on the general case of resonators made of *dispersive materials*, for which Eq. (2.2) defines a *nonlinear eigenvalue problem*. We review three approaches that are widely used to compute the QNMs of dispersive resonators.

**Time-domain approach.** Figure 3 provides a simple picture of how computing QNMs with time-domain solvers. If the QNM of interest present the longest decay, then it will be the dominant field in the resonator at long times after an initial short excitation. This means that one may compute long-lived QNMs quite easily using time-domain approaches. So far, the finite-difference time-domain (FDTD) methods have been mainly used [Taf13,Ge14b,Kr14a] to compute the fundamental QNMs of photonic and plasmonic resonators. Technical difficulties may arise when several resonances overlap in space and frequency [Kr14a], as often happens for low-Q plasmonic systems, implying that efficient filtering techniques should be implemented to recover every individual QNMs in the temporal signal.

**Pole-search approach.** An alternative to time-domain methods uses the fact that the resonator response to any driving field diverges as the driving frequency approaches a QNM eigenfrequency. The QNMs can thus be computed by searching poles in the complex frequency plane of some representative analytic quantity, such as the electromagnetic field response to a driving source [Bai13] or the determinant of the discretized scattering matrix [Pow14,Zhe14]. Some iterative algorithms, such as the Newton method, are well suitable for the pole searching [Kra00]. They usually require an initial guess value as close as possible to the actual QNM eigenfrequency for good numerical efficiency, and compute QNMs one by one by iteratively exploring the complex plane around the QNM pole. A single pole-search QNM-solver freeware to compute and normalize QNMs for arbitrary resonator geometries is available [Bai13]. The freeware that relies on fitting the resonator response to a Padé approximated pole-response function can be used with any frequency-domain electromagnetic solver, including commercial softwares such as COMSOL Multiphysics. It is particularly relevant when only a few QNMs need to be computed. Alternatively, a non-iterative method, the so-called Cauchy Integration Method, has also been developed [Zol05,Del67]. Interestingly, this method is capable to find all the poles in a closed predetermined region of the complex frequency plane, but it needs an extra computational cost associated with the contour integration over the outer boundary of the closed region [Byk13,Pow14]. For a better accuracy, the poles found with the Cauchy Integration Method can be further refined with the Newton's method [Pow14].

**Auxiliary-field eigenvalue approach.** A radically different approach consists in computing all the QNMs at one time by solving a linearized version of the eigenvalue problem, for which a myriad of efficient and stable numerical methods exist. A general approach consists in transforming the nonlinear eigenvalue problem into a linear one by introducing auxiliary fields to account for material dispersion. Several variants exist and, for the sake of simplicity, we limit ourselves to a generic presentation of auxiliary-field techniques [Jos91,Ram10,Zha13,Luo10,Che06,Taf13,Zi16a,Zi16b]. The latter have been initially used in the time-domain for modelling wave propagation in dispersive media [Jos91,Taf13] and then in the frequency domain for computing band diagrams of dispersive photonic crystals [Ram10,Zha13,Luo10,Che06].

For the sake of illustration, let us consider a medium with a dispersive permittivity described by the single-pole Lorentz model, $\boldsymbol{\varepsilon}(\omega) = \boldsymbol{\varepsilon}_\infty \left(1 - \frac{\omega_p^2}{\omega^2 - \omega_0^2 + i\omega\gamma}\right)$, and a nondispersive permeability $\mu = \mu_0$. We introduce two auxiliary fields, the polarization $\boldsymbol{P} = -\boldsymbol{\varepsilon}_\infty \frac{\omega_p^2}{\omega^2 - \omega_0^2 + i\omega\gamma} \boldsymbol{E}$ and the current density $\boldsymbol{J} = -i\omega\boldsymbol{P}$. With elementary algebraic manipulations, we can reformulate Eq. (2.2) into an extended eigenvalue problem

$$\begin{bmatrix} 0 & -i\mu_0^{-1}\nabla\times & 0 & 0 \\ i\boldsymbol{\varepsilon}_\infty^{-1}\nabla\times & 0 & 0 & -i\boldsymbol{\varepsilon}_\infty^{-1} \\ 0 & 0 & 0 & i \\ 0 & i\omega_p^2\boldsymbol{\varepsilon}_\infty & -i\omega_0^2 & -i\gamma \end{bmatrix} \begin{bmatrix} \widetilde{\boldsymbol{H}}_m(\boldsymbol{r}) \\ \widetilde{\boldsymbol{E}}_m(\boldsymbol{r}) \\ \widetilde{\boldsymbol{P}}_m(\boldsymbol{r}) \\ \widetilde{\boldsymbol{J}}_m(\boldsymbol{r}) \end{bmatrix} = \widetilde{\omega}_m \begin{bmatrix} \widetilde{\boldsymbol{H}}_m(\boldsymbol{r}) \\ \widetilde{\boldsymbol{E}}_m(\boldsymbol{r}) \\ \widetilde{\boldsymbol{P}}_m(\boldsymbol{r}) \\ \widetilde{\boldsymbol{J}}_m(\boldsymbol{r}) \end{bmatrix}, \qquad (3.1)$$

which is linear. The approach can be straightforwardly extended to multiple-pole Lorentz models by increasing the number of auxiliary fields. Note that the Drude model, a particular case for which $\omega_0 = 0$, requires a single auxiliary-field. QNM eigensolvers based on auxiliary fields method have been initially implemented with finite-difference methods [Zi16a,Zi16b]. The latter may introduce inaccuracies for complex geometries, which may lead to the prediction of spurious modes when discretizing curved metallic surfaces on a rectangular grid for instance. QNM eigensolvers based on finite element methods are thus preferable. In [Yan17], a general freeware using the COMSOL Multiphysics platform was developed, and numerical tests indicate that a high accuracy is achieved even for complex nanoantenna geometries placed in non-uniform environments.

Finally, let us note that, for dispersive materials, the pole-search and auxiliary-field eigenvalue rely on an analytic continuation of $\boldsymbol{\varepsilon}(\omega)$ and $\boldsymbol{\mu}(\omega)$ at complex frequencies. This stringent requirement is usually met by using physical models, which provide fully analytic expressions for the material parameters, as for instance the Drude model for free carriers in metals or highly-doped semiconductors. Alternatively, one may fit material parameters measured at real frequencies to ad hoc analytic expressions, such as multiple-pole Lorentz expansions [Zha13,Pow17,Ram10], which guaranty causality and Hermitian symmetry. Systematic and effective procedures for fitting experimental data exist, see for instance [Gar17] that uses Hermitian functions in the form of polynomial fractions. Similar considerations applied to the time-domain approaches for which a model of the frequency-dependent permittivity is also needed for analyzing dispersive media [Jos91,Taf13].

**Numerical methods.** For resonators with complex shapes and materials, analytic solutions are not available and the continuous Maxwell's operators of Eqs. (2.2) or (3.1) have to be approximated and transformed by a discretization scheme, which satisfies the outgoing-wave condition at $|\boldsymbol{r}| \to \infty$. A first option is to calculate the QNMs from a Fredholm-type integral equation [Las13,Ber16,Pow17,Pow14,Zhe14], in which case the outgoing-wave condition is perfectly fulfilled by construction. However, since the QNM resonance frequencies (i.e., the unknown) enters the outgoing-wave condition, the integral equation is a nonlinear problem and the computation of the complex eigenfrequencies should be performed with care, even for the simple case of resonators placed in a uniform background.

An alternative way is to surround the resonator by perfectly-matched layers (PMLs). This solution offers a precious advantage. The new eigenstates do not grow exponentially away from the resonator; they are even exponentially damped inside the PMLs. Thus the modes become square-integrable and easy to normalize, see Section 4 for details. QNMs of PML-truncated spaces have been first computed with the pole-search approach. Virtually any frequency-domain method may be used, such as the finite-element method (FEM) [Bai13], the finite-difference method [Zi16a], the Fourier modal method [Lal04,Arm08].

Whatever the solution chosen to fulfil the outgoing-wave boundary conditions, in practice, the original continuous operator in Eqs. (2.2) or (3.1) is discretized in a finite space and becomes a discrete operator (i.e., a matrix with a finite dimension). The latter always has a discrete eigenvalue spectrum and, if linear as in the case of Eq. (3.1), its eigenmodes form a complete set in the whole finite computational space [Yan17], see Section 5.3 for numerical examples. Because of discretization inaccuracies, however, the discretized operator represents an accurate version of the actual operator only for a finite spectral interval $\mathcal{F}$ around some operating frequency, see Fig. 7(a). This has two important consequences on the numerical spectrum. First, only a sub-set of the true QNMs of the continuous operator is accurately recovered, approximately the sub-set composed of states with $\mathrm{Re}(\widetilde{\omega}_m) \in \mathcal{F}$. Second, a new set of numerical modes, which are different from the actual QNMs, appears in the numerical spectrum. These modes should not be considered as spurious modes since they contribute to form the complete finite basis of the discretized operator.

For numerical methods which rely on PMLs to fulfil the outgoing-wave conditions, the numerical modes are often called PML modes [Vi14a,Vi14b,Yan17]. There are two types of PML modes. The first ones behave like Fabry-Perot modes formed by waves that are bouncing back and forth between the outer PML boundaries, thereby depending strongly on the PML parameters. Their eigenfrequencies usually trace around a straight line in the complex frequency plane [Vi14a], see Fig. 7. Note that, in the limiting case where the PML becomes infinitely large (i.e., the unmapped domain is negligible), these modes recover the continuum modes of the PML medium. The second type of PML modes originates from the modes, whose outgoing waves are not sufficiently damped in the PML. These modes are strongly affected by the PML and fail to accurately approximate the true QNMs lying outside the spectral interval $\mathcal{F}$ in Fig. 7(a).

Figure 7(b) shows the frequencies and decay rates of the eigenmodes computed for a bowtie silver nanoantenna in air. The computation is performed with a freeware relying of the auxiliary-field method and implemented in the COMSOL environment [Yan17]. The two subsets of QNMs and PML modes are clearly identified. As shown in the bottom panel, note that a typical feature of PML modes is that their field intensities are dominantly located in the PML.

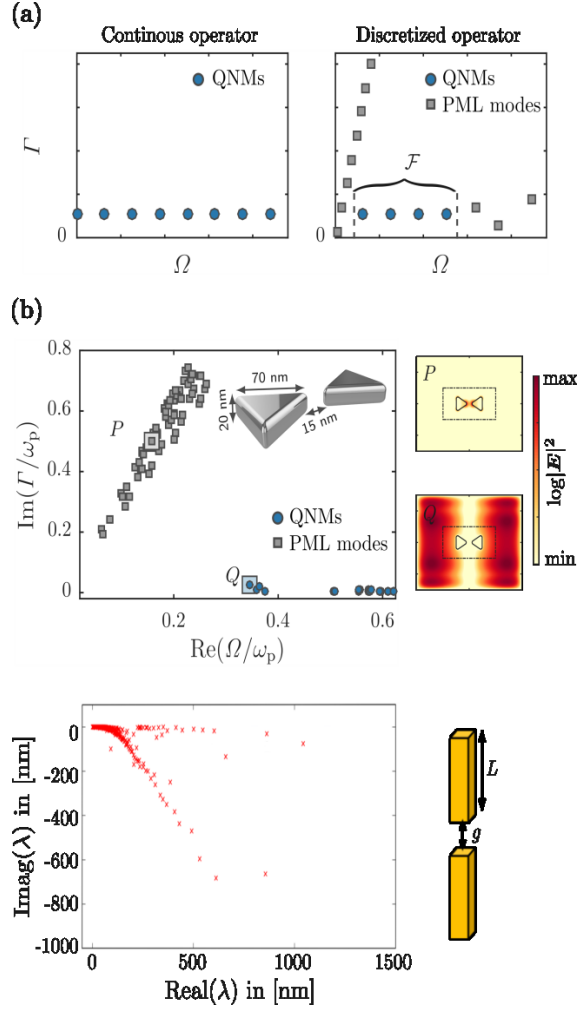

**Fig. 7. QNMs and PML modes of dispersive nanoantennas computed with auxiliary-fields.** In general, QNM eigenfrequencies have small imaginary parts and the PML-modes are arranged on a tilted "line" in the complex plane and the angle of the tild depends on the PML coefficients. **(a)** Difference between the eigenspectra of the actual continuous operator (left) and the discretized operator defined on a finite domain bounded by PMLs (right). In general, if the discretization is accurate and the outgoing-wave condition is faithfully matched over a broad spectral interval around the operating frequencies, and the spectrum of the discretized operator offers a relevant representation that captures all the important QNMs involved in the resonator dynamics in the spectral interval $\mathcal{F}$. Only QNMs with $\Omega > 0$ are represented for the sake of clarity. **(b)** Distribution of the eigenstate energies and decay rates computed with finite-element (top, adapted from [Yan17]) and finite-difference (bottom, adapted from [Zi16b]) QNM solvers. The bowtie is placed in air and is made of a Drude metal with $\lambda_p = 2\pi c/\omega_p = 138$ nm and $\gamma = 0.0023\omega_p$. Its parameters are given in the inset. Intensity distributions of the normalized QNM labelled "Q" and the PML-mode labelled "P" are also shown. The dash lines represent the air-PML interfaces. A golden rectangular nanorod dimer is considered for the finite-difference computation. The cross-section of the rods is 28 nm×28 nm, the rod length is $L = 100$ nm, and the gap is $g = 22$ nm. A Drude model is used as a constitutive relation for gold, and the background medium is homogeneous with an index of refraction of 1.5.

## 4. QNM normalization

The derivation of analytical expressions for the excitation coefficients $\alpha_m(\omega)$ or $\beta_m(t)$ is concomitant with the delicate question of QNM normalization. Normalization is also directly related to important physical quantities such as the mode volume. As discussed in Section 2, since QNMs decay in time and satisfy outgoing-wave boundary conditions, they exponentially diverge in space at large distances from the resonator. *QNMs are no longer of finite energy and standard normalizations based on energy consideration therefore cannot be applied*. Let us first emphasize that, because of the presence of absorption, textbooks tell us that unconjugated products of the form $\widetilde{\boldsymbol{E}}_m \cdot \widetilde{\boldsymbol{E}}_m$ instead of $\widetilde{\boldsymbol{E}}_m \cdot \widetilde{\boldsymbol{E}}_m^*$ should be used for the normalization [Sny83,Mor53, Har93]. This being said, *the key issue is the need to cope with open systems and diverging fields*: the integral of $\widetilde{\boldsymbol{E}}_m \cdot \widetilde{\boldsymbol{E}}_m$ diverges as the integration domain tends to infinity. The real challenge is to define in what sense the integral can be properly defined with a finite value, and how it may be computed for the general case of resonators with complex geometries that require numerical computations for their analysis.

After early works on 1D optical cavities in the 90's [Leu94a,Leu94b], the normalization issue stayed dormant for a long time. Then, in the last five years, different authors addressed independently the problem for 3D plasmonic and photonic resonators [Kri12,Sau13,Bai13,Ge14,Vi14a,Doo14,Mu16b]. These works led to four different approaches to normalize QNMs, which are summarized in Section 4.3. Since these normalization methods have been vividly debated in the recent literature [Kri15,Sau15,Doo16,Mu16b,Mul17,Kr17a], we attempt hereafter to help the interested reader into the winding path of QNM normalization. Works comparing the mathematical soundness and numerical performance of the different normalization methods are rare [Kri15,Sau15,Mu16b,Mul17,Kr17a], difficult to perform because the subject is technical, and not empty of mistakes. The objective of this Section is not to go into tough mathematical details, but rather to set the different approaches in a historical perspective and to discuss their similarities and differences.

### 4.1 Historical perspective and recent debates

The majority of early works on optical resonators has focused on systems composed of non-magnetic materials, $\boldsymbol{\mu}(\boldsymbol{r}) = \mu_0$. Except when stated otherwise, this particular case is considered in the following discussion to ease the comparison between different works. In addition, this choice results in simpler formulas since the magnetic field can be eliminated from the equations.

Initial works on QNM normalization considered 1D Fabry-Perot cavities and proposed the following expression for the norm [Leu94a,Leu94b]

$$\langle \widetilde{\boldsymbol{E}}_m | \widetilde{\boldsymbol{E}}_m \rangle_{1D} = \int \widetilde{\boldsymbol{E}}_m \cdot \left( \varepsilon + \frac{\partial \omega \varepsilon}{\partial \omega} \right) \widetilde{\boldsymbol{E}}_m dx, \qquad (4.1)$$

where the integral runs over the whole space, from $x = -\infty$ to $x = +\infty$ and the derivative is taken at the QNM frequency $\widetilde{\omega}_m$. To avoid the divergence of the integral in Eq. (4.1) and obtain a well-defined value, a proper integration contour has to be chosen in the complex plane. The path of integration has to go first along the real axis and then along a line making a positive angle with the real axis [Leu94a], with a sufficiently large angle so that the exponentially growing fields is transformed into exponentially damped fields. If one properly specifies the integration domain, the integral in Eq. (4.1) becomes the sum of two contributions

$$\langle \widetilde{\boldsymbol{E}}_m | \widetilde{\boldsymbol{E}}_m \rangle_{1D} = \int_\Omega \widetilde{\boldsymbol{E}}_m \cdot \left( \varepsilon + \frac{\partial \omega \varepsilon}{\partial \omega} \right) \widetilde{\boldsymbol{E}}_m dx + \int_{\Omega_c} \widetilde{\boldsymbol{E}}_m \cdot \left( \varepsilon + \frac{\partial \omega \varepsilon}{\partial \omega} \right) \widetilde{\boldsymbol{E}}_m dx, \qquad (4.2)$$

with $\Omega$ a real finite domain and $\Omega_c$ a complex domain that goes to infinity. Equation (4.2) corresponds to the primary definition of the QNM norm proposed in [Leu94a,Leu94b] and related works by the same authors. The latter also proposed an alternative definition that avoids the analytic continuation in the complex $x$-plane. Indeed, for 1D systems, the field outside the resonator is simply a plane wave. The integral in the complex infinite domain $\Omega_c$ can thus be easily calculated by hand; it is simply proportional to the field squared at the border $x_0$ of the domain $\Omega$. The QNM norm can thus be rewritten as [Leu94a]

$$\langle \widetilde{\boldsymbol{E}}_m | \widetilde{\boldsymbol{E}}_m \rangle_{1D} = \int_\Omega \widetilde{\boldsymbol{E}}_m \cdot \left( \varepsilon + \frac{\partial \omega \varepsilon}{\partial \omega} \right) \widetilde{\boldsymbol{E}}_m dx + i \frac{\varepsilon_0 c n}{\widetilde{\omega}_m} \widetilde{\boldsymbol{E}}_m(x_0)^2, \qquad (4.3)$$

where $n$ is the refractive index of the background medium around the resonator, assumed to be non-dispersive for simplicity. A detailed demonstration of the equivalence between Eqs. (4.2) and (4.3) is provided in [Lee09].

The authors in [Kri12] were the first to introduce an analogue of Eq. (4.3) for 3D photonic-crystal geometries. Referring to the analysis made in [Leu94a] for 1D resonators, they first proposed to transform the $\widetilde{\boldsymbol{E}}_m(x_0)^2$ term in Eq. (4.3) into a surface integral over the border of the 3D domain $\Omega$ [Kri12]. They further extended their proposal to 3D plasmonic resonators [Ge14]. It seems, however, that the difficulty raised when moving from 1D to 3D was overlooked. If the equivalence between integrating in the complex plane [Eq. (4.2)] and adding a surface term [Eq. (4.3)] is clear in 1D [Lee09], the situation is much more complex in 3D. In particular, as shown in [Mu16b,Mul17], the limit of the integral in Eq. (4.3) does not exist in 3D as the integration domain tends to infinity. This result has been vividly debated in [Kri15,Kr17a], where the authors argued that, although the norm oscillates and diverges as a function of the integration domain size, good numerical accuracy may be achieved in practice. This issue was also addressed in [Sau15] by evidencing that Eq. (4.3) or related expressions based on a surface integral over the border of the domain are incorrect for resonators placed on a substrate. The authors contend that a better option to is to generalize Eq. (4.2) instead of Eq. (4.3). As we shall now see, safe expressions f or the QNM norm fortunately do exist as well.

The 1D QNM norm of Eq. (4.2) with a path of integration in the complex plane has been generalized in [Sau13] for the most general case of 3D resonators, possibly located over a substrate and possibly connected to open (non-periodic) waveguides. Taking advantage of the decisive work performed on perfectly matched layers (PMLs) for solving numerically Maxwell's equations, it was proposed to consider a mapped space in which a finite domain $\Omega$ of the real space is surrounded by PMLs in a surrounding domain $\Omega_{PML}$ [Sau13]. If the PML damping factor is large enough, the exponentially growing field of the QNM is transformed into an exponentially damped field. For ideal PMLs with an infinite thickness, the outgoing-wave boundary conditions are strictly maintained and the complex frequencies of the initial QNMs of the actual open system are fully retrieved into the new analytically continued set. In other words, the spectra of the original and mapped operators are identical. Within this approach, the norm is defined as [Sau13]

$$\langle \widetilde{\boldsymbol{E}}_m | \widetilde{\boldsymbol{E}}_m \rangle_{PML} = \iiint_\Omega \widetilde{\boldsymbol{E}}_m \cdot \left( \varepsilon + \frac{\partial \omega \varepsilon}{\partial \omega} \right) \widetilde{\boldsymbol{E}}_m d^3\boldsymbol{r} + \iiint_{\Omega_{PML}} \widetilde{\boldsymbol{E}}_m \cdot \left( \varepsilon + \frac{\partial \omega \varepsilon}{\partial \omega} \right) \widetilde{\boldsymbol{E}}_m d^3\boldsymbol{r}, \qquad (4.4)$$

where the whole integration domain $\Omega \cup \Omega_{PML}$ comprises the real finite domain $\Omega$ and the PML domain $\Omega_{PML}$. It has been shown that the norm defined by Eq. (4.4) is *independent of the permittivity and permeability parameters used to implement the PML domain* and thus defines a unique normalization [Sau13]. It is noteworthy that the field inside the PML is crucial for the normalization.

Equation (4.4) can easily be extended to resonators composed of dispersive magnetic materials, so that the most general expression for the QNM norm becomes [Sau13]

$$\langle \widetilde{E}_m | \widetilde{E}_m \rangle_{PML} = \iiint_{\Omega \cup \Omega_{PML}} \left[ \widetilde{E}_m \cdot \frac{\partial \omega \varepsilon}{\partial \omega} \widetilde{E}_m - \widetilde{H}_m \cdot \frac{\partial \omega \mu}{\partial \omega} \widetilde{H}_m \right] d^3 r = 1. \qquad (4.5)$$

Except when stated differently, the QNMs discussed in all following Sections of this Review are normalized according to Eq. (4.5). A few remarks are worth being made: 1/ The integrand in Eq. (4.5) has been used for a long time to normalized the QNMs of lossy but closed cavities, for which the divergence issue at large distance is inexistent, see chapter 9 in [Har93]; 2/ the derivatives in Eq. (4.5) are taken at the QNM frequency $\widetilde{\omega}_m$; 3/ perhaps counterintuitively, they do not result from a Taylor expansion of $\varepsilon$ and $\mu$, so that the norm is exact even for strongly dispersive materials; 4/ the dimension of a normalized QNM electric field is not that of a classical electric field, but $L^{-1}C^{-1}$, so that the quantity $\varepsilon_0 \widetilde{E}_m^2$ has the same dimension as the inverse of a volume.

The proposal to use PMLs to define the QNM norm is more than a simple generalization of Eq. (4.2) as it addresses the important issue of how to implement the analytic continuation in the complex coordinate plane for complex geometries that require numerical computations. Interconnecting theoretical and computational aspects by explicitly using PMLs in theoretical derivations [Sau13] constitutes a decisive step towards practical and efficient QNM solvers for photonic and plasmonic resonators with arbitrary geometries.

**4.2 Resonator spectral response close to the resonance**

A totally different approach to normalize a QNM has been recently proposed in [Bai13]. Its principle is based on the calculation of the response of a resonator driven at a frequency close to the resonance. This calculation is also instructive to appreciate the physical relevance of the QNM normalization.

Let us use the Lorentz reciprocity formula [Eq. (A3-3)], which relates two solutions of Maxwell's equations and whose derivation can be found in Annex 3. For the first solution $[E_1, H_1]$, we consider the scattered field $[E_S, H_S]$ that is solution to Maxwell's equations at real frequency $\omega$ for a current-source distribution $J_1 = -i\omega \Delta \varepsilon(r, \omega) E_b(r)$, see Annex 2. For the second solution $[E_2, H_2]$, we consider the QNM labeled $m$ that is solution to the source-free Maxwell's equations at complex frequency $\widetilde{\omega}_m$. To avoid the problematics associated with the exponentially diverging QNM field for $|r| \to \infty$, we apply the Lorentz reciprocity formula to the regularized space obtained by surrounding the resonant system with PMLs. Under this condition, the surface integral of the Lorentz reciprocity formula in Eq. (A3-3) is equal to zero and we obtain an important relation linking the scattered field to every individual QNM

$$\iiint_{\Omega \cup \Omega_{PML}} \left[ E_S \cdot \left( \omega \varepsilon(\omega) - \widetilde{\omega}_m \varepsilon(\widetilde{\omega}_m) \right) \widetilde{E}_m - H_S \cdot \left( \omega \mu(\omega) - \widetilde{\omega}_m \mu(\widetilde{\omega}_m) \right) \widetilde{H}_m \right] d^3 r =$$

$$-\omega \iiint_{V_r} \Delta \varepsilon(\omega) E_b(\omega) \cdot \widetilde{E}_m d^3 r, \qquad (4.6)$$

where the permittivity and permeability $\varepsilon$ and $\mu$ consistently incorporate the anisotropic PML materials, and $V_r$ denotes the resonator domain for which $\Delta \varepsilon \neq 0$.

Since QNMs are poles of the scattering operator, by definition, the scattered field admits a pole singularity (we assume that it is a simple pole of order one) for the frequency $\widetilde{\omega}_m$. Thus, for a frequency $\omega$ close to $\widetilde{\omega}_m$, a single term can be retained in the QNM expansion in Eq. (1.1),

$$[\boldsymbol{E}_S(\boldsymbol{r},\omega), \boldsymbol{H}_S(\boldsymbol{r},\omega)] \approx \alpha_m(\omega)[\widetilde{\boldsymbol{E}}_m(\boldsymbol{r}), \widetilde{\boldsymbol{H}}_m(\boldsymbol{r})]. \tag{4.7}$$

By injecting this form into Eq. (4.6) and multiplying both sides of the equation by $(\omega - \widetilde{\omega}_m)$, one gets an expression of $(\omega - \widetilde{\omega}_m)\alpha_m(\omega)$ valid in the vicinity of $\widetilde{\omega}_m$. We then obtain $\lim_{\omega \to \widetilde{\omega}_m} (\omega - \widetilde{\omega}_m)\alpha_m(\omega) =$

$-\widetilde{\omega}_m \iiint_{V_r} \Delta\boldsymbol{\varepsilon}(\boldsymbol{r},\widetilde{\omega}_m) \boldsymbol{E}_b(\boldsymbol{r},\widetilde{\omega}_m) \cdot \widetilde{\boldsymbol{E}}_m(\boldsymbol{r}) d^3\boldsymbol{r} / \iiint_{\Omega \cup \Omega_{PML}} \left[ \widetilde{\boldsymbol{E}}_m \cdot \frac{\partial \omega \boldsymbol{\varepsilon}}{\partial \omega} \widetilde{\boldsymbol{E}}_m - \widetilde{\boldsymbol{H}}_m \cdot \frac{\partial \omega \boldsymbol{\mu}}{\partial \omega} \widetilde{\boldsymbol{H}}_m \right] d^3\boldsymbol{r}$, where the QNM norm $\langle \widetilde{\boldsymbol{E}}_m | \widetilde{\boldsymbol{E}}_m \rangle_{PML} = \iiint_{\Omega \cup \Omega_{PML}} \left[ \widetilde{\boldsymbol{E}}_m \cdot \frac{\partial \omega \boldsymbol{\varepsilon}}{\partial \omega} \widetilde{\boldsymbol{E}}_m - \widetilde{\boldsymbol{H}}_m \cdot \frac{\partial \omega \boldsymbol{\mu}}{\partial \omega} \widetilde{\boldsymbol{H}}_m \right] d^3\boldsymbol{r}$ automatically emerges, bringing its physical relevance to the front.

With the normalization choice in Eq. (4.5), $\langle \widetilde{\boldsymbol{E}}_m | \widetilde{\boldsymbol{E}}_m \rangle_{PML} = 1$, we finally obtain the residue of $\alpha_m(\omega)$ at the QNM pole $\widetilde{\omega}_m$

$$\mathrm{Res}\left(\alpha_m(\omega)\right) = \lim_{\omega \to \widetilde{\omega}_m} (\omega - \widetilde{\omega}_m)\alpha_m(\omega) = -\widetilde{\omega}_m \iiint_{V_r} \Delta\boldsymbol{\varepsilon}(\boldsymbol{r},\widetilde{\omega}_m) \boldsymbol{E}_b(\boldsymbol{r},\widetilde{\omega}_m) \cdot \widetilde{\boldsymbol{E}}_m(\boldsymbol{r}) d^3\boldsymbol{r}. \tag{4.8}$$

We will see that the literature proposes several expressions for $\alpha_m(\omega)$, discussed in Sec. 5.1. To be correct these expressions should verify the condition (4.8). The latter is also important as it provides a simple and general way to compute and normalize QNMs with any kind of frequency-domain Maxwell's equations solver, see the paragraph "Pole-response normalization" in the next section.

### 4.3 Summary of the different approaches for normalizing QNMs

The different approaches that have been recently proposed to normalize the QNMs of 3D plasmonic and photonic resonators can be classified into three different frameworks: PML-based normalization [Sau13,Vi14a], finite-domain normalization [Kri12,Ge14,Doo14,Mu16b] and pole-response normalization [Bai13]. The first method is strongly connected with the numerical implementation and uses perfectly matched layers (PMLs) to regularize the divergence of the QNM field. The second method uses a finite domain $\Omega$ and defines a norm that includes both a volume integral over $\Omega$ and a surface integral over the border of $\Omega$. The third method, finally, simply relies on the fact that QNMs are the poles of the scattered field and does not require any integral calculation. A comparison between the three normalization approaches, performed numerically on the particular case of a nanoscale patch antenna lying on a substrate, is given in [Sau15], showing in particular that the PML-based and pole-response normalization methods are remarkably stable and accurate.

**PML-based normalization.** With Maxwell's solvers that use PMLs to satisfy outgoing-wave boundary conditions, this approach is very effective to compute the QNMs and then to normalize them by directly computing the integral of Eq. (4.5) in the whole computational domain, including the PML domain. Note that in any numerical implementation, the PML thickness is necessarily finite. Thus on the outer PML boundary, the field that is exponentially decaying is not exactly zero and a residual inaccuracy can appear. However, the error can be reduced by increasing the PML thickness, and a very good accuracy is generally achieved [Vi14a,Yan17]. Note that this approach requires knowing exactly the permittivity and permeability distribution in the PML, which is not always possible with commercial softwares.

**Finite-domain normalization.** QNM normalization can also be performed by considering a "purely real" finite domain without using any regularization of the QNM field [Kri12,Ge14,Doo14,Mu16b]. In addition to the volume integral over the real domain Ω (first term in Eq. (4.4)), the normalization then relies on a surface integral term that avoids the divergence by compensating for the exponential increase of the volume integral as the size of the integration domain Ω increases. Two different solutions can be found in the literature. The first one involves the QNM field on the border of the domain Ω [Kri12,Ge14]. Although it has been proven to be numerically accurate for several particular cases, this proposal seems to suffer from mathematical flaws, as discussed in Section 4.1 and in [Kri15,Mu16b,Mul17]. Another expression for the surface integral that involves the derivatives of the QNM fields has been proposed in [Doo14,Mu16b]. This method has been shown to be mathematically equivalent to the PML-based approach of Eq. (4.4) [Mu16b]. However, the normalization using the derivatives of the QNM field may be difficult to implement in practice. For resonators with arbitrary shapes, for which analytical solutions do not exist, the divergence cancellation is sensitive to numerical inaccuracies [Mul17,Sau15]. Moreover, only the QNMs of resonators with uniform (free-space) backgrounds can be normalized [Sau15].

**Pole-response normalization.** Another approach [Bai13] to normalize a QNM relies on the fact that QNMs are poles of the scattering operator. As shown in Section 4.2, the field $E_s(r,\omega)$ scattered by any resonator (taken at $\omega \approx \widetilde{\omega}_m$) is proportional to the QNM field $\widetilde{E}_m(r)$ with a known proportionality factor $\alpha_m(\omega)$, see Eq. (4.8). This pole-response approach is very general: 1/ The scattered field at $\omega \approx \widetilde{\omega}_m$ can be computed with virtually any Maxwell's equations solver operating in the frequency domain, 2/ its computation does not require to implement PMLs to regularize the QNM field, and 3/ any driving field $E_b(r,\omega)$ may be used, an incident plane wave or the field created by a near-field dipole source. The pole-response approach has led to the development of a freeware [http] to compute and normalize QNMs in the general case of resonators made of dispersive and anisotropic materials. It has already been successfully applied to a variety of photonic-crystal microresonators and plasmonic nanoresonators [Sau14,Yan15,Gub17,Sel16,Dez17c,Fag17], eventually placed on a substrate [Fag15] or coupled to a periodic waveguide [Fa17b].

Resonators coupled to periodic waveguides deserves specific considerations. Indeed since PMLs cannot be used to satisfy outgoing-wave boundary conditions in periodic waveguides, the PML-based normalization method becomes meaningless. To faithfully take into account the divergence of the QNM field in the periodic waveguide, a first solution consists in regularizing the divergent field in the waveguide with advanced theorems in complex analysis. The method initially developed in [Kr14b] for analyzing a single cavity coupled to a periodic waveguide has been also applied to the normalization of two cavities coupled by a waveguide [Las15]. An alternative solution is to use the pole-response normalization. Since the latter only relies on the resonator spectral response close to the pole (a general property independent of the specific geometry under consideration), it can be applied without modification [Fa17b].

# 5. QNM expansions for modeling electromagnetic resonances

The question of the spectral decomposition of waves as a superposition of QNMs, see Eqs. (1.1) or (1.2), has a venerable history, which probably started in quantum mechanics [Sie39,Mor71,Mor73,Hoe79]. Seemingly, the first theoretical studies in electromagnetism can be traced back to the seventies [Bau71,Bau76]. From the beginning, two fundamental questions have drawn considerable attention. How can we properly normalize states whose field diverges as $|r| \to \infty$ ? Are QNM expansions such as Eqs. (1.1) or (1.2) complete? Another more practical, yet central question is: how to calculate the modal excitation coefficients, the $\alpha_m(\omega)$'s in the frequency domain or the $\beta_m(t)$'s in the time domain?

The normalization issue has been discussed in Section 4 and this Section addresses the two other questions. Two different frameworks have been developed in the literature in order to cope with the derivation of QNM excitation coefficients. They are reviewed in Sections 5.1 and 5.2 and the main results are summarized in Table 1. The completeness issue is discussed in Section 5.3. Finally, Section 5.4 illustrates the theoretical discussions with the analytical example of the 1D Fabry-Perot resonator. The Section is carried on at a more technical level than the rest of the article. It can be skipped over at first reading, but the main analytical formulas for the excitation coefficients that are summarized in Table 1 are worth being known.

Early works have primarily addressed the problem of QNM expansions in the simple case of 1D Fabry-Perot cavities (the analogue in quantum mechanics being the model of a particle in a box with finite potential barriers) [Sie39,Mor71,Mor73,Hoe79,Leu94a,Leu94b]. Since 1D systems can present particular properties, which may not exist for 3D geometries, it was important to go beyond these first studies. In the context of optical resonators, recent works tackled more complex 3D systems [Mul10,Kri12,Sau13,Bai13,Vi14a,Ge14,Doo14,Mu16b,Yan17,Pow17,Zsc18]. Two main approaches can be found in the literature for calculating the QNM excitation coefficients, leading to different expressions for the $\alpha_m$'s and thus to different modal expansions. We provide hereafter an overview of the main results. Presently, the literature lacks from a comprehensive comparison between these two different expansions, which have been derived in very different theoretical frameworks and tested in different contexts. We believe that both approaches lead to mathematically correct expansions, as illustrated in Figs. 8 and 9 with a few promising numerical results showing the high accuracy reached with present QNM solvers. We also attempt in Section 5.1.3 to make a first bridge between the two approaches by retrieving the results of one approach with the theoretical concepts of the other.

### 5.1 QNM expansions: two approaches

The first approach for calculating the excitation coefficients closely follows the routine used for establishing a modal formalism for optical waveguides [Sny83,Mar91]. It comes from the simple geometrical idea of projecting the resonator response (for instance the scattered field) into a vector space formed by the QNMs. Of particular importance in the approach are the QNM norms and their orthogonality relation, which can be derived for both non-dispersive and dispersive resonators by applying the Lorentz reciprocity theorem [Har93,Sau13,Yan17]. If the QNM operator is linear with respect to the frequency (as it is the case with non-dispersive resonators), an alternative is to use bi-orthogonality relations between the eigenstates of the initial and adjoint QNM operators [Vi14a,Han02]. We refer to this approach as the *orthogonality-decomposition approach*.

The second approach relies on the theory of functions of a complex variable, a branch of mathematical analysis with many physical applications. It is assumed that, inside the volume of inhomogeneity defining the resonator, a representative function of the resonator response, say $\mathbf{G}(\omega)$ that might denote the Green's tensor, is a meromorphic function whose poles are the QNMs frequencies. According to Mittag-Leffler's theorem, it can be expanded as $\mathbf{G}(\omega) = \sum_{m=1}^{\infty} \text{Res}[\mathbf{G}(\widetilde{\omega}_m)](\omega - \widetilde{\omega}_m)^{-1}$ [Mor71,Mor73,Doo14,Mu16a,Mu16b], where $\text{Res}[\mathbf{G}(\widetilde{\omega}_m)]$ is the residue of $\mathbf{G}(\omega)$ at the QNM pole $\widetilde{\omega}_m$, assumed to be a simple pole of order one. Once the residues of the Green's tensor are known, the computation of the excitation coefficients $\alpha_m(\omega)$ is then straightforward. We refer to this approach as the *residue-decomposition approach*.

### 5.1.1 Orthogonality-decomposition approach

The following derivation strictly follows [Sau13]; the only difference is that a scattered-field formulation (Annex 2) is used here for the sake of consistency. We project the scattered field at any frequency $\omega$ onto the entire set of QNMs

$$\begin{bmatrix} E_S(r,\omega) \\ H_S(r,\omega) \end{bmatrix} = \sum_m \alpha_m(\omega) \begin{bmatrix} \widetilde{E}_m(r) \\ \widetilde{H}_m(r) \end{bmatrix}$$ and insert the expansion into Eq. (4.6) to obtain a set of linear equations that can be solved and determine the $\alpha_m(\omega)'s$

$$\begin{bmatrix} A_{11}(\omega) & & A_{1N}(\omega) \\ & & \\ A_{N1}(\omega) & & A_{NN}(\omega) \end{bmatrix} \begin{bmatrix} (\omega-\widetilde{\omega}_1)\alpha_1 \\ \ldots \\ (\omega-\widetilde{\omega}_N)\alpha_N \end{bmatrix} = -\omega \iiint_{V_r} \Delta\boldsymbol{\varepsilon}(r,\omega) E_b(r,\omega) \cdot \begin{bmatrix} \widetilde{E}_1 \\ \ldots \\ \widetilde{E}_N \end{bmatrix} d^3r, \quad (5.1)$$

with

$$A_{mn} = (\omega-\widetilde{\omega}_n)^{-1} \iiint_{\Omega \cup \Omega_{PML}} \{\widetilde{E}_n(\omega\boldsymbol{\varepsilon}(\omega) - \widetilde{\omega}_m\boldsymbol{\varepsilon}(\widetilde{\omega}_m))\widetilde{E}_m - \widetilde{H}_n(\omega\boldsymbol{\mu}(\omega) - \widetilde{\omega}_m\boldsymbol{\mu}(\widetilde{\omega}_m))\widetilde{H}_m\} d^3r. \quad (5.2)$$

The general form of the orthogonality relation can be found by applying the Lorentz reciprocity formula, Eq. (A3.3), to two different QNMs labeled $n$ and $m$, at frequencies $\widetilde{\omega}_n$ and $\widetilde{\omega}_m$. It reads as [Sau13]

$$\iiint_{\Omega \cup \Omega_{PML}} [\widetilde{E}_n \cdot (\widetilde{\omega}_n\boldsymbol{\varepsilon}(\widetilde{\omega}_n) - \widetilde{\omega}_m\boldsymbol{\varepsilon}(\widetilde{\omega}_m))\widetilde{E}_m - \widetilde{H}_n \cdot (\widetilde{\omega}_n\boldsymbol{\mu}(\widetilde{\omega}_n) - \widetilde{\omega}_m\boldsymbol{\mu}(\widetilde{\omega}_m))\widetilde{H}_m] d^3r = 0. \quad (5.3)$$

Let us first consider the case of non-dispersive materials. If the permittivity and permeability $\boldsymbol{\varepsilon}$ and $\boldsymbol{\mu}$ are frequency-independent, Eq. (5.3) simplifies

$$\iiint_{\Omega \cup \Omega_{PML}} [\widetilde{E}_n \cdot \boldsymbol{\varepsilon}\widetilde{E}_m - \widetilde{H}_n \cdot \boldsymbol{\mu}\widetilde{H}_m] d^3r = \delta_{nm}, \quad (5.4)$$

for QNMs normalized with Eq. (4.5). The matrix $A$ in Eq. (5.1) is then equal to the identity matrix and we are led to a simple expression for $\alpha_m(\omega)$

$$\alpha_m(\omega) = -\frac{\omega}{\omega-\widetilde{\omega}_m} \iiint_{V_r} \Delta\boldsymbol{\varepsilon}(r) E_b(r,\omega) \cdot \widetilde{E}_m d^3r. \quad (5.5)$$

Note that $\lim_{\omega \to \widetilde{\omega}_m} (\omega - \widetilde{\omega}_m)\alpha_m(\omega)$ recovers the residue expression of Eq. (4.8).

For dispersive materials, the off-diagonal components of the matrix $A$ are nonzero. The computation of $\alpha_m(\omega)$ thus requires inverting the matrix $A$, which is truncated due to the finite number of QNMs retained in the computation. Figure 8(a) provides a serious test of the accuracy reached with Eq. (5.1) for the extinction cross-section of a silver nanosphere. It emphasizes the necessity to include the off-diagonal terms in matrix $A$ for analyzing dispersive media, especially when high accuracy is required.

Thus, the excitation coefficient of the $m^{\text{th}}$ QNM does not only depend on $\widetilde{E}_m$, but also on the fields of all the other QNMs. This complication is completely removed once the dispersion relation is specified [Yan17]. For instance, for the important practical case of dispersive materials that can be modeled with a single (or a *N*-pole in general) Lorentz-pole permittivity, $\boldsymbol{\varepsilon}(\omega) = \boldsymbol{\varepsilon}_\infty(1 - \omega_p^2(\omega^2 - \omega_0^2 + i\omega\gamma)^{-1})$, it has been recently shown that an auxiliary-field formulation of Maxwell's equations allows deriving a closed-form expression for the coefficients $\alpha_m(\omega)$ [Yan17]. The vector that describes the scattered field is enlarged from $[E_S, H_S]$ to $[E_S, H_S, P_S, J_S]$ (see Eq. (3.1) for the definitions of the auxiliary fields $P_S$ and $J_S$). Projecting $[E_S, H_S, P_S, J_S]$ into the enlarged QNM vector $[\widetilde{E}_m, \widetilde{H}_m, \widetilde{P}_m, \widetilde{J}_m]$ by using Lorentz reciprocity theorem results in

$$\alpha_m(\omega) = \iiint_{V_r} \left[ \boldsymbol{\varepsilon}_b(\boldsymbol{r}) - \boldsymbol{\varepsilon}_\infty(\boldsymbol{r}) - \frac{\widetilde{\omega}_m}{\omega - \widetilde{\omega}_m} \Delta\boldsymbol{\varepsilon}(\boldsymbol{r}, \widetilde{\omega}_m) \right] \boldsymbol{E}_b(\boldsymbol{r}, \omega) \cdot \widetilde{\boldsymbol{E}}_m d^3\boldsymbol{r}, \qquad (5.6)$$

which provides highly accurate predictions, as shown in Fig. 9b. More details on the derivation, in particular the orthogonality relation between two QNMs enlarged with the auxiliary fields, can be found in [Yan17].

### 5.1.2 Residue-decomposition approach

The residue-decomposition approach focuses on the Green's tensor $\mathbf{G}(\boldsymbol{r}, \boldsymbol{r}', \omega)$, defined as $\nabla \times \boldsymbol{\mu}(\boldsymbol{r}, \omega)^{-1} \nabla \times \mathbf{G}(\boldsymbol{r}, \boldsymbol{r}', \omega) - \omega^2 \boldsymbol{\varepsilon}(\boldsymbol{r}, \omega) \mathbf{G}(\boldsymbol{r}, \boldsymbol{r}', \omega) = \mathbf{I}\delta(\boldsymbol{r} - \boldsymbol{r}')$. According to Mittag-Leffler's theorem, the Green's tensor can be expanded as a series of poles [Mor71,Mor73,Mul10,Mu16a,Mu16b]

$$\mathbf{G}(\boldsymbol{r}, \boldsymbol{r}', \omega) = \sum_m \frac{\text{Res}[\mathbf{G}(\boldsymbol{r}, \boldsymbol{r}', \widetilde{\omega}_m)]}{\omega - \widetilde{\omega}_m}, \qquad (5.7)$$

The validity of Eq. (5.7) is directly related to the issue of the completeness of the QNM set, as will be discussed in Section 5.3. Two slightly different expressions for the Green's tensor expansion are found in the literature [Mul10,Doo14,Mu16b]. The first one is

$$\mathbf{G}(\boldsymbol{r}, \boldsymbol{r}', \omega) = -\sum_{m=1}^{\infty} \frac{\widetilde{\boldsymbol{E}}_m(\boldsymbol{r}) \otimes \widetilde{\boldsymbol{E}}_m(\boldsymbol{r}')}{(\omega - \widetilde{\omega}_m)\widetilde{\omega}_m}, \qquad (5.8)$$

and a second possible expansion is

$$\mathbf{G}(\boldsymbol{r}, \boldsymbol{r}', \omega) = -\sum_{m=1}^{\infty} \frac{\widetilde{\boldsymbol{E}}_m(\boldsymbol{r}) \otimes \widetilde{\boldsymbol{E}}_m(\boldsymbol{r}')}{(\omega - \widetilde{\omega}_m)\omega}, \qquad (5.9)$$

which can be deduced from Eq. (5.8) with the sum rule $\sum_{m=1}^{\infty} \frac{\widetilde{\boldsymbol{E}}_m(\boldsymbol{r}) \otimes \widetilde{\boldsymbol{E}}_m(\boldsymbol{r}')}{\widetilde{\omega}_m} = 0$ [Mu16b]. Once the Green's tensor has a known QNM expansion, it is straightforward to derive the QNM decomposition of the scattered field $\boldsymbol{E}_S(\boldsymbol{r}, \omega) = i\omega \iiint \mathbf{G}(\boldsymbol{r}, \boldsymbol{r}', \omega) \boldsymbol{J}_s(\boldsymbol{r}', \omega) d^3\boldsymbol{r}'$, with $\boldsymbol{J}_s(\boldsymbol{r}, \omega) = -i\omega\Delta\boldsymbol{\varepsilon}(\boldsymbol{r}, \omega)\boldsymbol{E}_b(\boldsymbol{r}, \omega)$. By injecting Eq. (5.8) in the latter expression, one obtains

$$\alpha_m(\omega) = -\frac{\omega^2}{\widetilde{\omega}_m(\omega - \widetilde{\omega}_m)} \iiint_{V_r} \Delta\boldsymbol{\varepsilon}(\boldsymbol{r}, \boldsymbol{\omega}) \boldsymbol{E}_b(\boldsymbol{r}, \omega) \cdot \widetilde{\boldsymbol{E}}_m d^3\boldsymbol{r}. \qquad (5.10)$$

By using the Green's tensor expansion of Eq. (5.9), we obtain a slightly different expression for $\alpha_m(\omega)$, where $\omega^2/\widetilde{\omega}_m$ is replaced by $\omega$, see Eq. (5.11) in Table 1. Note that Eq. (5.11) takes the same form as Eq. (5.5) except that $\Delta\boldsymbol{\varepsilon}$ becomes frequency-dependent, while Eq. (5.5) applies to non-dispersive resonators.

The residue-decomposition approach has been mainly tested with fully analytical problems (spherical resonators, essentially) so far [Mul10,Doo14,Mu16a,Mu16b]. For instance, in recent works, it has been successfully used to compute the QNMs of a dielectric nanosphere (refractive index 1.5) from the sole knowledge of the complete initial set of QNMs for the same sphere made of Gold [Mu16a], as shown in Fig. 8(c).

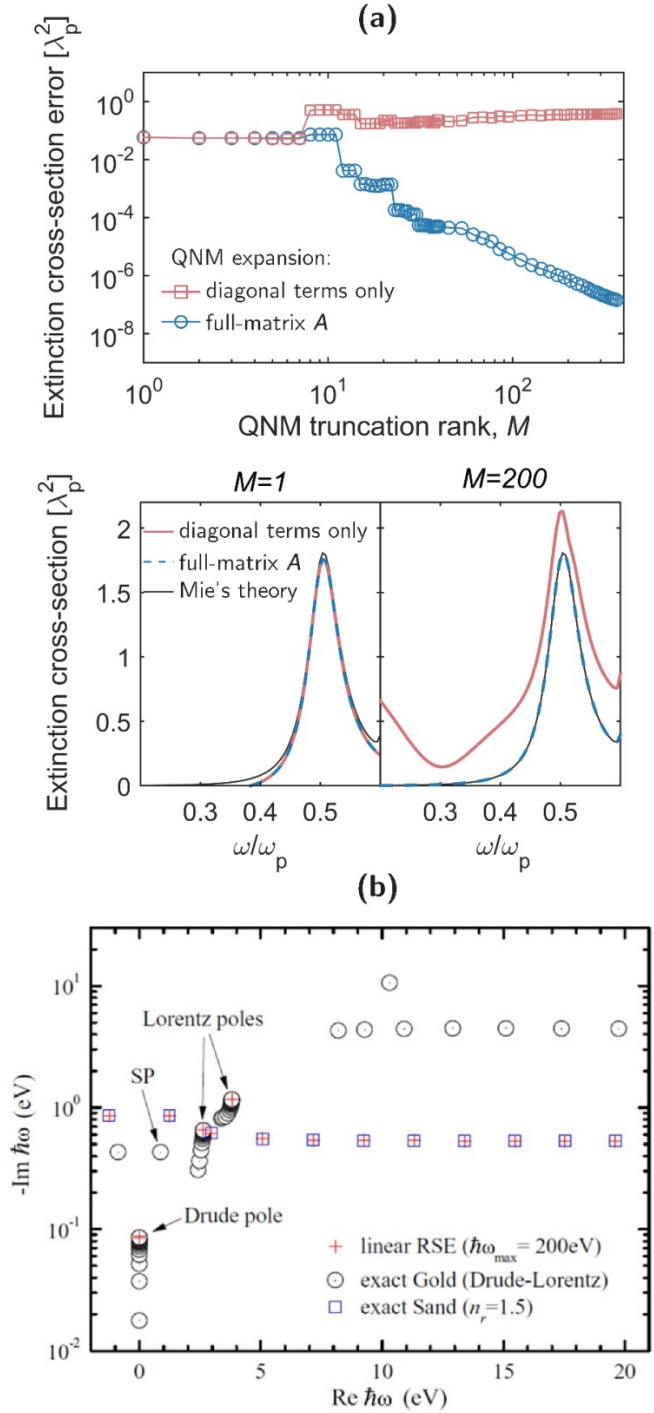

**Fig. 8. Numerical examples of highly-accurate results obtained with QNM expansions for simple spherical particles for which exact analytical solutions exist.** (a) Convergence test of the QNM expansion of Eq. (5.1) for dispersive resonators and importance of the off-diagonal terms in the matrix $\boldsymbol{A}$. The resonator is a silver nanosphere in air with a 50-nm diameter and a Drude electric permittivity, $\varepsilon_{Ag} = 1 - \omega_p^2/(\omega^2 + i\omega\gamma)$, with $\gamma = 0.0023\omega_p$ and $\lambda_p \equiv 2\pi c/\omega_p$ =138 nm. Top: Average numerical error for the extinction cross-section, computed with the QNM expansion in the frequency interval $0.2\omega_p - 0.6\omega_p$, versus the number $M$ of QNMs

retained in the computation. The modal excitation coefficients are computed by taking into account (dashed blue) or neglecting (solid red) the off-diagonal components in $A$. Bottom: Extinction cross-section spectra obtained with the QNM expansion for $M$ = 1 and 200. The extinction cross-section computed from Mie theory (black curve) is used as the reference. (b) Test of QNM expansions of Eqs. (5.8) and (5.9). Blue squares: Complex energies $\hbar\widetilde{\omega}_m$ for $l$ = 1 TM modes and for a 200-nm-radius sand (refractive index 1.5) sphere. Black circles with dots: Complex energies $\hbar\widetilde{\omega}_m$ for the same sphere made of Gold (Drude model + 2 Lorentz-poles). The red crosses represent the sand-sphere QNMs predicted by a change of basis from known gold-sphere QNMs using the residue-decomposition approach. Note the quantitative predictions of the accumulation points close to the poles of the permittivity. After [Mu16a].

### 5.1.3 Summary of the different formulas for $\alpha_m(\omega)$

Table 1 summarizes the main formulas that are available for the $\alpha_m$'s in the literature and that have been derived in Sections 5.1.1 and 5.1.2. As long as the QNM basis constitutes a complete basis, the formulas all provide an exact representation of the resonator response, as confirmed by numerical data shown in Fig. 8. The existence of different but mathematically-correct formulas is due to the fact that the QNM basis is overcomplete [Pea81], see the Eqs. (2.23) and (2.24) in [Leu94a] and the related discussions. Overcompleteness can help to achieve more stable, robust, or compact decompositions, but such studies remain to be done and are beyond the scope of the Review.

The orthogonality- and residue-decomposition approaches seem to come from radically different frameworks. To bridge the apparent gap, it is interesting to derive the residue of the Green tensor $\text{Res}[\mathbf{G}(\mathbf{r},\mathbf{r}',\omega_m)]$ used in the residue-decomposition approach from Lorentz-reciprocity arguments, especially using Eq. (4.8). For that purpose, let us first note that the scattered field can be expressed as a function of the Green's tensor, $\mathbf{E}_S(\mathbf{r},\omega) = i\omega \iiint \mathbf{G}(\mathbf{r},\mathbf{r}',\omega)\mathbf{J}_S(\mathbf{r}',\omega)d^3\mathbf{r}'$, with $\mathbf{J}_S(\mathbf{r},\omega) = -i\omega\Delta\boldsymbol{\varepsilon}(\mathbf{r},\omega)\mathbf{E}_b(\mathbf{r},\omega)$. Thus, $\text{Res}[\mathbf{E}_S(\mathbf{r},\widetilde{\omega}_m)] = \text{Res}[\alpha_m(\widetilde{\omega}_m)]\widetilde{\mathbf{E}}_m = i\widetilde{\omega}_m \iiint \text{Res}[\mathbf{G}(\mathbf{r},\mathbf{r}',\omega_m)]\mathbf{J}_S(\mathbf{r}',\widetilde{\omega}_m)d^3\mathbf{r}$. Then, by injecting $\text{Res}[\alpha_m(\widetilde{\omega}_m)]$ given by Eq. (4.8) within the orthogonality-decomposition framework into the later equation, one easily recovers Eq. (5.8).

| $\begin{bmatrix}\mathbf{E}_S(\mathbf{r},\omega)\\\mathbf{H}_S(\mathbf{r},\omega)\end{bmatrix} = \sum_m \alpha_m(\omega)\begin{bmatrix}\widetilde{\mathbf{E}}_m(\mathbf{r})\\\widetilde{\mathbf{H}}_m(\mathbf{r})\end{bmatrix}$ ||
|---|---|
| *Orthogonality decomposition* | *Residue decomposition* |
| Dispersive materials (general case) [Sau13] | Dispersive or non-dispersive materials [Doo14,Mu16a] |
| $A\begin{bmatrix}(\omega-\widetilde{\omega}_1)\alpha_1\\ \ldots \\ (\omega-\widetilde{\omega}_N)\alpha_N\end{bmatrix} = -\omega\Delta\varepsilon(\omega)\begin{bmatrix}O_1(\omega)\\ \ldots \\ O_N(\omega)\end{bmatrix}$, (5.1) | $\alpha_m(\omega) = -\frac{\omega^2}{\widetilde{\omega}_m(\omega-\widetilde{\omega}_m)}\Delta\varepsilon(\omega)O_m(\omega)$, (5.10) |
| Lorentz-Drude materials [Yan17] | $\alpha_m(\omega) = -\frac{\omega}{\omega-\widetilde{\omega}_m}\Delta\varepsilon(\omega)O_m(\omega)$, (5.11) |
| $\alpha_m(\omega) = \left[\varepsilon_b - \varepsilon_\infty - \frac{\widetilde{\omega}_m}{\omega-\widetilde{\omega}_m}\Delta\boldsymbol{\varepsilon}(\widetilde{\omega}_m)\right]O_m(\omega)$, (5.6) | |
| Non-dispersive materials [Sau13,Vi14a,Yan17] | |
| $\alpha_m(\omega) = -\frac{\omega}{\omega-\widetilde{\omega}_m}\Delta\varepsilon O_m(\omega)$, (5.5) | |

**Table 1. Summary of the main available formulas for the excitation coefficients $\alpha_m$ of the scattered-field expansion.** In all formulas, $O_m(\omega) = \iiint_{V_r} \boldsymbol{E}_b(\boldsymbol{r},\omega) \cdot \widetilde{\boldsymbol{E}}_m d^3\boldsymbol{r}$ represents the overlap integral between the driving (background) field $\boldsymbol{E}_b$ in the absence of the resonator and the QNM field $\widetilde{\boldsymbol{E}}_m$. The volume $V_r$ is the resonator inclusion domain for which $\Delta\varepsilon(\boldsymbol{r},\omega) = \varepsilon(\boldsymbol{r},\omega) - \varepsilon_b(\boldsymbol{r},\omega) \neq 0$, see Annex 2. For the sake of simplicity, we have assumed an isotropic and spatially uniform $\Delta\varepsilon$ within the resonator volume but this assumption is not mandatory. In the literature, there exists other formulas [Via14a,Ge14,Pow14,Pow17], which slightly differ from those presented in the Table because of mathematical details.

Equation (5.1) deserves a specific attention. It is the only one in the Table that does not provide a fully analytical expressions, since the calculation of the $\alpha_m$'s requires a matrix inversion. Surprisingly, the values of the coefficients depends on the number $N$ of QNMs retained in the matrix, although they stabilize as $N$ increases. The extra computation cost might appear as a drawback, compared to the other decompositions. In fact, the most important question concerns the convergence performance of the QNM expansion, and in this regard, there is no reason to suspect that the decomposition of Eq. (5.1) offers a convergence rate slower than the others. Which decomposition does provide the most accurate prediction by retaining the smallest number of QNMs in the expansion? This open question remains to be answered.

**5.2 Total-field expansion for dipolar emission**

So far, we focused our attention on QNM expansions of the scattered field. This is natural because scattered fields are outgoing waves and are directly imprinted by the signature of the natural modes of resonators. However, for analyzing light emission by molecules or atoms coupled to resonators, it may be relevant to study QNM expansions of the total field, which also satisfies outgoing-wave conditions in that particular case.

Let us consider an emitter with an electric dipole moment $\boldsymbol{p}$ located at position $\boldsymbol{r} = \boldsymbol{r}_0$ in a medium of refractive index $n$. Its spontaneous decay rate $\gamma$ takes the classical expression $\gamma = \frac{2}{\hbar}\text{Im}(\boldsymbol{p}^* \cdot \boldsymbol{E}_{tot}(\boldsymbol{r}_0,\omega))$ [Nov06], where $\boldsymbol{E}_{tot}$ is the total electric field emitted by the dipole at frequency $\omega$. In view of the expression of $\gamma$, it is appropriate to expand the total field, rather than the scattered field,

$$[\boldsymbol{E}_{tot}(\boldsymbol{r},\omega), \boldsymbol{H}_{tot}(\boldsymbol{r},\omega)] = \sum_m a_m(\omega)[\widetilde{\boldsymbol{E}}_m(\boldsymbol{r}), \widetilde{\boldsymbol{H}}_m(\boldsymbol{r})]. \tag{5.12}$$

The total electric field presents a singularity at $\boldsymbol{r} = \boldsymbol{r}_0$ so that the QNM expansion of Eq. (5.12) does not converge everywhere. However, since $\text{Im}(\boldsymbol{p}^* \cdot \boldsymbol{E}_{tot}(\boldsymbol{r}_0,\omega))$ is not singular, the expansion may provide accurate predictions for $\gamma$. A closed-form expression of the $a_m(\omega)$ coefficients in Eq. (5.12) can be easily found by repeating the derivations that lead to $\alpha_m(\omega)$ for scattered fields in Sec. 5.1, replacing $\Delta\varepsilon(\boldsymbol{r},\omega)\boldsymbol{E}_b(\boldsymbol{r})$ by $\boldsymbol{p}\delta(\boldsymbol{r}-\boldsymbol{r}_0)$. In this way, all the formulas of $\alpha_m(\omega)$ summarized in Table 1 can be transformed into new formulas for the $a_m$'s. For example, Eqs. (5.11) and (5.6) that are both valid for dispersive resonators become

$$a_m(\omega) = -\frac{\omega}{\omega-\widetilde{\omega}_m}\boldsymbol{p} \cdot \widetilde{\boldsymbol{E}}_m(\boldsymbol{r}_0). \tag{5.13}$$

Equation (5.13) is remarkably simple and its incorporation into the expression of $\gamma$ eventually leads to a convenient formula for the normalized decay rate

$$\gamma/\gamma_0 = -\frac{3}{4\pi^2}\sum_m \text{Im}\left(\frac{\omega}{\omega-\widetilde{\omega}_m}\frac{(\lambda/n)^3}{2\widetilde{V}_m}\right), \tag{5.14}$$

where we have introduced the complex mode volume $\widetilde{V}_m$ defined in Eq. (6.8), $\lambda = 2\pi c/\omega$ is the emission wavelength in vacuum, and $\gamma_0 = \omega^3 p^2 n/(3\pi\varepsilon_0\hbar c^3)$ is the radiative decay rate of the dipole in a uniform background of refractive index $n$. In Section 6, we will discuss the physical meaning of Eq. (5.14) in relation with the Purcell effect and of complex-valued mode volumes.

### 5.3 Completeness

The completeness of the QNM set is a critical issue, as it teaches us how accurate a given QNM expansion, necessarily truncated in practice, could be. It can be examined from two different viewpoints. First, one can study simple geometries for which all QNMs can be computed analytically. Secondly, one can consider complex geometries for which one should resort to numerical computations. In the latter case, PML modes appear among the discrete set of eigenstates, see Section 3, and they should be included in the expansions to obtain accurate results [Vi14a,Yan17], especially when the QNMs alone cannot form a complete basis.

Let us first consider simple geometries for which the QNMs are known analytically. The completeness of the QNM set has commanded considerable attention since the earlier works on QNM expansions. It is conveniently examined from the point of view of the residue-decomposition approach that invokes Mittag-Leffler's theorem [Mor71,Mor73]. Note that earlier works pointed out that the validity of this theorem could be questionable in some cases: "When this representation is valid, it is very useful […]. The convergence of this representation is a crucial question." [Mor73]. To analyze the validity conditions of Eq. (5.7), we apply the Cauchy's integral theorem and consider the contour integral $\oint_C \mathbf{G}(\mathbf{r},\mathbf{r}',\omega)/(\omega-\omega')d\omega$, where $C$ represents the circle of center $\omega=0$ with an infinitely large radius in the complex plane, $\omega'$ being a complex-valued constant. The contour integral can be easily evaluated by extracting the contributions from all singularities, including isolated poles and branch cuts. We straightforwardly see two conditions for Eq. (5.7) to be valid: (i) $\mathbf{G}(\mathbf{r},\mathbf{r}',\omega)$ should be analytic everywhere in the complex-$\omega$ plane excluding the QNM poles and (ii) the contour integral, $\oint_C \mathbf{G}(\mathbf{r},\mathbf{r}',\omega)/(\omega-\omega')\,d\omega$ should vanish as the radius of $C$ tends to infinity.

For 1D systems and 3D cavities surrounded by uniform backgrounds, $\mathbf{G}(\mathbf{r},\mathbf{r}',\omega)$ has no branch cuts and condition (i) is satisfied. However, for 2D systems and 3D resonators lying on a substrate, $\mathbf{G}(\mathbf{r},\mathbf{r}',\omega)$ might have branch cuts, so that condition (i) is violated. The second condition is better examined for simple geometries, such as 1D systems, 2D cylindrical and 3D spherical cavities in uniform backgrounds, taking advantage of the analytic solutions of $\mathbf{G}(\mathbf{r},\mathbf{r}',\omega)$, for which it has been shown that condition (ii) is satisfied for positions $\mathbf{r}$ and $\mathbf{r}'$ both inside the cavities that are defined by an outmost discontinuity in the refractive index [Leu94a,Leu94b]. Thus, combining the conditions for (i) and (ii), we conclude that the QNM set forms a complete basis inside the volume of 1D systems and 3D spherical cavities in uniform media. This is the celebrated theoretical results obtained by Leung and coworkers in the 90's [Leu94a, Leu94b], see also the recent study [Man17]. Nowadays, people widely conjecture that, for arbitrary resonators, the QNMs form a complete basis inside the volume of resonators, as long as $\mathbf{G}(\mathbf{r},\mathbf{r}',\omega)$ has no branch cuts. No rigorous mathematical proof exists, but the conjecture is supported by many numerical examples [Mul10,Mu16a,Mu16b,Vi14a,Pow14,Pow17,Yan17]. In a very recent work [Abd18] published during the proofreading of the present review, the possibility to ensure completeness outside the resonator volume with expansions based on regularized QNMs has been demonstrated mathematically for 1D open resonators and confirmed by numerical simulations. This appears to be a very important theoretical finding that may represent a new departure running counter the longstanding belief that QNM expansions are complete only inside resonators.

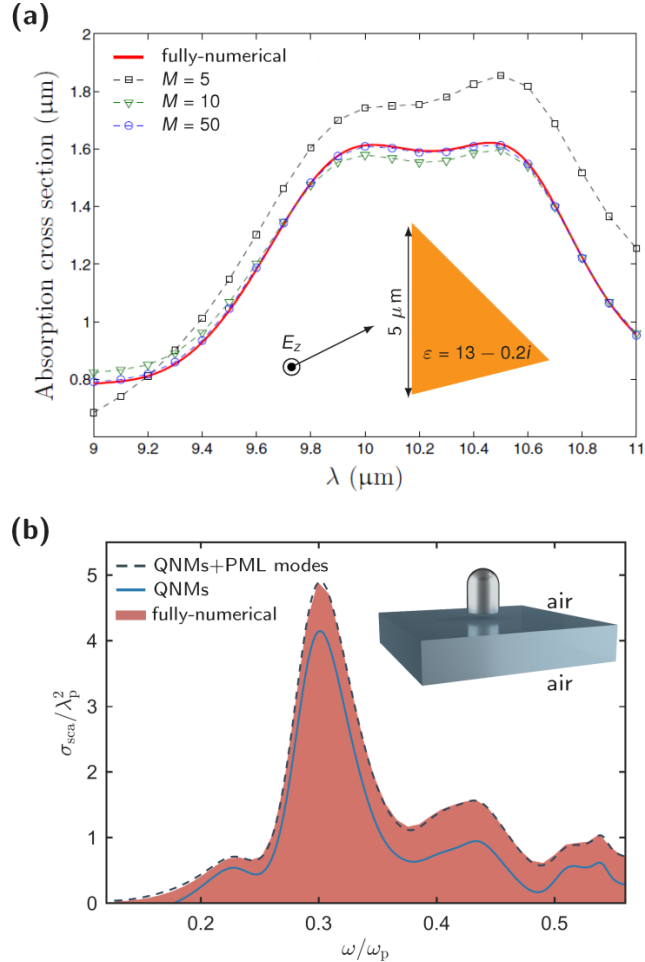

**Fig. 9 Completeness with QNMs and PML modes**. **(a)** Absorption cross-section spectrum of a 2D triangular rod in a vacuum for TE polarization. $M$ represents the total number of QNMs and PML modes retained in the expansion. As $M$ increases, the deviation with "fully-numerical" data (solid-red curve) obtained with a frequency-domain method vanishes. After [Vi14a]. **(b)** Scattering cross-section spectrum of a Drude-silver nanobullet (plasma frequency $\lambda_p \equiv 2\pi c/\omega_p =$138 nm, 60-nm diameter, 90-nm height) lying on a semiconductor slab. The solid-blue and dashed-black curves are computed by retaining 20 QNMs only and 20 QNMs plus 200 PML modes, respectively, and the shadowed pink curves are the data obtained with the frequency-domain solver of COMSOL Multiphysics. After [Yan17].

Let us now consider complex geometries that cannot be handled analytically. In practice, because of the inevitable discretization used in the computation, the original continuous operator, after discretization of the finite space, becomes a discrete operator (a matrix with a finite dimension). When the matrix is linear, e.g., for dielectric resonators [Vi14a] or dispersive resonators in the auxiliary-field formulation [Yan17], the eigenstates, composed of QNMs and numerical modes that we refer to as PML modes (see Section 3), form a complete basis in the entire computational domain, inside and outside the resonator. Complicated theoretical issues on the completeness of QNM expansions in open systems with continuous operators restricted to inside the resonator [Leu94a] are thus avoided. As a result, for resonators whose Green's tensor exhibits branch cuts, such as 2D resonators [Vi14a] and 3D resonators [Yan17] on a substrate, one promising way to recover the accuracy of the modal expansions is to include the PML modes, as shown in Fig. 9.

Completeness is a powerful mathematical property. However, we should keep in mind that *approximate* expansions, relying on the excitation of a few dominant resonant modes, as it is mostly the case in practice, may already be very useful to experiment interpretations, initial designs, or optimizations. Intuition and understanding are then readily made available, in sharp contrast with more widespread classical scattering theories that do not intrinsically rely on the natural modes of the resonator.

### 5.4 Example: 1D Fabry-Perot resonators

To illustrate how the reconstruction may be applied in practice, it is convenient to consider again the non-dispersive 1D Fabry-Perot cavity considered in Section 2, a layer of length $L$ and refractive index $n$, surrounded by a medium with refractive index $n_1$. Using the expressions for the normalized QNMs of Eq. (A1-3), one can check that the orthogonality relation of Eq. (5.4) is verified and we find that $\int_{-\infty}^{+\infty}[\tilde{E}_m \cdot \varepsilon \tilde{E}_n - \tilde{H}_m \cdot \mu \tilde{H}_n]dz = \frac{\sin[(m-n)\pi/2]}{(m-n)\pi/2} = \delta_{mn}$ for QNMs with the same symmetry and 0 otherwise. We further assume that the 1D cavity is illuminated by an incident plane wave at a frequency $\omega$ with an electric field given by $E_b(z) = E_0 \exp\left(ik_0 n_1 \left(z + \frac{L}{2}\right)\right)$, with $E_0$ an arbitrary constant and $k_0 = \omega/c$. Then, using Eq. (5.5), we easily derive the closed-form expression for the excitation coefficient $\alpha_m^{(FP)}$

$$\alpha_m^{(FP)}(\omega) = \frac{-\omega}{\omega - \tilde{\omega}_m} \int_{-L/2}^{L/2} \Delta\varepsilon(z) \tilde{E}_m(z) \cdot E_b(z) \, dz = E_0 \frac{-k_0}{k_0 - \tilde{k}_m} \frac{2\sqrt{\varepsilon_0}(n^2 - n_1^2)}{n\sqrt{L}(n^2 \tilde{k}_m^2 - n_1^2 k_0^2)}$$

$$\times \begin{cases} \pm \left[n\tilde{k}_m \cos\left(k_0 n_1 \frac{L}{2}\right) \sin\left(\tilde{k}_m n \frac{L}{2}\right) - n_1 k_0 \sin\left(k_0 n_1 \frac{L}{2}\right) \cos\left(\tilde{k}_m n \frac{L}{2}\right)\right], \ (m \text{ even}) \\ \mp i \left[n_1 k_0 \cos\left(k_0 n_1 \frac{L}{2}\right) \sin\left(\tilde{k}_m n \frac{L}{2}\right) - n\tilde{k}_m \sin\left(k_0 n_1 \frac{L}{2}\right) \cos\left(\tilde{k}_m n \frac{L}{2}\right)\right], \ (m \text{ odd}) \end{cases} \quad (5.15)$$

The total field in the cavity simply reads as $E(z,\omega) = E_b(z,\omega) + \sum_{m=-\infty}^{+\infty} \alpha_m^{(FP)}(\omega) \tilde{E}_m(z)$ and the transmission coefficient of the dielectric layer $t = E(z = L/2)/E_0$. Figure 10 illustrates the validity of the QNM expansion by comparing the field inside the cavity computed with either the well-known Airy formulas for Fabry-Perot resonators [Yeh88] (reference method) or the QNM expansion. The agreement is excellent only inside the resonator.

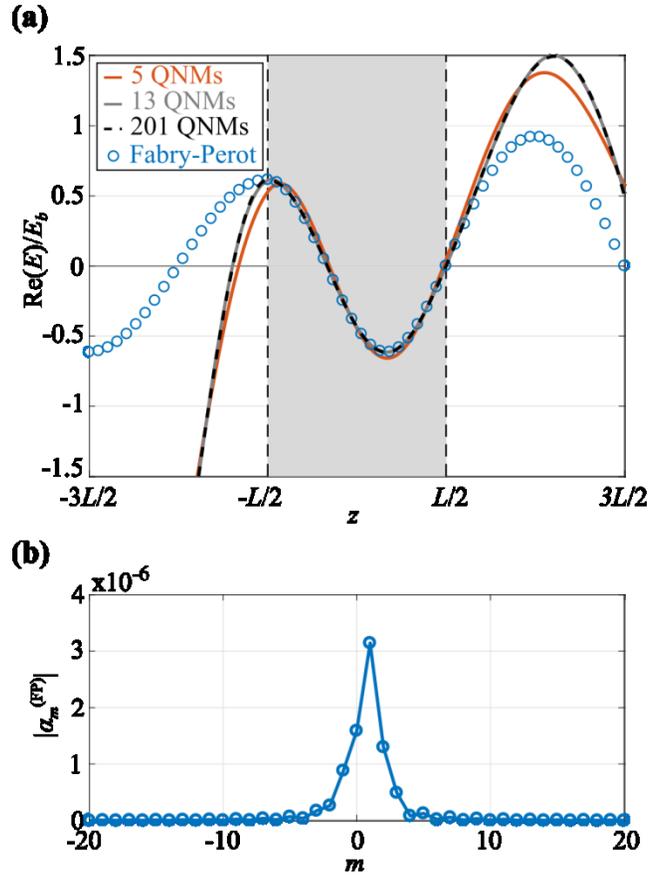

**Fig. 10. Illustration of the QNM-expansion approach for the simple case of 1D Fabry-Perot resonators.** As in Fig. 5, we consider a dielectric layer of length $L$ and refractive index $n = 1.5$ embedded in air ($n_1 = 1$). **(a)** Real part of the electric field inside and outside the dielectric layer at $\lambda/n_1 = 4L/3$ (out-of-resonance) as predicted by the QNM expansion with varying number of QNMs considered (centered on $m = 0$). The blue circles indicate the prediction from the exact Fabry-Perot model (Airy formula). The electric field inside the dielectric layer is accurately reproduced with 10 and more QNMs in the expansion, but the field outside the dielectric layer is very different from expectations. Completeness of the QNM decomposition is guaranteed only "inside the resonator" [Leu94a]. **(b)** Amplitude of the excitation coefficients $|\alpha_m^{(FP)}|$ corresponding to the different QNMs, which indicates that several QNMs are excited by the incident plane wave, and thus, should be taken into account to reconstruct the electric field.

## 6. Local density of states, Purcell factor, and mode volume

The density of electromagnetic states (DOS) is a quantity of central importance for understanding many optical phenomena that can be described in terms of dipole currents. Actually, for non-uniform systems, it is more appropriate to introduce the local density of states (LDOS) that gives the density of states at a given position $r$ in the system. For example, Fermi's golden rule states that the spontaneous emission rate of atoms (key to fluorescence and lasing phenomena) is proportional to the LDOS in the weak coupling regime [Xu00,Nov06]. Other examples of physical quantities that can be derived from the knowledge of the LDOS are the Casimir force [Kam68] and heat transfer [Mul02]. The LDOS is often defined as the imaginary part of the trace of the Green tensor [Eco83,Nov06]. One advantage of such a generic definition is that the LDOS can be computed for any system by solving directly Maxwell's equations. The physical picture that is commonly used to understand the LDOS concept is to count the states available in the system in a way similar to the "particle in a box" approach used in solid-state physics to define the electron and phonon density of states. Within this picture, the LDOS provides a local measure of the eigenfrequencies spectrum. It is therefore essential to link the LDOS and the eigenstates of the system. In the following, we consider both the LDOS and the partial (or projected) LDOS that appears in the spontaneous emission rate of an atom; the partial LDOS solely takes into account the modes that can be coupled to the dipole of the emitter transition.

The link between the LDOS and the eigenstates spectrum is commonly obtained in the realm of ideal closed and nonabsorbing systems (the "particle in a box") whose operators are Hermitian. Energy dissipation is then introduced as a perturbation. As we will see, this phenomenological approach that flatters our physical intuition is correct for resonators with large $Q$-factors. Unfortunately, it is limited since the intuition gained from an ideal system becomes largely unsubstantiated for real non-Hermitian systems with low quality factors, casting doubts upon the applicability of fundamental concepts to plasmonic nanoresonators, such as the mode volume and the Purcell factor [Koe10].

This Section aims at clarifying what can be really kept from the understanding of the low-loss limit case and what cannot be applied to comprehend real lossy systems. In particular, we will show how to define the mode volume $V_m$ and the Purcell factor $F_m$ for a low-$Q$ resonant mode $m$. The Purcell factor characterizes the maximum spontaneous emission enhancement of a single emitter coupled to a resonant mode in the weak-coupling regime [Pur46]. It is an important figure of merit in cavity quantum electrodynamics, initially used to interpret experiments involving atoms in microwave or optical cavities [Har89], quantum dots coupled to solid-state optical microcavities [Ger98], and more recently fluorescent emitters in plasmonic nanoresonators [Azo00,Ang06,Kuh06,Rus12,Aks14,Bid16].

We first recall in Section 6.1 the relationship between the LDOS and the modes of the system in the limit of weak energy dissipation and recall the classical expressions of $V_m$ and $F_m$ valid in this asymptotic case. We then apply the QNM formalism in Section 6.2 to generalize these asymptotic results to the general case of resonant systems with non-negligible energy dissipation (low quality factors).

## 6.1 Resonators with large $Q$'s: the textbook case

All textbook approaches to bridge the LDOS and the eigenstates spectrum rely on the assumption that leakage and absorption can be treated as a perturbation. They start by considering a lossless and non-dispersive version of the system, in which material absorption is simply neglected, $Im(\varepsilon) = 0$, and leakage is removed by enclosing the geometry in a finite box (so-called quantization box) with periodic boundary conditions or perfect-metallic walls. Within these assumptions, the Maxwell operator of Eq. (2.2) becomes Hermitian and semidefinite under the inner product $\langle \mathbf{E}, \mathbf{F} \rangle = \int_V \boldsymbol{E}^* \cdot \boldsymbol{\varepsilon} \boldsymbol{F} d^3 \boldsymbol{r}$, where the superscript $*$ stands for complex conjugate. Thus, the eigenfrequencies $\widetilde{\omega}_m = \Omega_m$ are real, they form a discrete set ($m = 1, 2, ...$), and the eigenstates $\widetilde{\boldsymbol{E}}_m$ form a complete orthonormal basis. Since the number of states $dN(\omega)$ having energies between $\hbar\omega$ and $\hbar\omega + \delta E$ is, by definition, $dN(\omega) = \text{DOS}(\omega)\delta E$, we have for a discrete spectrum

$$\text{DOS}(\omega) = \sum_m \delta(\omega - \Omega_m). \tag{6.1}$$

Therefore, integrating the DOS over the frequency simply amounts to count the number of modes in the frequency integration interval. The concept of LDOS accounts for the fact that all positions are not equivalent. For instance, if an atom is adsorbed on a metal surface, the density of electronic states is perturbed around it. A similar effect holds for photons close to an interface. If we normalize the eigenmodes such that $\iiint_V \tilde{\mathbf{E}}_m^* \cdot \varepsilon \tilde{\mathbf{E}}_m d^3\mathbf{r} = 0.5$, then the LDOS $\varrho(\mathbf{r}, \omega)$ can be simply expressed as

$$\varrho(\mathbf{r},\omega) = \sum_m 2\varepsilon_0 |\tilde{\mathbf{E}}_m(\mathbf{r})|^2 \delta(\omega - \Omega_m). \tag{6.2}$$

Thus, the LDOS counts the number of modes weighted by the electric-field intensity of each mode at position $\mathbf{r}$ in a similar way as the density of charge in systems composed of point-like charge carriers. The sum of Dirac delta distributions may appear unphysical and comes from the assumption of a lossless system. When infinitesimal absorption and/or leakage is added, the Maxwell operator is no longer Hermitian and the eigenfrequencies become complex, $\tilde{\omega}_m = \Omega_m - i\Gamma_m/2$. One can show from perturbation theory that the real part of the frequency $\Omega_m$ is unchanged to first order [Joh13]. Furthermore, the Dirac terms in Eq. (6.2) become Lorentzian functions and finally the phenomenological modal expansion of the LDOS in the limit of weak energy dissipation reads as

$$\varrho(\mathbf{r},\omega) = \frac{1}{\pi}\sum_m \frac{\Gamma_m/2}{(\omega-\Omega_m)^2 + (\Gamma_m/2)^2} 2\varepsilon_0 |\tilde{\mathbf{E}}_m(\mathbf{r})|^2. \tag{6.3}$$

We end up with a very intuitive picture: every lossy mode positively contributes to the LDOS with a Lorentzian peak. The schemes in Fig. 11 illustrate this phenomenological transition from the lossless to the weakly lossy case. Adding new modes in the system simply consists in increasing the LDOS. We will see in Section 6.2 that this approach fails in the presence of non-negligible energy dissipation, in particular in spectral regions where two or several modes overlap and Fano interferences takes place. Note that we have only discussed the electric part of the LDOS; a more complete definition should include a magnetic part, see for instance [Joh13,Jou03].

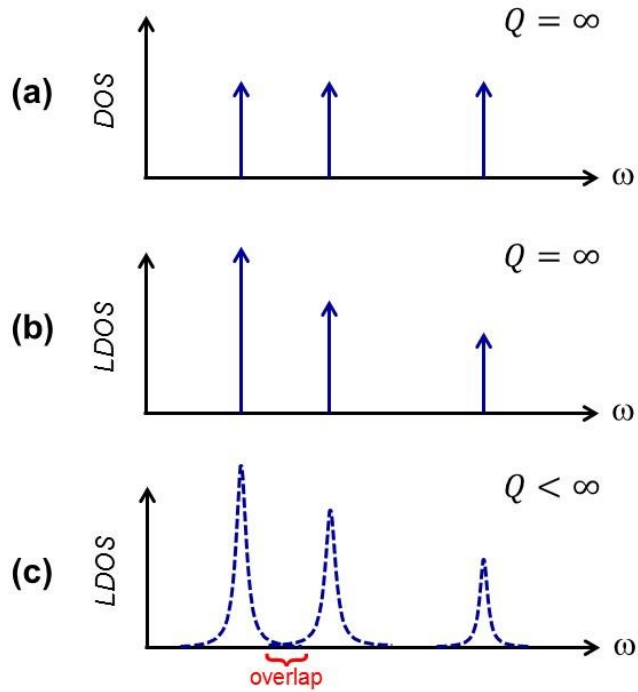

**Fig. 11. Textbook intuitive presentation of the LDOS spectrum for resonators with small losses. (a)** Ideal lossless (infinite $Q$'s) resonator (the analogue of the "particle in a box" case used in quantum mechanics): the DOS

is composed of a series of Dirac functions, see Eq. (6.1). **(b)** Associated LDOS weighted by the electric energy density, see Eq. (6.2). It is the analogue of the probability density of finding a particle in a volume element. **(c)** In the limit of high-$Q$ resonators, the LDOS becomes a series of Lorentzians, see Eq. (6.3). Each mode positively contributes to the LDOS. This phenomenological presentation, generally adopted in many textbook, is valid as $Q \to \infty$, but completely fails in spectral regions where two or several modes overlap, as discussed in Section 6.2.

We now address the emission of a dipole source in order to define the mode volume and the Purcell factor. For an electric-dipole transition, only the electric part of the LDOS matters. Indeed, the spontaneous decay rate $\gamma$ of an emitter with a dipole moment $\boldsymbol{p} = p\boldsymbol{u}$ is proportional to the electric LDOS projected on the dipole direction $\boldsymbol{u}$ [Nov06,Joh13] and

$$\gamma = \frac{\pi\omega}{\hbar\varepsilon_0}|\boldsymbol{p}|^2\varrho_u(\boldsymbol{r}_0,\omega) = |\boldsymbol{p}|^2\frac{2\omega}{\hbar}\sum_m \frac{\Gamma_m/2}{(\omega-\Omega_m)^2+(\Gamma_m/2)^2}\left|\widetilde{\boldsymbol{E}}_m(\boldsymbol{r}_0)\cdot\boldsymbol{u}\right|^2, \tag{6.4}$$

with $\boldsymbol{r}_0$ the position of the emitter and $\omega$ the emission frequency. Normalizing Eq. (6.4) by the decay rate $\gamma_0$ in a homogeneous medium with the same refractive index $n$ as the refractive index at the emitter position leads to

$$\frac{\gamma}{\gamma_0} = \sum_m F_m \frac{\Omega_m^2}{\omega^2}\frac{(\Gamma_m/2)^2}{(\omega-\Omega_m)^2+(\Gamma_m/2)^2}, \tag{6.5}$$

with $\gamma_0 = n\omega^3|\boldsymbol{p}|^2/(3\pi\hbar\varepsilon_0 c^3)$. The factor $F_m$ corresponds to the maximum spontaneous emission enhancement achieved when the $m^{\text{th}}$ mode is matched in frequency with the emitter, $\omega = \Omega_m$. It is the well-known Purcell factor

$$F_m = \frac{3}{4\pi^2}\left(\frac{\lambda_m}{n}\right)^3\frac{Q_m}{V_m}, \tag{6.6}$$

where $\lambda_m = 2\pi c/\Omega_m$ is the resonance wavelength in vacuum, $Q_m = \Omega_m/\Gamma_m$ is the quality factor of the mode and $V_m$ is the mode volume defined as

$$V_m = \frac{1}{2\varepsilon_0 n^2|\widetilde{\boldsymbol{E}}_m(\boldsymbol{r}_0)\cdot\boldsymbol{u}|^2}. \tag{6.7}$$

Note that, since $\widetilde{\boldsymbol{E}}_m$ is normalized with energy considerations $\iiint_V \widetilde{\boldsymbol{E}}_m^* \cdot \boldsymbol{\varepsilon}\widetilde{\boldsymbol{E}}_m d^3\boldsymbol{r} = 0.5$ in this Subsection dedicated to large $Q$'s, $V_m$ has the dimension of a volume. For the sake of consistency with the general derivation in Section 6.2, we have chosen to define the mode volume with the electric-field intensity at the emitter position. In the literature, the mode volume is sometimes defined at the maximum of the electric-field intensity. In that case, an additional factor appears in Eq. (6.5) that takes into account the position and polarization mismatches between the emitter and the mode.

Equations (6.3) to (6.7) are classical and can be found in many textbooks. However, we emphasize once again that they are valid only in the limit of non-overlapping resonances with high $Q$-factors. Let us now discuss how these familiar expressions are modified in the general case of a low-$Q$ resonator.

### 6.2 General case of resonators with arbitrary $Q$'s

The perturbation description is valid in the limit of large $Q$'s only. With increasing energy dissipation, either by leakage or absorption, the resonance peaks broaden and start overlapping, as sketched in Fig. 11(c). In the spectral regions where the overlap is important, several modes contribute to the LDOS and Fano resonances may arise from the interference between different modes. Because of these interferences, the LDOS can no longer be seen as a sum of Lorentzian positive contributions. Indeed, the dipolar source feeds every individual mode at the emission frequency and the total emission is given by the coherent superposition of the modes. However, energy does not sum up and destructive or constructive interference occurs in different directions. This simple reasoning evidences that the total emission cannot be decomposed as a sum of the emission (positive quantities) in every individual mode, as suggested by Eqs. (6.3)-(6.5). Similar arguments hold for cavities that are limited by absorption rather than leakage: indeed the total absorption is not the sum of the absorptions of every individual mode. It is therefore evident that the rigorous introduction of a set of QNMs for low-$Q$ systems is more involved than a simple perturbation of the lossless case.

The generalization of the modal expansion of the spontaneous emission rate of Eq. (6.5) to low-$Q$ dispersive systems requires expanding the field radiated by an electric dipole in the QNM basis. Simple analytical expressions are easily obtained by using the total-field expansion presented in Section 5.2 and by introducing a complex mode volume $\tilde{V}_m$ [Sau13]

$$\tilde{V}_m = \frac{1}{2\varepsilon_0 n^2 (\tilde{\boldsymbol{E}}_m(\boldsymbol{r}_0)\cdot\boldsymbol{u})^2}, \tag{6.8}$$

which is inversely proportional to the square of the modal electric field at the emitter position and projected along the dipole direction. $\tilde{V}_m$ represents a generalization of the usual mode volume $V_m$ and likewise depends on the emitter position. Note that, for conservative systems ($Q_m \to \infty$), the QNM field is real and Eq. (6.8) reduces to Eq. (6.7). The meaning of the complex nature of $\tilde{V}_m$ will be clarified below.

From Eq. (5.14), a meaningful expression of the normalized decay rate $\gamma/\gamma_0$ is straightforwardly derived [Sau13,Sau14,Mu16a,Pow14,Ge14,Ge15,Yan17, Zam15]

$$\frac{\gamma}{\gamma_0} = \sum_m F_m \frac{\Omega_m^2}{\omega^2} \frac{(\Gamma_m/2)^2}{(\omega-\Omega_m)^2+(\Gamma_m/2)^2} \left[1 + 2Q_m \frac{\omega-\Omega_m}{\omega} \frac{\mathrm{Im}(\tilde{V}_m)}{\mathrm{Re}(\tilde{V}_m)}\right], \tag{6.9}$$

where

$$F_m = \frac{3}{4\pi^2}\left(\frac{\lambda_m}{n}\right)^3 Q_m \mathrm{Re}\left(\frac{1}{\tilde{V}_m}\right), \tag{6.10}$$

with $\lambda_m = 2\pi c/\Omega_m$ the emission wavelength at the resonance frequency, and $n$ the refractive index of the medium in which the emitter is located. The generalized Purcell factor $F_m$ takes the same form as the textbook expression valid in the limit of weak energy dissipation, see Eq. (6.6), except that the mode volume $V_m$ becomes a complex number $\tilde{V}_m$.

For the sake of completeness, we also provide a modal expansion of the LDOS for dissipative resonant systems. The total LDOS is a sum over the three possible polarizations, $\varrho(\boldsymbol{r},\omega) = \varrho_x(\boldsymbol{r},\omega) + \varrho_y(\boldsymbol{r},\omega) + \varrho_z(\boldsymbol{r},\omega)$, and from Eq. (6.9), we have

$$\varrho_l(\boldsymbol{r},\omega) = \frac{\varepsilon_0}{\pi}\sum_m \frac{\Gamma_m |\tilde{\boldsymbol{E}}_m\cdot\boldsymbol{u}_l|^2}{(\omega-\Omega_m)^2+\left(\frac{\Gamma_m}{2}\right)^2}\cos(2\arg[\tilde{\boldsymbol{E}}_m\cdot\boldsymbol{u}_l])\left[1 - 2Q_m\frac{\omega-\Omega_m}{\omega}\tan(2\arg[\tilde{\boldsymbol{E}}_m\cdot\boldsymbol{u}_l])\right], \tag{6.11}$$

with $l = x, y, z$. Equation (6.11) reduces to Eq. (6.3) for $Q_m \to \infty$.

In summary, Eqs. (6.8), (6.10) and (6.11) generalize the classical expressions of the Purcell factor, mode volume and LDOS for systems with strong energy dissipation. A visible consequence of energy dissipation is the appearance of the phase of the QNM electric field with the term $\arg[\widetilde{\boldsymbol{E}}_m \cdot \boldsymbol{u}]$ or a complex QNM volume. Another consequence, less visible albeit fundamental, is the fact that the QNMs involved in these expressions obey the normalization condition of Eq. (4.5) and not the usual normalization of normal modes. In marked contrast with the classical expressions in Section 6.1, the presence of the phase leads to two counter-intuitive effects:

- QNMs do not contribute to the LDOS with a Lorentzian peak [Sau13], as evidenced by the imaginary part of the mode volume in Eq. (6.9). This is directly related to the phase of the modal electric field since $\mathrm{Im}(V_m)/\mathrm{Re}(V_m) = -\tan(2\arg[\widetilde{\boldsymbol{E}}_m(\boldsymbol{r}_0) \cdot \boldsymbol{u}])$. For systems that are driven by a single resonance in a given spectral range, mode volumes with large imaginary parts are dominantly observed for low-$Q$ plasmonic nanoantennas [Fri12,Sau13], for which the asymmetry of the LDOS on the red and blue sides of the resonance can be understood from the large metal dispersion over the broad spectral range covering the resonance linewidth. Non-Lorentzian line shapes with large ratios $\mathrm{Im}(V_m)/\mathrm{Re}(V_m) = 0.42$ have also be observed for high-$Q$ photonic-crystal cavities (made of non-dispersive materials) [Fag17], with specific design for which the QNM of the cavity is essentially formed by an ultra-slow Bloch mode whose band edge is close the resonance frequency. It is the highly dispersive nature of the Bloch mode that explains the surprising observation of asymmetric response for large $Q$'s.
- In multimode dissipative systems, the complexity is augmented. Not only the response of the individual QNMs is not Lorentzian, but in addition, the contribution of every QNM to the LDOS is not necessarily positive, even if the emitter is spectrally matched with the QNM [Sau13]. *Adding new QNMs may lead to a decrease of the total LDOS*, so that calculating the LDOS does not merely consist in counting the modes as for Hermitian systems. Actually, using the relation $\mathrm{Re}(V_m^{-1}) = 2\varepsilon_0 n^2 |\widetilde{\boldsymbol{E}}_m(\boldsymbol{r}_0) \cdot \boldsymbol{u}|^2 \cos(2\arg[\widetilde{\boldsymbol{E}}_m(\boldsymbol{r}_0) \cdot \boldsymbol{u}])$, it can be easily seen that the generalized Purcell factor $F_m$ is negative for $\pi/4 < \arg[\widetilde{\boldsymbol{E}}_m(\boldsymbol{r}_0) \cdot \boldsymbol{u}] < 3\pi/4$. In that case, complex Fano-like LDOS spectra can be observed.

Figure 12 provides an illustration of this complexity for a system formed by two coupled photonic-crystal cavities, which are created by removing two rods in a 2D finite-size photonic crystal formed by a 9×10 array of semiconductor rods in air [Dig06,Sau13]. In the spectral range of interest, there are two dominant QNMs1 and 2 with electric-field distributions localized in the lower and upper cavities, respectively. The solid blue and dashed green curves represent the contributions $\gamma_1/\gamma_0$ and $\gamma_2/\gamma_0$, calculated with Eq. (6.9). The sum shown with the bold red curve correctly predicts the total normalized decay rate calculated with the Green-tensor formalism (black circles), evidencing that the coupled system is bi-modal. Consistently with our previous discussion, the contribution of mode 1 (for which $\mathrm{Im}(V_1)/\mathrm{Re}(V_1) = 2.3$) to the total decay rate may be either positive or negative and is not Lorentzian at all. The origin of the large value of $\mathrm{Im}(V_1)/\mathrm{Re}(V_1)$ in a system with neither material nor geometric dispersion is due to the evanescent coupling between the two cavities that results in a remote energy delocalization. We note that increasing the quality factor of both modes (e.g. by increasing the size of the photonic-crystal array around the cavities) results in a transition towards the weak energy dissipation regime, for which both modes will have a positive and Lorentzian contribution to the total LDOS.

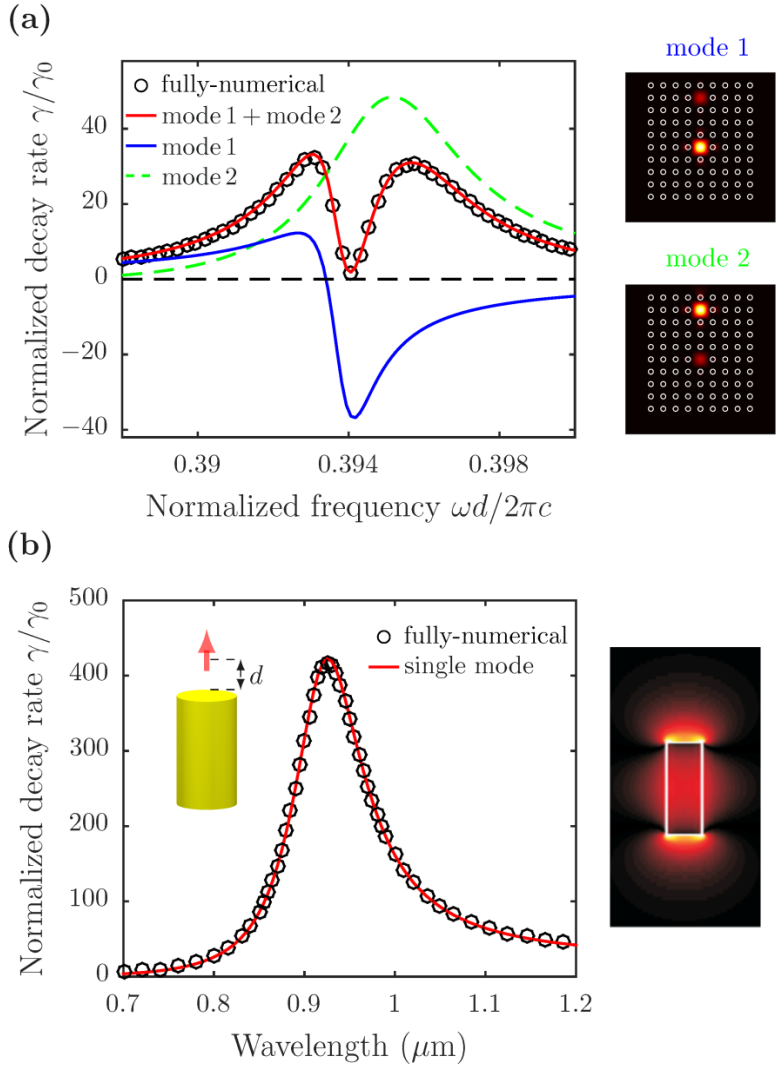

**Fig. 12. Non-Lorentzian spectral behavior of the LDOS in the presence of energy dissipation (radiative leakage or absorption). (a)** Coupled cavity system. Left: normalized decay rate of an emitter located in the upper cavity. Full numerical calculations (circles) are compared to a bi-QNM reconstruction using Eq. (6.9). The dip in the spectrum comes from the peculiar contribution of mode 1 to the total emission, which highlights the significant impact of the phase of its electric field. Right: normalized electric-field intensities of the two QNMs supported by the coupled cavities. After [Sau13]. **(b)** Plasmonic nanorod. Left: enhanced spontaneous emission factor of an emitter located on-axis 10 nm above the metal rod (see arrow in left figure). Full numerical calculations (circles) are compared to a single QNM reconstruction. Right: normalized electric-field intensities of the QNM supported by the nanorod. Adapted from [Sau13] and [Kr14a].

## 7. Cavity perturbation theory

Predicting how the presence of a foreign object near a resonant cavity or a small change in the cavity shape modifies its optical response is paramount in many fields, ranging from physics and material science to medicine. Measurement techniques based on cavity perturbation are used, for instance, to determine the dielectric and magnetic parameters of materials or to test the functionalities of microwave circuit components [Kle93]. In the optical domain, owing to their long lifetimes or highly confined resonances, microcavities or metallic nanoparticles are used to effectively convert ultra-small refractive-index changes into frequency shifts of the resonance [Lal07,Ste08]. Even single-molecule sensitivity has recently been achieved [Vol08,Zij12].

In practice, the change in the resonant system response induced by a perturbation translates into a modification of the initial eigenfrequency $\widetilde{\omega}$ of the unperturbed cavity, characterized by a frequency shift $\text{Re}(\Delta\widetilde{\omega})$ and a resonance broadening $\text{Im}(\Delta\widetilde{\omega})$, as illustrated in Fig. 13. Cavity perturbation theory describes methods for deriving analytical formulae for $\Delta\widetilde{\omega}$ assuming that electromagnetic fields in the perturbed cavity differ only slightly from the initial unperturbed fields.

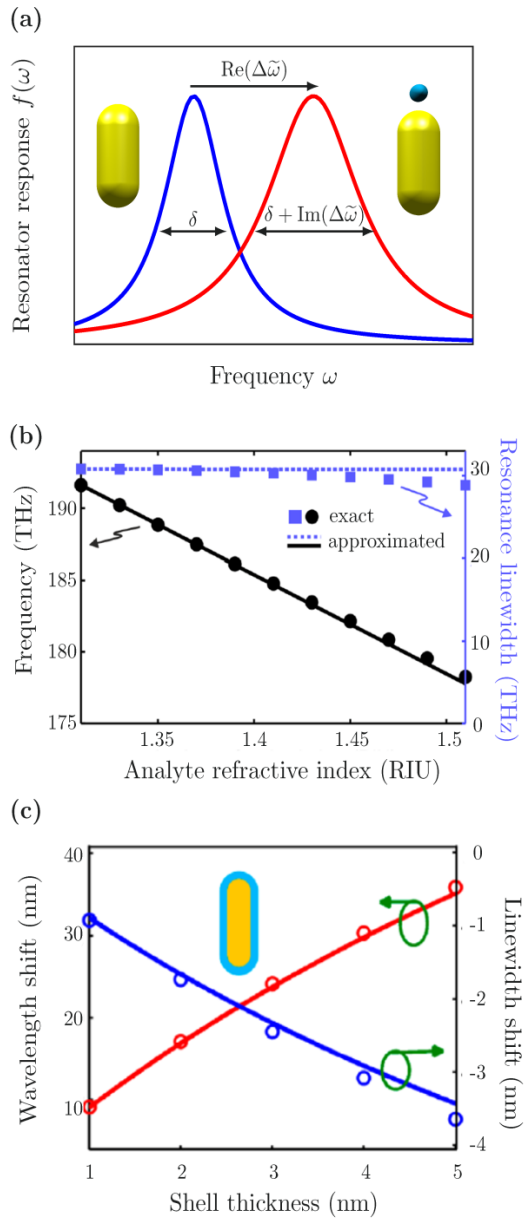

**Fig. 13. Changes in the response of a resonator due to a perturbation**. **(a)** The original resonance of the unperturbed cavity (blue) with a complex frequency $\widetilde{\omega}$ is changed into a new resonance (red) with a complex frequency $\widetilde{\omega} + \Delta\widetilde{\omega}$. $\mathrm{Re}(\Delta\widetilde{\omega})$ represents the frequency shift and $\mathrm{Im}(\Delta\widetilde{\omega})$ represents the spectral width change, often corresponding to a broadening of the resonance. A narrowing of the original resonance can also be observed. **(b) Grating case**: Resonance frequency (black) and linewidth (blue) as a function of the refractive index in the sensing volume for a periodic array of gold rod antennas. Solid and dotted lines: prediction of perturbation theory. Dots and squares: full numerical calculations. RIU denotes refractive index units. After [Wei16]. **(c) Nanoantenna case**: Resonance shifts and linewidth changes induced by a thin dielectric shell around a plasmonic nanorod (inset) in aqueous environment. Full numerical calculations (circles) are compared to perturbation theory predictions (solid curves). After [Ya15b].

The failure of a correct normalization for the modes of open resonators has durably impacted the development of cavity perturbation theory. The initial works have all assumed that the cavity mode can be normalized through energy consideration, $\iiint_V \left[\varepsilon|\widetilde{\boldsymbol{E}}|^2 + \mu|\widetilde{\boldsymbol{H}}|^2\right]d^3\boldsymbol{r} = 1$, to derive first-order frequency-shift formulae of the form [Kle93]

$$\Delta\widetilde{\omega} = -\widetilde{\omega}\iiint_{V_p}\Delta\varepsilon(\boldsymbol{r},\widetilde{\omega})\left|\widetilde{\boldsymbol{E}}(\boldsymbol{r})\right|^2 d^3\boldsymbol{r}, \tag{7.1}$$

where $\widetilde{\omega}$ and $\widetilde{\boldsymbol{E}}(\boldsymbol{r})$ represent the unperturbed QNM frequency and electric field, the integral is performed over the perturbation volume $V_p$, and $\Delta\varepsilon$ is the change of permittivity distribution introduced by the perturbation. The literature on photonic-crystal cavities [Koe05,Ram09] or microwave resonators [Kle93] have consistently relied on this kind of formula involving stored energy considerations.

Common sense dictates that the maximum change in resonant frequency is achieved when the perturbation is placed at the maximum of the resonant mode intensity, thereby making Eq. (7.1) appear physically sound. Our intuition is yet misleading in this case. The range of validity of Eq. (7.1) is, as a matter of fact, very limited. For low-Q resonators, such as plasmonic particles, it is generally not valid at all, and for high-Q resonators, it correctly predicts the spectral shift $\mathrm{Re}(\Delta\widetilde{\omega})$, but fails for the bandwidth change $\mathrm{Im}(\Delta\widetilde{\omega})$. This can easily be understood by considering dielectric perturbations with real $\Delta\boldsymbol{\varepsilon}$. Since the volume integral in Eq. (7.1) is a real number, $\mathrm{Im}(\Delta\widetilde{\omega})/\mathrm{Im}(\widetilde{\omega}) = \mathrm{Re}(\Delta\widetilde{\omega})/\mathrm{Re}(\widetilde{\omega})$, regardless of the mode profile and of the perturbation position. This would imply that the normalized shift should always be equal to the normalized bandwidth change, and that a blue (resp. red) shift should always be accompanied by a broadening (resp. a narrowing) of the resonance. This is illogical, and indeed, Eq. (7.1) has never been used to predict quality factor changes even for high-Q resonances.

As discussed in Sections 4 and 6, a leaky mode cannot be normalized with energy considerations based on $\widetilde{\boldsymbol{E}}\cdot\widetilde{\boldsymbol{E}}^*$ since $\widetilde{\boldsymbol{E}}(\boldsymbol{r})$ exponentially diverges as $|\boldsymbol{r}|\rightarrow\infty$. This normalization is valid only for infinite-Q resonances, for which the electric field components are purely real. The impact of leakage or absorption on the perturbed resonance is indeed encoded in $\mathrm{Im}(\widetilde{\boldsymbol{E}})$, implying that any expression like Eq. (7.1) that ignores the phase of the mode cannot be used to consistently predict the effect of a perturbation on the mode itself.

Equipped with the correct QNM normalization, see Eq. (4.5), it is possible to remove the mistake brought by the energy normalization. In general, all the perturbed QNMs of the perturbed resonators can be expanded into the complete set of the unpertubed QNMs, and the expansion coefficients can be computed by solving a linear matrix eigenvalue problem [Mul10,Doo14,Doo16,Via16]. This rigorous expansion approach works well provided that the basis is complete and remains valid well beyond the perturbation regime. For instance, it may be used to compute the QNMs of a gold sphere from the sole knowledge of the QNMs of the same sphere made of silica [Mu16a], amazingly. We restrict ourselves to first-order perturbations hereafter, and assume that the frequency shift $\Delta\widetilde{\omega}$ can be accurately predicted solely by considering the unperturbed QNM $\widetilde{E}(r)$. The application of the correct QNM normalization leads to the formula [Ya15b]

$$\Delta\widetilde{\omega} = -\widetilde{\omega} \iiint_{V_p} \Delta\varepsilon(r,\widetilde{\omega})\, \widetilde{E}(r) \cdot \widetilde{E}(r)\, d^3r, \qquad (7.2)$$

where the $\widetilde{E} \cdot \widetilde{E}^*$ product in Eq. (7.1) is simply replaced by the unconjugated product $\widetilde{E} \cdot \widetilde{E}$. Although this replacement may appear minor, Eq. (7.2) considerably enlarges the validity range of cavity perturbation theory, owing to the fact that the QNM phase is properly taken into account. The predictive force of Eq. (7.2) has been thoroughly tested in [Ya15b]. It was shown that, by including local field corrections induced by the perturbation, the closed-form expression of Eq. (7.2) is highly accurate for various plasmonic nanoresonators used in plasmonic sensing technologies [Ste08], such as dimers composed by two identical gold nanorods and gold nanocones, and for perturbations with different shapes (local tiny spheroids and extended thin shells covering the resonator surface), refractive indices and positions.

For the sake of illustration, let us consider the asymptotic case of perturbations formed by deep-subwavelength spheroids of permittivity $\Delta\varepsilon + \varepsilon_b$ that are introduced into a background material of permittivity $\varepsilon_b$. In the static limit, the spheroid can be considered as a point-like particle positioned at $r_p$ with an isotropic electric dipolar polarizability $\alpha = V_p \frac{3\Delta\varepsilon}{\Delta\varepsilon + 3\varepsilon_b}$. Taking into account these local field corrections [Ya15b], we straightforwardly get $\Delta\widetilde{\omega} = -\widetilde{\omega}\, V_p \frac{3\Delta\varepsilon}{\Delta\varepsilon + 3\varepsilon_b} \widetilde{E}^2(r_p)$ from Eq. (7.2). The present issue is very similar to that encountered in Section 6 for the emission of an electric dipole. Modifying slightly the definition of the mode volume to consider the isotropic response of the induced dipole, and supposing that both $\Delta\varepsilon$ and $\Delta\varepsilon + \varepsilon_b$ are real, one finds that

$$\mathrm{Re}(\Delta\widetilde{\omega}) \approx -\mathrm{Re}(\widetilde{\omega}) \frac{3}{2} \frac{\Delta\varepsilon}{\Delta\varepsilon + 3\varepsilon_b}\left[\mathrm{Re}\left(\frac{V_p}{\widetilde{V}_I}\right) - \frac{1}{2Q}\mathrm{Im}\left(\frac{V_p}{\widetilde{V}_I}\right)\right], \qquad (7.3a)$$

$$\mathrm{Im}(\Delta\widetilde{\omega}) \approx -\mathrm{Re}(\widetilde{\omega}) \frac{3}{2} \frac{\Delta\varepsilon}{\Delta\varepsilon + 3\varepsilon_b}\left[\frac{1}{2Q}\mathrm{Re}\left(\frac{V_p}{\widetilde{V}_I}\right) + \mathrm{Im}\left(\frac{V_p}{\widetilde{V}_I}\right)\right], \qquad (7.3b)$$

where the isotropic complex mode volume

$$\widetilde{V}_I = \frac{1}{2\varepsilon_b \widetilde{E}^2(r_p)} \qquad (7.4)$$

is introduced for convenience. Equations (7.3a) and (7.3b) evidence that the complex mode volume plays an important role in predicting the resonance frequency change. In general, $\mathrm{Re}\left(\frac{V_p}{\widetilde{V}_I}\right) \gg \frac{1}{2Q}\mathrm{Im}\left(\frac{V_p}{\widetilde{V}_I}\right)$, and the resonance shift is mainly impacted by the ratio of the spheroid and QNM volumes. However, since $\mathrm{Im}(\widetilde{E})/\mathrm{Re}(\widetilde{E}) = O(1/Q)$, both terms in the right side of Eq. (7.3b) are of the same order of magnitude even for resonators with large $Q$'s. Presently, it is important to notice that, irrespectively of the sign of $\Delta\varepsilon$, the bracketed term in Eq. (7.3b) can be either positive or negative, implying that it is equally possible to broaden or narrow the resonance by perturbing. This is fixed by the sign of $\text{Re}\left(\frac{V_p}{\widetilde{V}_I}\right) + 2Q\text{Im}\left(\frac{V_p}{\widetilde{V}_I}\right)$, which depends on the QNM field profile and the location of the perturbation.

The closed-form expressions of Eqs. (7.2), (7.3a) and (7.3b) are simple and just require the knowledge of the unperturbed mode. They provide intuitions that may help early designs or interpretation of experimental results, avoiding tedious fully-vectorial electromagnetic computations for various parameters, such as the location, size, or refractive index of the perturbation, usually performed to theoretically predict $\Delta\widetilde{\omega}$. They may also be decisive for applications relying on the cross-action of electromagnetic resonant fields with other physical effects, e.g. optomechanical cooling [Kip08], plasmonic trapping [Gie12], photonic bistability [Kau12], plasmon-enhanced Raman scattering [Roe16], which require repeated computations of the full photon Green function to iteratively model the nonlinear dynamics, although the nonlinear permittivity changes are small.

## 8. Strong coupling and superradiance

Light emission is not an intrinsic property of matter and can be tailored by modifying the environment of the emitting source. As we saw in Section 6, QNMs are helpful to understand how the spontaneous emission rate of a quantum emitter changes in the weak coupling regime, i.e., when the coupling of the emitter with the environment is weak enough so that the emission frequency is unchanged. There are two other important situations where light-matter interactions can be suitably described with QNM expansions.

The first one refers to the strong coupling regime, for which the characteristic time of energy transfer between the emitter and the optical modes of the environment (often a cavity) can become smaller than the characteristic damping times of the coupled system [Har89]. The photons can then be "reabsorbed" by the emitter before being lost and spontaneous emission becomes a reversible process; the energy oscillates in time between photonic and electronic energies, a phenomenon known as Rabi oscillations. Strong coupling is in general characterized by a modification of the emission frequency: the energy levels of the coupled system are different from those of the emitter and the optical environment taken individually.

The second important situation corresponds to collective (or cooperative) effects in spontaneous emission by an ensemble of $N$ quantum emitters [Gro82]. Superradiance is one example of such collective phenomena that occur when $N$ emitters interact with the same light field [Dic54]. When the ensemble is dense (i.e., the wavelength is larger than the distance separating the emitters), the emitters collectively and coherently respond to the excitation, resulting in a boosted emission rate that is $N$ times larger than the one of the set of emitters taken independently. Superradiance has been observed in various systems, including atomic clouds, quantum dot arrays and J-aggregates.

The objective of this Section is to show how QNMs may be used to study these two emblematic regimes of light emission (strong coupling and collective spontaneous emission), in particular when the emitters are coupled to a low-$Q$ cavity such as a plasmonic nanoresonator. We have seen in Section 6 that, for systems with large energy dissipation, the weak coupling regime is not correctly modeled with normal modes normalized with energy considerations and perturbed by a weak coupling to a reservoir. A correct description of this interaction regime was obtained thanks to a proper normalization of QNMs and the definition of complex mode volumes. We will see in this Section that similar concepts can be used to describe strong coupling and collective spontaneous emission.

For the sake of simplicity, we assume hereafter that the optical resonator supports a single mode in the spectral range of interest, characterized by a QNM with a complex frequency $\widetilde{\omega}_m = \Omega_m - i\Gamma_m/2$ and an electric field $\widetilde{\boldsymbol{E}}_m(\boldsymbol{r})$. We further adopt a classical description for the two-level system with a classical point-like scatterer positioned at $\boldsymbol{r}_e$ and characterized by a resonant electric-dipolar polarizability matrix $\boldsymbol{\alpha}(\omega)$. Any exciting electric field $\boldsymbol{E}(\boldsymbol{r})$ induces an electric dipole $\boldsymbol{p} = \varepsilon_0 \boldsymbol{\alpha}(\omega) \boldsymbol{E}(\boldsymbol{r}_e)$. For a two-level system with a transition dipole linearly polarized and placed in a medium of refractive index $n$, the polarizability matrix is $\boldsymbol{\alpha}(\omega) = \alpha(\omega) \boldsymbol{u} \otimes \boldsymbol{u}$ with $\boldsymbol{u}$ the unit vector along the polarization direction aligned with one axis of the coordinate system. The scalar polarizability near resonance $\alpha(\omega)$ is given by

$$\alpha(\omega) = \frac{3\pi c^3}{\omega^3} \frac{\gamma_e/n}{\omega_e - \omega - i\gamma_e/2}, \tag{8.1}$$

where $\omega_e$ is the transition frequency and $\gamma_e$ is the decay rate of the excited state in the medium of refractive index $n$. The resonant character of the two-level system can be described by a dipolar QNM with a complex pole $\widetilde{\omega}_e = \omega_e - i\gamma_e/2$. The prefactor $3\pi c^3 \gamma_e/(n\omega^3)$ is chosen so that the absorption cross-section associated to this polarizability is strictly equal to zero for any frequency [Boh83]. Hence, the polarizability in Eq. (8.1) does include neither non-radiative nor dephasing processes. Non-radiative processes could be taken into account simply by replacing $\gamma_e$ in the denominator of Eq. (8.1) by $\gamma_e + \gamma_e^{nr}$ while keeping only $\gamma_e$ in the numerator. Moreover, choosing other forms of the polarizability matrix allows to model different types of transitions. For instance, an isotropic polarizability matrix $\boldsymbol{\alpha}(\omega) = \alpha(\omega) \boldsymbol{I}$ with $\boldsymbol{I}$ the identity matrix would correspond to a classical $J = 0 \rightarrow J = 1$ atom with three transitions.

## 8.1 Strong coupling between QNMs and two-level systems

When an emitter is coupled to a cavity mode, the coupled system can no longer be described by the initial QNMs of the individual resonators with eigenfrequencies $\widetilde{\omega}_e$ and $\widetilde{\omega}_m$, but by new, hybrid eigenstates. In general, the so-called strong coupling is defined as the regime when the frequency splitting between the hybrid modes is larger than their linewidths [Ger03,Tor15]. Single-emitter strong coupling was first observed with microwave cavities [Rem87] and then with atoms in optical cavities [Tho92]. Solid-state single-emitter strong coupling was achieved with epitaxial quantum dots in photonic crystal microcavities [Yos04,Rei04] and more recently with molecules or colloidal quantum dots in plasmonic nanoresonators [Chi16,San16]. Strong coupling can also be observed with $N > 1$ emitters, see [Tor15] for a comprehensive review. Such regime is less pure from a fundamental point of view but is easier to observe since the frequency splitting is proportional to $\sqrt{N}$ [Tor15].

To find the hybridized eigenstates, we write the self-consistent equation that gives the induced dipole moment of the two-level system in the absence of incident field. The induced moment is given by $p = \varepsilon_0 \alpha(\omega) \boldsymbol{E}(\boldsymbol{r}_e) \cdot \boldsymbol{u}$, with $\boldsymbol{E}(\boldsymbol{r}_e)$ the field radiated by the two-level system and backscattered by the resonator at $\boldsymbol{r}_e$. With the single-QNM-resonator assumption, the field $\boldsymbol{E}(\boldsymbol{r}_e)$ is proportional to the QNM field and using the total-field expansion presented in Section 5.2, $\boldsymbol{E}(\boldsymbol{r}_e)$ can be written as $\boldsymbol{E}(\boldsymbol{r}_e) \approx a_m \widetilde{\boldsymbol{E}}_m(\boldsymbol{r}_e)$, where the excitation coefficient $a_m = -\frac{\omega p \widetilde{\boldsymbol{E}}_m(\boldsymbol{r}_e) \cdot \boldsymbol{u}}{\omega - \widetilde{\omega}_m}$ is given by Eq. (5.13). Two coupled equations, $p = \varepsilon_0 \alpha(\omega) E$ and $E = -\frac{\omega (\widetilde{\boldsymbol{E}}_m(\boldsymbol{r}_e) \cdot \boldsymbol{u})^2}{\omega - \widetilde{\omega}_m} p$, have to be solved to get the induced dipole moment $p$ and the field at the emitter position projected along the unit vector $\boldsymbol{u}$, $E = \boldsymbol{E}(\boldsymbol{r}_e) \cdot \boldsymbol{u}$.

Using the polarizability of Eq. (8.1), these equations are recast into a homogenous linear system,

$$\begin{pmatrix} \widetilde{\omega}_e - \omega & -\kappa_{12}(\omega) \\ -\kappa_{21}(\omega) & \widetilde{\omega}_m - \omega \end{pmatrix} \begin{pmatrix} p \\ E \end{pmatrix} = 0, \tag{8.2}$$

with $\kappa_{12}(\omega) = \varepsilon_0 \frac{3\pi c^3}{n\omega^3} \gamma_e$ and $\kappa_{21}(\omega) = \omega(\widetilde{\mathbf{E}}_m(\mathbf{r}_e) \cdot \mathbf{u})^2$. Equation (8.2) is the most general form of the coupling between two oscillators with frequency-dependent and complex coupling constants, $\kappa_{12}(\omega) \neq \kappa_{21}(\omega)$. The complex frequencies of the hybridized eigenstates are the zeros of the determinant and verify

$$(\widetilde{\omega}_e - \omega)(\widetilde{\omega}_m - \omega) - \kappa_{12}(\omega)\kappa_{21}(\omega) = 0. \tag{8.3}$$

The product $\kappa_{12}(\omega)\kappa_{21}(\omega)$ varies as $1/\omega^2$ and Eq. (8.3) can be solved numerically by iterative zero-searching methods. Closed-form expressions are easily derived by assuming that the variation of the product of the coupling constants over the spectral range of interest (typically given by $\Gamma_m$) is small and we set $\kappa_{12}(\omega)\kappa_{21}(\omega) \approx \kappa_{12}(\Omega_m)\kappa_{21}(\Omega_m) \equiv g^2$. By introducing the expressions of the mode volume $\widetilde{V}_m = \frac{1}{2\varepsilon_0 n^2 [\widetilde{\mathbf{E}}_m(\mathbf{r}_e) \cdot \mathbf{u}]^2}$ and of the Purcell factor $F_m = \frac{3}{4\pi^2} \left(\frac{\lambda_m}{n}\right)^3 Q_m Re\left(\frac{1}{\widetilde{V}_m}\right)$ both defined in Section 6.2, one gets

$$g^2 = \Omega^2 \left[1 - i\frac{\mathrm{Im}(\widetilde{V}_m)}{\mathrm{Re}(\widetilde{V}_m)}\right], \tag{8.4}$$

where $\Omega$ is the usual expression of the Rabi frequency, $\Omega^2 = F_m \gamma_e \Gamma_m / 4$, which characterizes the coupling strength [Ger03]. The two complex eigenfrequencies of the hybridized states are $\widetilde{\omega}_{\pm} = \frac{\widetilde{\omega}_e + \widetilde{\omega}_m}{2} \pm \sqrt{g^2 + \left(\frac{\widetilde{\omega}_e - \widetilde{\omega}_m}{2}\right)^2}$. When the resonance frequencies of the two-level system and the resonator are matched, $\Omega_m = \omega_e$, one simply obtains

$$\widetilde{\omega}_{\pm} = \omega_e - i\frac{\gamma_e + \Gamma_m}{4} \pm \sqrt{\Omega^2 - \left(\frac{\gamma_e - \Gamma_m}{4}\right)^2 - i\Omega^2 \frac{\mathrm{Im}(\widetilde{V}_m)}{\mathrm{Re}(\widetilde{V}_m)}}. \tag{8.5}$$

With the usual mode normalization based on energy considerations, the volume becomes a real number inversely proportional to $|\widetilde{\mathbf{E}}_m(\mathbf{r}_e) \cdot \mathbf{u}|^2$ and the imaginary term under the square root sign does not show up: the coupling strength $g$ is real-valued [Rai95,Ger03,Nov10,Tor15]. The additional term is proportional to $\mathrm{Im}(\widetilde{V}_m)/\mathrm{Re}(\widetilde{V}_m)$ and directly arises from the presence of energy dissipation, as discussed in Sections 6 and 7. If the resonator quality factor is large enough, the imaginary part of the coupling strength can be neglected and we have the usual clear boundary between the weak and strong coupling regimes, as depicted in Fig. 14(a). For $\Omega^2 < \left(\frac{\gamma_e - \Gamma_m}{4}\right)^2$, the square root is purely imaginary and the emission frequency is unaltered by the coupling, whereas the decay rates are modified. This is the weak coupling regime. Reversely, for $\Omega^2 > \left(\frac{\gamma_e - \Gamma_m}{4}\right)^2$, the two hybrid modes have different frequencies but the same decay rate. The additional term due to the presence of non-negligible energy dissipation results in a blurring of the boundary between the two regimes, with a non-zero frequency splitting whatever the sign of $\Omega^2 - \left(\frac{\gamma_e - \Gamma_m}{4}\right)^2$. This situation is promoted with low-$Q$ resonators such as plasmonic nanoantennas [Sau13] or photonic-crystal cavities with highly-dispersive intra-cavity elements [Fa17b].

Figure 14 illustrates the impact of the ratio $\text{Im}(\tilde{V}_m)/\text{Re}(\tilde{V}_m)$ on the splitting of the hybrid eigenstates, showing a strong decrease of the linewidth of the upper hybrid mode as this ratio increases. As a consequence, the value of the coupling strength $\Omega$ necessary to observe strong couplings is slightly decreased, see the position of the dashed vertical lines in Fig. 14. We define here the boundary between weak and strong couplings as the point when the frequency splitting is larger than the half sum of the linewidths. The data shown in Fig. 14 have been obtained by applying Eq. (8.5) to an emitter with a linewidth $\gamma_e = 1$ GHz coupled to a plasmonic resonator with a typical $Q$-factor of 10. For these values, the weak-to-strong coupling transition takes place for a coupling strength of the order of 50 THz, which corresponds to a mode volume $\text{Re}(\tilde{V}_m) \sim 5\, \lambda^3/10^6$, a typical value for plasmonic nanoresonators that have been used recently to demonstrate strong coupling with a few emitters [Chi16,San16]. Finally, note that the values of the ratio $\text{Im}(\tilde{V}_m)/\text{Re}(\tilde{V}_m)$ that we consider in Fig. 14 can be encountered in realistic plasmonic systems, such as gold nanorods dimers [Sau13].

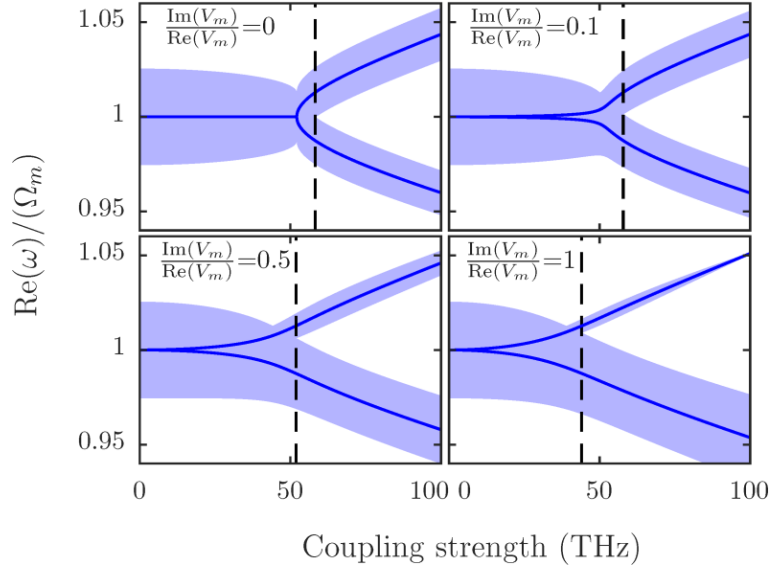

**Fig. 14. Strong coupling between an emitter and a plasmonic nanoresonator that are matched in frequency, $\Omega_m = \omega_e$.** The figures display the real parts of the frequencies $\tilde{\omega}_\pm$ of the two hybrid modes (enlarged by their linewidths) as a function of the real part of the coupling strength, i.e., the Rabi frequency $\Omega$. **(a)** $\text{Im}(\tilde{V}_m)/\text{Re}(\tilde{V}_m) = 0$, 0.1, 0.5 and 1. Note that $\text{Im}(\tilde{V}_m)/\text{Re}(\tilde{V}_m) = 0$ corresponds to the result of the usual formalism in which energy dissipation is introduced as a perturbation [Ger03]. The curves are calculated by applying Eq. (8.5) to an emitter with a linewidth $\gamma_e = 1$ GHz coupled to a plasmonic resonator with a typical $Q$-factor of 10. The vertical dashed lines mark the limit between the weak coupling (left) and strong coupling (right) regimes.

The approach developed in this Section is purely classical and does not include saturation or non-linear effects but it leads to the correct splitting for the energy levels [Nov10,Tor15]. Deriving a semi-classical description with a quantized description of the two-level system can be performed with the same classical QNM expansion and does not require further development in QNM theory [Yan15]. In contrast, a fully quantum description requires a field quantization compatible with the use of QNMs, taking into account their orthogonality. This is a more challenging task. For that purpose, one could use the field-quantization formalism proposed in the 90's for absorptive and dispersive media [Dun98,Gru96,Mat95]. Such an approach has been recently used to study the interaction of two quantum emitters with a plasmonic nanoresonator [Ge15] and non-linear effects at the single plasmon level with metallic nanodimers [Alp16].

**8.2 Resonance-assisted superradiance**

Collective effects in an ensemble of quantum emitters, e.g. cold molecules or atoms, are becoming increasingly important in modern science and technology [Gro82,Rai82]. Of particular interest are high cooperativity regimes reached by coupling the ensemble with an electromagnetic resonance. The latter reinforces cooperativity by promoting long-range interactions and exploiting cavity quantum electrodynamics. Recent exemplary advances encompass the generation of coherent visible radiation by many emitters placed near plasmonic nanoparticles [Ber03,Nog09,Boh12] or hybrid quantum systems combining cold-atom clouds with photonic-crystal resonances [Tho13,Gob14].

Even for steady-state cases, the theoretical analysis of quantum hybrids represents a major challenge in computational electrodynamics, requiring the repeated calculation of the full photon Green's tensor of the resonator for different frequencies and atomic positions [Pus09] to compute sequentially, one by one, the complex eigenfrequencies $\widetilde{\omega}_m$ by iterative pole-searching methods. The challenge increases in complexity when studying the dynamics by iteratively solving coupled equations for the Maxwell's fields and carrier-population operators, or when computing ensemble-averaged responses to interpret experiments for which the exact location and orientation of atoms are unknown.

It is possible to extend the preceding strong-coupling QNM formalism to model quantum hybrids composed by a large number $N$ of oscillators coupled by a microcavity or a nanoresonator, without any restriction on the resonator shape, material properties, or polarization orientations. The QNM expansion brings analyticity in the theoretical treatment and thus considerably reduces the computational loads. As a consequence, a direct computation of all sub/super-radiant states of the hybrid system becomes feasible in a small timescale by solving a generalized eigenproblem and some essential properties of hybrids, which are robust to spatial and polarization disorders, become predictable. For instance, it can be shown that the mean decay rate $\langle \Gamma_{sup} \rangle$ of the superradiant states, where $\langle ... \rangle$ denotes an ensemble average, scales linearly with the mean Purcell Factor $\langle F_p \rangle$, evidencing that ultra-bright hybrid states in large ensembles of emitters coupled via an electromagnetic resonance arise from the combination of Dicke and Purcell effects [Fau17], and not from direct dipole-dipole interactions like in the classical Dicke effect [Dic54].

# 9. Light scattering by resonant systems

**9.1 Scattering and absorption properties of complex nanoparticles**

The recent advent of complex electromagnetic nanoparticles not only offers completely new interaction regimes at the nanoscale, but also provides new opportunities to manipulate the light absorption and scattering over the visible and near-infrared ranges [Boh83]. A first great step forward has been made with the realization of composite, metallo-dielectric nanoparticles, to tune the resonance frequency, almost at will, over a broad spectral range typically covering the visible and near-infrared [Old98]. The optical properties can be also tailored by promoting the excitation of resonances of orders higher than the standard electric dipole, either by breaking the symmetry [Wan06,Mir09] or carefully designing the nanoparticle shapes [Sta13]. Yet additional exotic properties can be achieved by promoting electromagnetic interactions, between the nanoparticles themselves in small nanoparticle aggregates [Dan04], or with guided modes for nanoparticles buried in structured substrates, such as metallo-dielectric thin-film stacks supporting guided modes [Mor12,Vyn12,Jou16], or with interfaces [Che11].

As a natural consequence, the theoretical analysis of light interaction with small particles has experienced numerous developments over the years, among which the well-known Mie theory, which provides analytical solutions for spherical particles of any size in terms of resonant modes [Boh83]. Analytical expressions for the polarizability of other regular objects (e.g. ellipsoids) and some hybrids can also be obtained. However, in the general case of objects with non-regular shape, exact analytical models are not possible and one has to resort to full-wave numerical calculations to retrieve quantities such as the radiation pattern and the scattering/absorption cross-sections. A complete description of these properties, up to recently, was obtained by repeated calculations at real frequencies for various excitation fields. Fortunately, the concept of QNM can be exploited as well for this purpose, providing a large reduction of computational cost [Bai13,Pow17,Per16] and a great physical insight especially when a few resonances spectrally overlap and result in a Fano response.

Hereafter, we adopt the scattered-field formalism described in Annex 2. The background medium excluding the particle has a permittivity $\varepsilon_b(\mathbf{r}, \omega)$ and hosts an incident (background) electromagnetic field $[\mathbf{E}_b, \mathbf{H}_b]$, which is usually a plane wave for probing the cross-sections. The inclusion of a particle modifies the background permittivity to $\varepsilon(\mathbf{r}, \omega) = \varepsilon_b(\mathbf{r}, \omega) + \Delta\varepsilon(\mathbf{r}, \omega)$, where $\Delta\varepsilon$ is null everywhere expect in the particle volume denoted as $V_r$. The particle induces a scattered field $[\mathbf{E}_s, \mathbf{H}_s]$, such that the total field $[\mathbf{E}, \mathbf{H}]$ is given by $[\mathbf{E}, \mathbf{H}] = [\mathbf{E}_b, \mathbf{H}_b] + [\mathbf{E}_s, \mathbf{H}_s]$.

The Poynting theorem [Jac99] gives analytical expressions for the absorbed ($P_\text{abs}$) and scattered ($P_\text{sca}$) powers as

$$P_\text{abs} = \frac{\omega}{2} \iiint_{V_r} \text{Im}(\varepsilon) |\mathbf{E}_s + \mathbf{E}_b|^2 \, d^3\mathbf{r}, \tag{9.1a}$$

$$P_\text{sca} = \frac{1}{2} \iint_{\Sigma_r} \text{Re}(\mathbf{E}_s \times \mathbf{H}_s^*) \cdot d\mathbf{S}, \tag{9.1b}$$

where $\Sigma_r$ denotes the surface that encloses $V_r$. We note that the scattered power $P_\text{sca}$ might be allocated to various output channels, depending on the complexity of the background. For example, in a uniform lossless background, $P_\text{sca}$ is merely carried by radiation modes that propagate to infinity while for particles deposited on or embedded in stratified structures, guided modes may carry part of the scattered power as well. Moreover, if the background medium is lossy, the scattered light would be partly damped by absorption during its propagation.

Equations (9.1) with the QNM expansions for the scattered field $[\mathbf{E}_s, \mathbf{H}_s]$ (see Section 5) could be directly applied to compute $P_\text{abs}$ and $P_\text{sca}$. Nevertheless, Ref. [Bai13] suggests that, as evidenced from numerical tests, a better accuracy of $P_\text{sca}$ is reached when a small number of the QNMs is used if $P_\text{sca}$ is expressed as a volume integral over $V_r$, like in Eq. (9.1a), instead of using a surface integral as in Eq. (9.1b). Let us slightly modify the equations for the scattered field $[\mathbf{E}_s, \mathbf{H}_s]$ of Eq. (A2-3): $\nabla \times \mathbf{E}_S = i\omega\boldsymbol{\mu}(\mathbf{r}, \omega)\mathbf{H}_S$ and $\nabla \times \mathbf{H}_S = -i\omega\varepsilon_b(\mathbf{r}, \omega)\mathbf{E}_S - i\omega\Delta\varepsilon(\mathbf{r}, \omega)(\mathbf{E}_b(\mathbf{r}) + \mathbf{E}_S)$. The equations tell us that $[\mathbf{E}_s, \mathbf{H}_s]$ can be seen as the field generated in the background medium by a current-source distribution $-i\omega\Delta\varepsilon(\mathbf{r}, \omega)(\mathbf{E}_b(\mathbf{r}) + \mathbf{E}_S)$. Then, applying the Poynting theorem [Jac99], we obtain $\frac{1}{2}\iint_{\Sigma_r} \text{Re}(\mathbf{E}_s \times \mathbf{H}_s^*) \cdot d\mathbf{S} = -\frac{\omega}{2}\iiint_{V_r} \text{Im}(\varepsilon_b)|\mathbf{E}_s|^2 d^3\mathbf{r} - \frac{\omega}{2}\iiint_{V_r} \text{Im}[\Delta\varepsilon \mathbf{E}_S^*(\mathbf{E}_S + \mathbf{E}_b)] \, d^3\mathbf{r}$, thereby leading to an alternative expression of $P_\text{sca}$

$$P_\text{sca} = -\frac{\omega}{2} \iiint_{V_r} \text{Im}(\epsilon_b) |\mathbf{E}_s|^2 d^3\mathbf{r} - \frac{\omega}{2} \iiint_{V_r} \text{Im}[\Delta\varepsilon \mathbf{E}_S^*(\mathbf{E}_S + \mathbf{E}_b)] \, d^3\mathbf{r}. \tag{9.2}$$

The summation of $P_{abs}$ and $P_{sca}$ with Eqs. (9.1a) and (9.2) then gives the extinction power

$$P_{ext} = \frac{\omega}{2} \iiint_{V_r} \text{Im}[\Delta\varepsilon \boldsymbol{E}_b^*(\boldsymbol{E}_s + \boldsymbol{E}_b)] \, d^3\boldsymbol{r} + \frac{\omega}{2} \iiint_{V_r} \text{Im}(\epsilon_b)[|\boldsymbol{E}_s + \boldsymbol{E}_b|^2 - |\boldsymbol{E}_s|^2] \, d^3\boldsymbol{r}. \quad (9.3)$$

When the background is lossless, $\text{Im}(\epsilon_b) = 0$, the second term on the right hand side of Eq. (9.3) could be dropped simply and Eq. (9.3) reduces to the one for lossless backgrounds as proposed in [Bai13].

The powers $P_{abs}$, $P_{sca}$, and $P_{ext}$ can therefore be computed by combining Eqs. (9.1a), (9.2), (9.3) and the QNM expansion $\boldsymbol{E}_s = \sum_m \alpha_m(\omega) \widetilde{\boldsymbol{E}}_m(\boldsymbol{r})$ (see Section 5). The absorption, scattering, and extinction cross-sections, denoted respectively by $\sigma_{abs}$, $\sigma_{sca}$, and $\sigma_{ext}$ are then obtained with their definitions $\sigma_{abs} = P_{abs}/S_0$, $\sigma_{sca} = P_{sca}/S_0$, and $\sigma_{ext} = P_{ext}/S_0$, where $S_0$ stands for the time-averaged Poynting vector of the incident plane wave. As shown in Section 5, once the QNMs are known, the modal excitation coefficients, $\alpha_m$'s, are efficiently computed by simply performing the overlap integral between the normalized QNM field and the background field, so the computational loads to compute the cross-section spectra for many incidence are very low.

Figure 15 shows the extinction cross-section spectra of a gold dolmen nanoresonator composed of three rods. The excellent agreement between the QNM-expansion prediction and COMSOL computational results evidences that the QNM formalism (with only three QNMs retained in the computation) allows for a quantitative prediction of the Fano-like spectra with the precise contributions from every individual QNM.

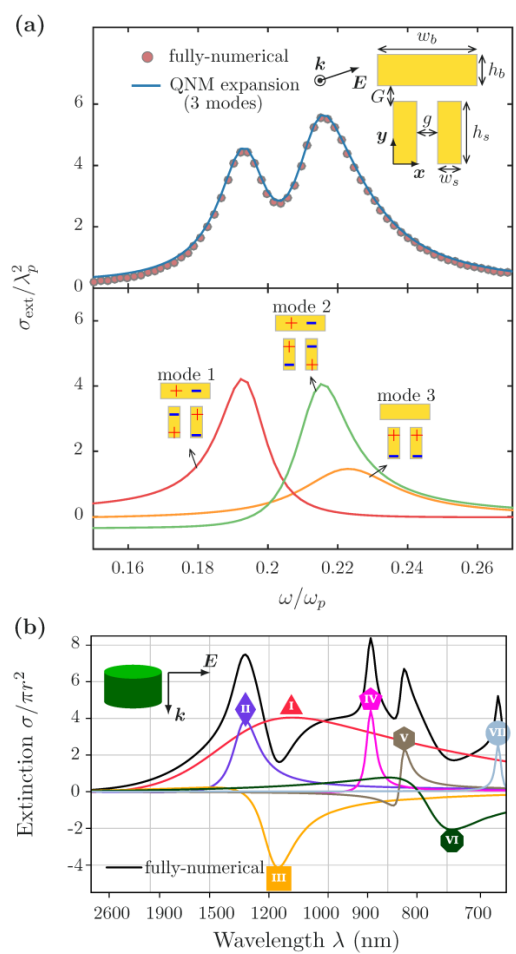

**Fig. 15. Extinction cross-section spectra interpreted with QNM expansions. (a)** Extinction cross-section spectra of dolmen nanoresonator composed of three gold rods in air. The results of a QNM expansion with only three QNMs (blue solid curve) are perfectly superimposed with those obtained with the frequency-domain solver of COMSOL Multiphysics (red dots). The gold permittivity is approximated by a Drude model, $\varepsilon = 1 - \omega_p^2/(\omega^2 + i\omega\gamma)$ with $\lambda_p = 2\pi c/\omega_p = 150$ nm and $\gamma = 0.0128\omega_p$. The incident plane wave is polarized along the $2\hat{x} + \hat{y}$ direction and propagates along the $\hat{z}$ direction. The geometrical parameters of the Dolmen nanoresonator are $h_s = 100$ nm, $w_s = g = G = 30$ nm, $h_b = 50$ nm and $w_b = 128$ nm. The rod thickness is 20 nm. Adapted from [Fag17]. **(b)** Extinction cross section of a silicon nanodisk modelled with a 8-pole Drude-Lorentz model. The black curve corresponds to the numerical solution and the colored curves represents the contributions from every individual QNMs. The disk has a radius of 242 nm, a height of 220 nm and rounded edges with a radius of 50 nm. Adapted from [Pow17].

In general, the scattered light might couple to multiple channels including free-space radiative and guided modes for scatterers deposited on stratified structures. Knowing the angular distribution of the power fluxes into these channels, the so-called free-space and guided-mode radiation diagrams, is of great relevance to interpret optical measurements and understand the modification of the optical properties of nanoparticles buried in structured substrates [Jou16]. This computation requires to know the far-zone field, and since QNM expansions only faithfully predict the scatterer near fields, near-to-far field transformation (NFFT) approaches [Bal05,Yan16,Dem96] or multipole decomposition techniques [Jac99] have to be used. Successful applications can be found in [Fag15,Pow17].

**9.2 Fano and temporal response**

Today, broadband lasers capable of delivering 15 fs pulses are used by many research groups for the study of the ultrafast processes and coherent dynamics in a large variety of systems. Since nanoresonators have $Q$-factors low enough to accommodate the laser bandwidth and additionally offer field enhancements and subwavelength confinements, the spatiotemporal control of strongly enhanced optical fields with nanometer and femtosecond resolutions has become a key challenge in nano-optics [Nov06]. Manipulating nanoscale dynamics with the near-field response of plasmonic nanostructures entails understanding of how the resonant character of nanostructures and propagation effects determine the temporal response in a chosen point in space.

In general, the near-field optical response of any complex structure can be described in terms of combinations of localized resonances. Additionally, since the spectral dependence is analytically known, the QNM formalism is particularly adapted to study resonators dynamics. We thus consider a resonator driven by an optical pulse, $\boldsymbol{E}_b(\boldsymbol{r},t)$, i.e., a wave packet that can be Fourier transformed $\boldsymbol{E}_b(\boldsymbol{r},t) = \int_{-\infty}^{+\infty} \boldsymbol{E}_b(\boldsymbol{r},\omega)\exp(-i\omega t)d\omega$, with $\boldsymbol{E}_b(\boldsymbol{r},\omega)$ the frequency spectrum of the pulse. Driven by the incident pulse, the resonator scatters a time-dependent electromagnetic field, $\boldsymbol{E}_{sca}(\boldsymbol{r},t)$. The temporal response being simply the Fourier transform of the spectral response, it follows that the scattered near-field can be expanded in the time domain as [Yan17]

$$\boldsymbol{E}_{sca}(\boldsymbol{r},t) = \text{Re}\left(\sum_m \beta_m(t)\widetilde{\boldsymbol{E}}_m(\boldsymbol{r})\right), \tag{9.4}$$

with

$$\beta_m(t) = \int_{-\infty}^{+\infty} \alpha_m(\omega)\exp(-i\omega t)d\omega, \tag{9.5}$$

where $\alpha_m(\omega)$ is weighted by the spectral power density of the driving pulse. For instance, if using Eq. (5.11) one gets

$$\beta_m(t) = \int_{-\infty}^{+\infty} \frac{\omega \exp(-i\omega t)}{\omega_m - \omega} \iiint \Delta\varepsilon(\boldsymbol{r},\omega)\boldsymbol{E}_b(\boldsymbol{r},\omega) \cdot \tilde{\boldsymbol{E}}_m(\boldsymbol{r}) d\boldsymbol{r}^3 d\omega. \tag{9.6}$$

Because of the linearity in the time-domain, the scattered field remains a simple sum of independent contributions from every single QNM retained in the expansion. These contributions are weighted by the time-dependent excitation coefficients $\beta_m(t)$. The latter depend on the resonant nature of the interaction through the $1/(\omega_m - \omega)$ terms and the coupling to the driving field through the inner product $\boldsymbol{E}_b(\boldsymbol{r},\tilde{\omega}_m) \cdot \tilde{\boldsymbol{E}}_m(\boldsymbol{r})$. Equations (9.4) and (9.6) provide a very simple and intuitive description of the temporal response of nanoresonators with decoupled time and space dependences. Additionally, the time integral of Eq. (9.6) can be easily computed with a Fast Fourier Transform algorithm [Fag17]. The total field induced by the optical pulse is simply computed as $\boldsymbol{E}(\boldsymbol{r},t) = \boldsymbol{E}_{sca}(\boldsymbol{r},t) + \boldsymbol{E}_b(\boldsymbol{r},t)$.

Figure 16 illustrates the relevance of the QNM expansion to analyze the near-field response of a Dolmen nanostructure composed of three gold nanorods illuminated by a 12.7-fs plane-wave Gaussian pulse. At visible frequency, the Dolmen supports three dominant QNMs with $Q$-values ranging from 10 to 20. The response computed at point $A = (-5, -10, 0)$ nm with a finite-difference time-domain method shows a clear signature of mode beating, with a beating period comparable to the driving pulse duration. The beating provides a rather complicated pattern that it is difficult to understand especially because it strongly depends on the near-field position. Conversely, in the frequency domain, the spectrum exhibits a double dip often qualitatively interpreted as a Fano response resulting from the interference of out-of-phase modes that are difficult to identify especially when they overlap spectrally and spatially [Zha08,Ver09,Lov13,Hsu14,Ala15,Tha17]. The main strength of the QNM expansion is to provide a quantitative interpretation that unambiguously reveals the role and impact of each QNM by providing a direct access to the few control knobs, the QNM excitation coefficients that are driving the temporal response. This method, therefore, provides a powerful tool to explore the effects of pulse shaping in nano-optics.

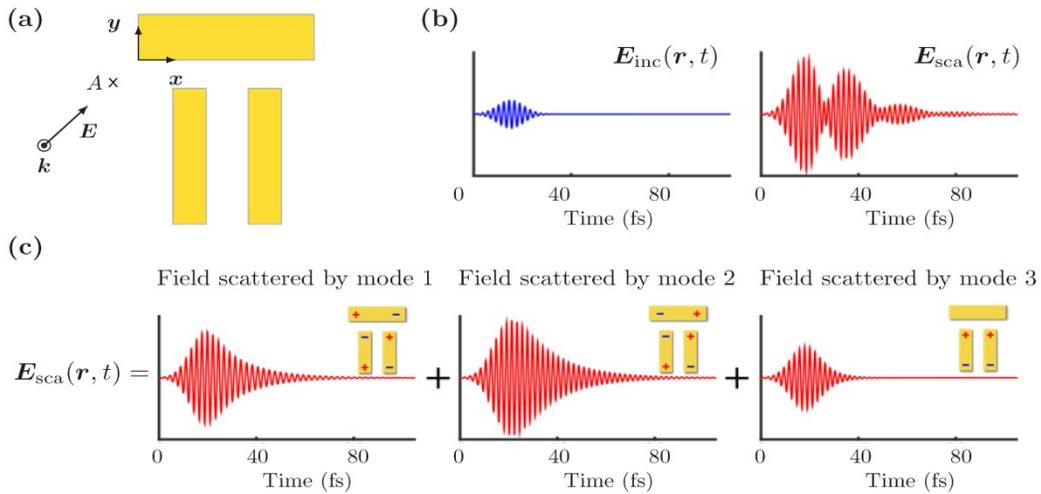

Fig. 16. Temporal response analysis with QNM expansions. (a) The same Dolmen nanoresonator as in Fig. 15 is considered for the illustration. It is illuminated by a 12.7 fs plane-wave Gaussian pulse with a central frequency $\omega_0 = 2.9 \times 10^{15}$ rad/s with a polarization $\boldsymbol{E}$ and a wavector $\boldsymbol{k}$. (b) Incident driving field $\boldsymbol{E}_{inc}$ and scattered

field $E_{sca}$ computed with a FDTD method at point $A$. **(c)** The complex beating response can be seen as resulting from three dominant QNMs, whose individual contributions sum up to accurately match the FDTD computational results, see [Fag17] for details. In the spectral domain, the beating reflects in a complex response shape (Fig. 15).

### 9.3 Resonances in complex media

The interaction of light with disordered media, which may be defined as structures whose permittivity randomly fluctuates in space, is a research topic that has received growing attention since the early 1980s, owing not only to its importance for the optical characterization of complex natural systems (biological tissues, porous media, foams, …) but also to the richness of wave phenomena that may be observed in such systems. The interference between waves experiencing multiple successive scattering events results in mesoscopic phenomena, similar to those found for electrons in condensed matter [Akk11], such as weak localization and speckle correlations, and can lead to new optical functionalities, like random lasing in media containing optical gain [Wie08] or nanoscale "hot spots" in metallo-dielectric composites [Sar00]. The possibilities to fabricate photonic structures with a well-controlled morphology (e.g., partially-disordered photonic crystal structures or amorphous, short-range correlated, disordered media) and to shape the incident light field boosted research in "mesoscopic optics", continuously providing new opportunities in the control of light propagation and confinement in disordered media [Wie13,Rot17].

At the heart of many optical phenomena in complex media lies the concept of resonance. Indeed, similarly to the case of individual scatterers, QNMs drive the light interaction with disordered media and their resulting optical features. By contrast, however, resonances in disordered media are formed by coherent multiple scattering and are thus an intrinsic property of an ensemble of scatterers rather than a property of only one. Figure 17 shows two practical examples where the concept of resonance is paramount in explaining the observed optical response.

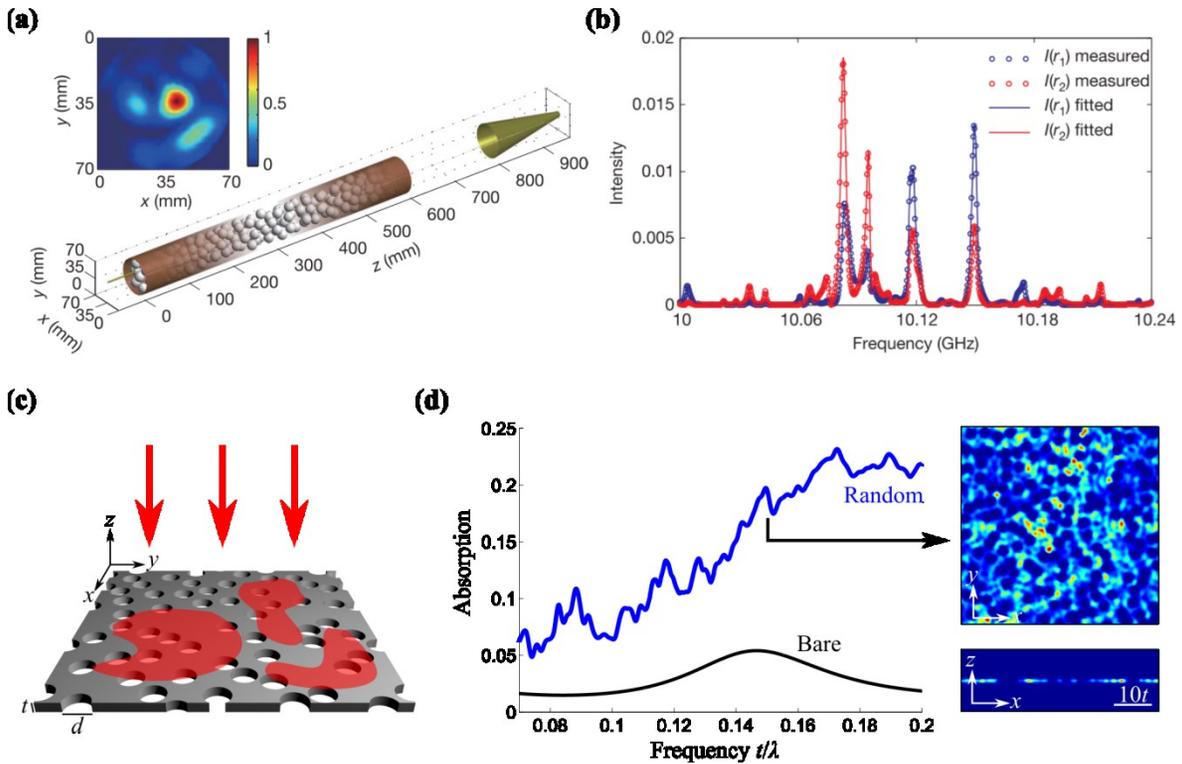

**Fig. 17. Use of the concept of resonance in disordered media. (a)-(b)** Microwave transmission through a quasi-1D disordered medium consisting of a random packing of alumina spheres contained in a copper tube of wavelength-scale diameter (adapted from [Wan11]). **(a)** Microwaves are launched from a horn and the transmitted field is measured at each point of the output surface. The inset shows the intensity speckle pattern normalized to its peak value. **(b)** The transmitted intensity spectra, measured at two positions $r_1$ and $r_2$, exhibit distinct sharp peaks, characteristic of a regime where the wave is exponentially-localized in the disordered medium. Assuming that the field can be expressed as a superposition of resonances with a Lorentzian lineshape, a fit of the transmitted intensity allows recovering the modes complex frequency and output profile. **(c)-(d)** Light trapping in a thin semiconductor film perforated by a random array of holes (adapted from [Vyn12]). **(c)** Light incident from free space couples to the film via its leaky modes illustrated with red areas. **(d)** Compared to the unpatterned film, the absorption efficiency is strongly enhanced. The spatial response at one frequency is due to several modes which overlap in frequency, resulting in an intricate intensity pattern (see the inset, where the intensity is taken at a thickness $t = 0.15\lambda$). The light trapping effect is clearly visible on the spatial intensity maps.

**Computation of modes in disordered media.** The simplest model of disordered optical medium is an ensemble of $N$ small objects placed in a uniform background, all behaving as point-like scatterers described by an electric dipole polarizability tensor $\boldsymbol{\alpha}_i(\omega)$ [Lax52]. This model nicely applies to a wide range of systems, from ensembles of small (dielectric or metallic) particles to atomic clouds [Lag96]. At frequency $\omega$, the field $\boldsymbol{E}_{exc}$ exciting the scatterer $j$ at position $\boldsymbol{r}_j$ depends on the field scattered by all other scatterers, so that

$$\boldsymbol{E}_{exc}(\boldsymbol{r}_j;\omega) = \boldsymbol{E}_{inc}(\boldsymbol{r}_j;\omega) + \frac{\omega^2}{c^2}\sum_{i\neq j}\boldsymbol{G}_0(\boldsymbol{r}_j,\boldsymbol{r}_i;\omega)\boldsymbol{\alpha}_i(\omega)\boldsymbol{E}_{exc}(\boldsymbol{r}_i;\omega), \qquad (9.7)$$

where $\boldsymbol{G}_0(\boldsymbol{r}_j,\boldsymbol{r}_i;\omega)$ is the free space Green tensor relating the field in $\boldsymbol{r}_i$ to the field in $\boldsymbol{r}_j$. The QNMs of the system are found by setting the driving field $\boldsymbol{E}_{inc}$ equal to zero, leading to a non-linear eigenvalue problem, which, in the case where the scatterers are assumed to be identical and described by a scalar and isotropic polarizability, $\boldsymbol{\alpha}_i(\omega) = \alpha(\omega)\boldsymbol{I}$, reads

$$\boldsymbol{\mathcal{M}}(\omega)\,\boldsymbol{E}_{exc} = \Lambda(\omega)\,\boldsymbol{E}_{exc}, \qquad (9.8)$$

with $\boldsymbol{\mathcal{M}}(\omega) = (1-\delta_{ij})\frac{6\pi c}{\omega}\boldsymbol{G}_0(\boldsymbol{r}_j,\boldsymbol{r}_i;\omega)$ the $3N \times 3N$ transformation matrix, composed of $N^2$ $3\times 3$ submatrices – one per pair $(i,j)$ of scatterers - that contains the electromagnetic interaction between scatterers, and $\Lambda(\omega) = \frac{6\pi c^3}{\omega^3}\alpha^{-1}(\omega)$ the eigenvalues that explicitly depend on frequency $\omega$. When the scatterers have a very high $Q$, the eigenvalue problem can be linearized, thereby reducing the problem to the evaluation of the eigenvalues of $\boldsymbol{\mathcal{M}}(\omega_e)$, where $\omega_e$ is the scatterer resonance frequency (see Appendix 5 for more details).

Due to its range of applicability, this so-called "random Green's matrix" approach [Rus00] has been applied almost exclusively to cold atomic clouds [Rus00,Pin04,Goa11,Ski14,Bel14,Sch16]. Its popularity is due to the fact that it can easily handle thousands of coupled resonators and analytical predictions for the eigenvalue distribution are possible in the limit of large matrix size [Gob11]. The numerical calculation of the QNMs of ensembles of low-$Q$ scatterers is possible as well [Seb02,And11,Sav11], but is much more demanding in computational time and memory resources when full-wave solvers such as FDTD or finite-elements are used.

Figure 18 shows an example of the eigenstate frequencies and decay rates (blue dots) in the complex-frequency plane for a 2D disordered ensemble of identical line-source scatterers, computed with the random Green's matrix approach. The electromagnetic interaction between scatterers leads to the formation of a set of collective modes whose resonance frequencies can be shifted by several linewidths and whose decay rates spread over several orders of magnitude. Superradiant states are located above the dashed horizontal line that indicates the linewidth of every individual resonant scatterer.

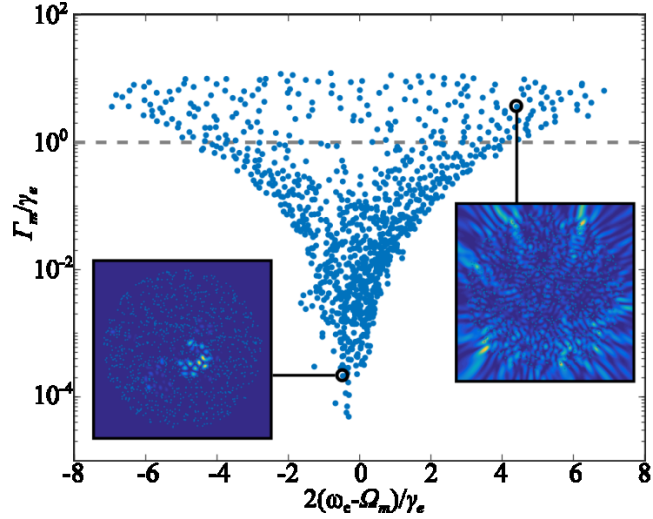

**Fig. 18.** Eigenstate frequencies and decay rates (blue dots) obtained by computing the complex eigenvalues of Eq. (A5-6) for a 2D disordered system consisting of a random (uncorrelated) distribution of 1000 resonant line dipole scatterers in a disk of radius $R = 8\lambda_e$. The electric field is normal to the line dipole axis (TM polarization), and the 2D free-space Green tensor is given by $G_0(\boldsymbol{r},\boldsymbol{r}';\omega) = \frac{i}{4} H_0^{(1)}\left(\omega \frac{|\boldsymbol{r}-\boldsymbol{r}'|}{c}\right)$. The scatterers are described by the 2D polarizability $\alpha(\omega) = \frac{2\gamma_e c^2}{\omega^2} \frac{1}{\omega_e - \omega - i\gamma_e/2}$, with $\omega_e = \frac{2\pi c}{\lambda_e}$ the resonance frequency and $\gamma_e$ the resonance linewidth. The interaction between the resonant scatterers yields collective modes with a broad distribution of resonant frequencies and linewidths. The insets show two QNM fieldmaps; the scattering centers are superimposed with gray dots. Long lifetime QNMs (subradiant modes) weakly couple to free space and tend to be localized in the system, while short lifetime QNMs (superradiant modes) are generally delocalized over the entire system.

**Application of resonant modes in disordered media: Anderson localization.** One of the most surprising wave phenomenon in disordered media is the celebrated Anderson localization [Lag09]. While waves are normally expected to diffuse freely and expand throughout a disordered medium, interference can make the whole diffusion process come to a halt: waves remain exponentially-localized around a point.

The extended (i.e., diffusive) and localized regimes in disordered media can be well identified in terms of resonances. As highlighted in a classical paper on electron transport [Tho74], the transition between the two regimes may be intuitively understood by referring to a single parameter $\delta = \frac{\delta\omega}{\Delta\omega}$, the ratio between the average resonance linewidth, $\delta\omega = \langle\text{Im}[\widetilde{\omega}_m]\rangle$ and the average level spacing between adjacent resonances, $\varDelta\omega = \langle\text{Re}[\widetilde{\omega}_m] - \text{Re}[\widetilde{\omega}_{m-1}]\rangle$. Here, the brackets $\langle...\rangle$ denote averaging over disorder realizations. When $\delta > 1$, several modes typically contribute to the response of the system at a given frequency, such that the light impinging onto the system excites several resonances and can spread freely throughout the system. By contrast, when $\delta < 1$, the optical response is typically dominated by a single resonant mode, to which light is necessarily confined, thereby corresponding to the localized regime. This implies that, in the localized regime, the transmission spectrum should be composed of a series of sharp, distinguishable peaks and that, at a fixed frequency, the intensity profile should typically be independent of the illumination (up to a prefactor due to the excitation) since it is dominated by a single QNM. These concepts have been demonstrated experimentally for microwaves in quasi-1D geometries, see Fig. 16(a)-(b), and numerically in 2D systems [Wan11,Pen14,Les14].

Although a system can be *statistically* in the localized regime, it is important to realize that the probability that two or more QNMs overlap in frequency may not be negligible. The co-existence of several localized modes at the same frequency in a specific disorder realization is known in the literature as "necklace state" [Pen87]. They have been observed at optical frequencies in 1D random thin-film stacks [Ber05] and were shown to play a key role on the transmission properties of 1D disordered systems [Bli08,Pen14] and on the crossover between diffusive and localized regimes in 2D dielectric photonic structures [Van09].

The modal picture also allows us to understand how localization is expected to depend on the dimensionality $d$ of the system under consideration. Neglecting the impact of wave interference on transport, one can show, via diffusion theory, that the degree of mode overlap scales as $\delta \propto L^{d-2}$, with $L$ the system size [van99]. Thus, in 1D, $\delta$ will unavoidably tend to values below 1 when increasing $L$, implying that localization should always take place, whereas in 3D, the resonances will tend to overlap increasingly more with increasing $L$ ($d=2$ being a critical dimension). Localization in 3D media is as a matter of fact expected to occur only if the scattering is strong enough. In other words, a transition between the extended and localized regimes is expected with increasing disorder strength.

The experimental observation of a localization transition for light in 3D media has been subject of intense research in the past decades, but has so far remained elusive [Ski16]. An important hint for this difficulty of observing a transition has been provided a few years ago with numerical studies of QNMs in random (uncorrelated) ensembles of strongly resonating dipole scatterers, computed with the Green's matrix approach, which revealed that the polarized nature of light prevents 3D Anderson localization to occur [Ski14]. This conclusion holds only for ensembles of very high-$Q$ scatterers, such as cold atoms or molecules, and efforts are currently being amplified to understand how an engineering of the scatterers shape and relative position (i.e., structural correlations) could facilitate the occurrence of localization. It is known, for instance, that certain disordered photonic structures can exhibit a photonic gap, in spite of lacking a long-range order, and localized modes at the gap edges [Ima10]. Clearly, the concept of resonant modes will continue to play a key role in this quest.

## 10. Summary

Modes are essential concepts in many areas described by fields and waves, and modal formalisms are thus essential too. We reviewed the significant progress made so far for analyzing light interaction with micro and nanoresonators with such formalisms. The observables related to the interaction, such as the scattering cross-section, the far-field radiation diagram, the modification of the spontaneous emission of quantum emitters and their quantum yield, can be analyzed by expanding the field scattered by the resonator in a basis composed by the natural resonances of the resonator, the so-called QNMs. The latter have complex frequencies related to an exponential damping of the field in time. In addition, since the QNM field has to satisfy outgoing-wave boundary conditions, it is exponentially diverging in space outside the resonator.

The longstanding critical issue of QNM normalization has been recently solved, and QNM solvers that correctly compute and normalize the modes for various geometries now exist, see Sections 3-4, thereby opening new perspectives. Sequential solvers are ideally suited for initial designs or for roughly analyzing the physics of resonators with a few dominant QNMs. Parallel ones are capable of accurately and efficiently computing many QNMs even for dispersive resonators, and allow to achieve much higher accuracies, similarly to what is usually achieved in integrated optics with formalisms based on guided and radiation modes.

The unusual normalization of QNMs that does not rely on energy considerations is not merely a mathematical curiosity; it has fundamental physical implications that have been addressed in Sections 6 to 9. In particular, we discussed in detail the importance of the phase of the QNM field, which is intimately related to the finite $Q$-factor of the resonance. This led us to introduce the concept of a complex mode volume, which consistently impacts various optical phenomena in resonant systems, especially for low-$Q$ resonances. Examples emphasized in this review are the perturbation of resonance, the spectral shape of the LDOS and the Purcell effect, and the strong coupling with quantum emitters. Particular emphasis was placed on the strong limitations of perturbation methods that are widely used to introduce absorption and leakage. Such classical methods rely on a normalization based on energy considerations with a real mode volume, which unfortunately disregards the phase of the QNM field. We again emphasize that the classical methods are largely inaccurate for resonances with low or moderate $Q$-factors.

After 30 years or more, research in this area continues unabated. This reflects the underlying importance of micro and nanoresonators to effectively alter and control light scattering and emission. It seems very likely that this situation will continue. We have discussed some of the areas that future research may develop. The research field now enters a critical phase that requires both theoretical and computational efforts. We have reached a stage with robust foundations, in which the analysis of electromagnetic resonance is routinely performed with a few dominant QNMs. To guaranty a major long-term impact of QNM theories in electromagnetism, one should pass onto a new stage in which the analysis can be made arbitrarily accurate by considering many QNMs. This will require new developments of QNM solvers and clarifications on the completeness of the QNM expansion will be extremely beneficial, especially for resonators placed in complex backgrounds.

The core of nano-cavity engineering consists in perturbing and hybridizing resonances of different natures [Pro03]. A non-exhaustive list encompasses photonic high-$Q$'s and plasmonic low-$Q$'s, plasmonic dark and bright modes, optical modes and quantum emitters. The temporal coupled-mode theory of resonators [Hau91,Shu04] is definitely beautiful due to its capacity to describe universal behaviors with simple equations. It has however some limitations. It is only valid for high $Q$-factors and weakly-coupled cavities. Additionally, it suffers from longstanding weaknesses, such as a phenomenological introduction of the coupling coefficients that are generally taken as fit parameters. Will it be possible to derive a simple and general formalism to hybridize resonances of different natures, similar to what has been done for spontaneous emission in the strong coupling regime? QNM formulations with correct normalization have the potential to bring major consistency to coupled-mode theory, but current literature contains scarce studies for very simple cases, e.g., 2D dielectric resonators [Via16] or 3D spheres [Mu16a]. Increasingly important applications of nanoresonators rely on the cross-action of Maxwell's resonant fields with other equations of physics, e.g. optomechanical cooling [Kip08], plasmonic trapping [Gie12], photonic switching [Kau12], plasmon-enhanced Raman scattering [Roe16]… Traditional modeling tools are not effective, as the full photon Green function needs to be repeatedly computed to iteratively model the nonlinear dynamics. Could we develop elegant QNM-based formalisms to analyze these largely unexplored areas? It is hoped that this review will stimulate new ideas and lead to new research.

## Acknowledgements

The author wishes to thank Kevin Cognée, André Nicolet, Boris Gralak, Guillaume Demesy, Jakob Rosenkrantz de Lasson for stimulating and often difficult discussions. The financial assistance of the French National Agency for Research (ANR) are gratefully acknowledged. This work was supported by the French National Agency for Research (ANR) under the projects "NanoMiX" (ANR-16-CE30-0008) and "Resonance" (ANR-16-CE24-0013). This study has been carried out with financial support from the French State, managed by the French ANR in the frame of "the Investments for the future" Programme IdEx Bordeaux – LAPHIA (ANR-10-IDEX-03-02).

## Annex

### Annex 1. QNMs of 1D non-dispersive Fabry-Perot resonators

This annex presents details on the derivation of the QNMs of 1D non-dispersive Fabry-Perot resonators, their normalization and discusses the limit of high-$Q$ resonances. This non-dispersive case is an excellent study case for understanding the main features of QNMs, their benefits and their limits.

Following Section 2.2, we consider a slab of length $L$ and frequency-independent refractive index $n$ embedded in a homogeneous medium of refractive index $n_1$, centered at $z = 0$, see Fig. 5. The QNM field distribution in the cavity should be composed of two counter-propagating plane waves, such that, after Eq. (2.3), we can write

$$\begin{cases} \tilde{E}_m^s(z) = 2A_s \cos(\tilde{k}_m n z) \text{ and } \tilde{H}_m^s(z) = 2iA_s \frac{n}{\mu_0 c} \sin(\tilde{k}_m n z) & (m \text{ even}) \\ \tilde{E}_m^a(z) = 2iA_a \sin(\tilde{k}_m n z) \text{ and } \tilde{H}_m^a(z) = 2A_a \frac{n}{\mu_0 c} \cos(\tilde{k}_m n z) & (m \text{ odd}) \end{cases}, \quad \text{(A1.1)}$$

where the $A$'s are the field amplitudes inside the resonator. These coefficients are obtained by normalizing the QNMs. For non-dispersive 1D media, the general form of the normalization relation, Eq. (4.5), reduces to

$$\int_{-\infty}^{+\infty} [\varepsilon \tilde{E}_m^2 - \mu \tilde{H}_m^2] dz = 1. \quad \text{(A1.2)}$$

Outside the resonator, $|z| > L/2$, we have $\varepsilon = \varepsilon_0 n_1^2$ and $\mu = \mu_0$, but since $n_1 \sqrt{\varepsilon_0} \tilde{E}_m = \sqrt{\mu_0} \tilde{H}_m$ for plane waves, the integral outside the resonator strictly equals zero and we are left with the integral in the resonator only, $\int_{-L/2}^{L/2} [\varepsilon \tilde{E}_m^2 - \mu \tilde{H}_m^2] dz = 1$. The integral can be solved straightforwardly for each parity, yielding $A_s^2 = (4\varepsilon_0 n^2 L)^{-1}$ and $A_a^2 = -(4\varepsilon_0 n^2 L)^{-1}$, such that

$$\begin{cases} \tilde{E}_m^s(z) = \frac{\pm\cos(\tilde{k}_m n z)}{n\sqrt{\varepsilon_0 L}} \text{ and } \tilde{H}_m^s(z) = \frac{\pm i \sin(\tilde{k}_m n z)}{\sqrt{\mu_0 L}} & (m \text{ even}) \\ \tilde{E}_m^a(z) = \frac{\mp\sin(\tilde{k}_m n z)}{n\sqrt{\varepsilon_0 L}} \text{ and } \tilde{H}_m^a(z) = \frac{\pm i \cos(\tilde{k}_m n z)}{\sqrt{\mu_0 L}} & (m \text{ odd}) \end{cases} \quad (A1.3)$$

The field of QNMs is defined up to a plus/minus sign - there is no arbitrary phase as when energy arguments are used for normalization. This, however, does not matter for the reconstruction of the field at real frequency since the QNM excitation coefficient $\alpha_m(\omega)$, which is defined in Eq. (1.1), also depends linearly of the QNM field. As a result, the reconstructed field has a prefactor in $A^2$.

As indicated in Section 2.2, the complex frequency in the non-dispersive case is simply $\tilde{\omega}_m \frac{n}{c} L = m\pi - \frac{i}{2}\log\left(\frac{1}{r^2}\right)$ with $r = \frac{n-n_1}{n+n_1}$. Due to a complex-value argument, the fields are not purely real or imaginary. Equation (A1.3) can in fact be rewritten as

$$\begin{cases} \tilde{E}_m^s(z) = \frac{\pm 1}{n\sqrt{\varepsilon_0 L}}\left[\cosh\left(\frac{\Gamma_m}{2}\frac{n}{c}z\right)\cos\left(\Omega_m\frac{n}{c}z\right) + i\sinh\left(\frac{\Gamma_m}{2}\frac{n}{c}z\right)\sin\left(\Omega_m\frac{n}{c}z\right)\right] & (m \text{ even}), \\ \tilde{E}_m^a(z) = \frac{\mp 1}{n\sqrt{\varepsilon_0 L}}\left[\cosh\left(\frac{\Gamma_m}{2}\frac{n}{c}z\right)\sin\left(\Omega_m\frac{n}{c}z\right) - i\sinh\left(\frac{\Gamma_m}{2}\frac{n}{c}z\right)\cos\left(\Omega_m\frac{n}{c}z\right)\right] & (m \text{ odd}), \end{cases} \quad (A1.4)$$

and similarly for $\tilde{H}_m^s$ and $\tilde{H}_m^a$, where we remind that $\tilde{\omega}_m = \Omega_m - i\Gamma_m/2$. As might be expected, both the real and imaginary parts of the electric field are therefore affected by loss (which might be leakage to open space and/or absorption). To first order in $\Gamma_m$, i.e. for high-$Q$ resonances, Eq. (A1.4) takes the form of

$$\begin{cases} \tilde{E}_m^s(z) = \frac{\pm 1}{n\sqrt{\varepsilon_0 L}}\left[\cos\left(\Omega_m\frac{n}{c}z\right) + i\frac{\Gamma_m}{2}\frac{n}{c}z\sin\left(\Omega_m\frac{n}{c}z\right)\right] & (m \text{ even}), \\ \tilde{E}_m^a(z) = \frac{\mp 1}{n\sqrt{\varepsilon_0 L}}\left[\sin\left(\Omega_m\frac{n}{c}z\right) - i\frac{\Gamma_m}{2}\frac{n}{c}z\cos\left(\Omega_m\frac{n}{c}z\right)\right] & (m \text{ odd}). \end{cases} \quad (A1.5)$$

In the absence of energy dissipation ($\Gamma_m = 0$), such as in an ideal closed system, the electric field is purely real. From Eq. (A1.5), one readily sees that the main impact of loss on the field inside the resonator is the presence of an imaginary part. Although this observation is made on a specific geometry (1D, non-dispersive, Fabry-Perot resonator), it is important to keep in mind that the presence of a non-zero imaginary part on the electric field in the resonator is a general result.

### Annex 2: Scattered field formulation for resonators

We consider the scattering of an incident field by a resonator. In the usual scattering-field formulation, the permittivity $\varepsilon(r,\omega)$ of the total system (resonator and background) is decomposed as $\varepsilon(r,\omega) = \varepsilon_b(r,\omega) + \Delta\varepsilon(r,\omega)$, where $\varepsilon_b(r,\omega)$ represents a background permittivity and $\Delta\varepsilon(r,\omega)$ is null everywhere except in the resonant structure. Note that $\varepsilon_b(r,\omega)$ does not necessarily correspond to a homogeneous medium and that the generalization to magnetic materials is straightforward. In the absence of the resonator, the background electromagnetic field $[E_b, H_b]$ satisfies the following equations

$$\nabla \times \boldsymbol{E}_b = i\omega\boldsymbol{\mu}(\boldsymbol{r},\omega)\boldsymbol{H}_b \text{ and } \nabla \times \boldsymbol{H}_b = -i\omega\boldsymbol{\varepsilon}_b(\boldsymbol{r},\omega)\boldsymbol{E}_b + \boldsymbol{J}(\boldsymbol{r},\omega), \qquad \text{(A2-1)}$$

where $\boldsymbol{J}(\boldsymbol{r},\omega)$ is the current distribution of the source generating the background field, potentially located at infinity in the case of illumination by a plane wave or a more complex beam. Note that the background field is in general different from the incident field. In the presence of the resonator, the total field $[\boldsymbol{E},\boldsymbol{H}]$ satisfies

$$\nabla \times \boldsymbol{E} = i\omega\boldsymbol{\mu}(\boldsymbol{r},\omega)\boldsymbol{H} \text{ and } \nabla \times \boldsymbol{H} = -i\omega\boldsymbol{\varepsilon}(\boldsymbol{r},\omega)\boldsymbol{E} + \boldsymbol{J}(\boldsymbol{r},\omega), \qquad \text{(A2-2)}$$

with the same source term $\boldsymbol{J}(\boldsymbol{r},\omega)$. The field $[\boldsymbol{E}_S,\boldsymbol{H}_S]$ scattered by the resonator is simply $[\boldsymbol{E}_S,\boldsymbol{H}_S] = [\boldsymbol{E},\boldsymbol{H}] - [\boldsymbol{E}_b,\boldsymbol{H}_b]$. It satisfies

$$\nabla \times \boldsymbol{E}_S = i\omega\boldsymbol{\mu}(\boldsymbol{r},\omega)\boldsymbol{H}_S \text{ and } \nabla \times \boldsymbol{H}_S = -i\omega\boldsymbol{\varepsilon}(\boldsymbol{r},\omega)\boldsymbol{E}_S - i\omega\Delta\boldsymbol{\varepsilon}(\boldsymbol{r},\omega)\boldsymbol{E}_b(\boldsymbol{r}). \qquad \text{(A2-3)}$$

Equation (A2-3) tells us that the field scattered by the resonant structure at frequency $\omega$ can be seen as the field radiated in the presence of the resonator by a current-source distribution $-i\omega\Delta\boldsymbol{\varepsilon}(\boldsymbol{r},\omega)\boldsymbol{E}_b(\boldsymbol{r})$, a known quantity that solely depends on the incident field and the chosen background.

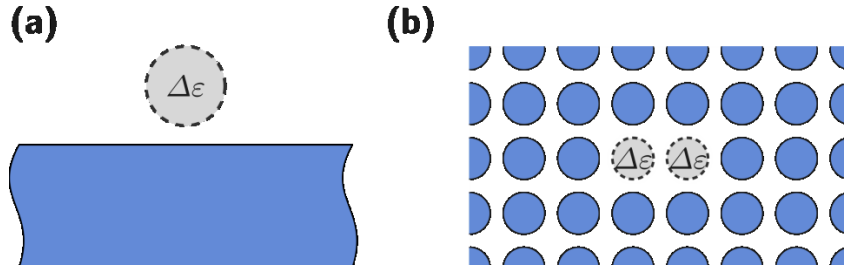

Fig. 19. **Sketch of permittivity change** $\Delta\varepsilon(\boldsymbol{r},\omega)$ **of the background permittivity** $\varepsilon_b(\boldsymbol{r},\omega)$. In the usual scattering-field formulation, the total permittivity $\varepsilon(\boldsymbol{r},\omega)$ of the system is decomposed as $\varepsilon(\boldsymbol{r},\omega) = \varepsilon_b(\boldsymbol{r},\omega) + \Delta\varepsilon(\boldsymbol{r},\omega)$. **(a)** A nanoparticle sitting on a substrate. **(b)** A photonic-crystal cavity formed by removing two lattice elements.

### Annex 3: Lorentz reciprocity formula

The following derivation is classical and is presented here for the sake of completeness. The system is characterized by the position- and frequency-dependent permittivity and permeability tensors $\boldsymbol{\varepsilon}(\boldsymbol{r},\omega)$ and $\boldsymbol{\mu}(\boldsymbol{r},\omega)$. The sole assumption consists in considering reciprocal materials, $\boldsymbol{\mu} = \boldsymbol{\mu}^T$ and $\boldsymbol{\varepsilon} = \boldsymbol{\varepsilon}^T$, where the superscript $T$ denotes matrix transposition. Lorentz reciprocity theorem relates two time harmonic solutions to Maxwell's equations, $(\boldsymbol{E}_1,\boldsymbol{H}_1)$ and $(\boldsymbol{E}_2,\boldsymbol{H}_2)$ with frequencies $\omega_1$ and $\omega_2$ and driving currents $\boldsymbol{j}_1$ and $\boldsymbol{j}_2$, respectively. It is derived by applying the divergence theorem to the vector $\boldsymbol{E}_2 \times \boldsymbol{H}_1 - \boldsymbol{E}_1 \times \boldsymbol{H}_2$ on a closed surface $S$ defining a volume $V$. Using the divergence theorem, we first transform the surface integral into a volume integral

$$\iint_S (\boldsymbol{E}_2 \times \boldsymbol{H}_1 - \boldsymbol{E}_1 \times \boldsymbol{H}_2) \cdot d\boldsymbol{S} = \iiint_V \nabla \cdot (\boldsymbol{E}_2 \times \boldsymbol{H}_1 - \boldsymbol{E}_1 \times \boldsymbol{H}_2) dV. \qquad \text{(A3-1)}$$

Then, by using the vector identity $\nabla \cdot (\boldsymbol{A} \times \boldsymbol{B}) = \boldsymbol{B} \cdot (\nabla \times \boldsymbol{A}) - \boldsymbol{A} \cdot (\nabla \times \boldsymbol{B})$ twice, we get

$$\iint_S (E_2 \times H_1 - E_1 \times H_2) \cdot dS = \iiint_V H_1 \cdot (\nabla \times E_2) - H_2 \cdot (\nabla \times E_1) - E_2 \cdot (\nabla \times H_1) + E_1 \cdot (\nabla \times H_2) dV. \tag{A3-2}$$

The fields $(E_1, H_1)$ and $(E_2, H_2)$ satisfy Maxwell's equations, $\nabla \times E_1 = i\omega_1 \mu(r, \omega_1) H_1$, $\nabla \times H_1 = -i\omega_1 \varepsilon(r, \omega_1) E_1 + j_1$, $\nabla \times E_2 = i\omega_2 \mu(r, \omega_2) H_2$ and $\nabla \times H_2 = -i\omega_2 \varepsilon(r, \omega_2) E_2 + j_2$. By using the assumption that the materials are reciprocal (i.e., $\varepsilon_1 E_1 \cdot E_2 = E_1 \cdot \varepsilon_1^T E_2 = E_1 \cdot \varepsilon_1 E_2$), we obtain the Lorentz reciprocity formula

$$\iint_S (E_2 \times H_1 - E_1 \times H_2) \cdot dS = \iiint_V \{j_2 \cdot E_1 - j_1 \cdot E_2\} d^3r + i \iiint_V \{E_1 \cdot (\omega_1 \varepsilon(\omega_1) - \omega_2 \varepsilon(\omega_2)) E_2 - H_1 \cdot (\omega_1 \mu(\omega_1) - \omega_2 \mu(\omega_2)) H_2\} d^3r, \tag{A3-3}$$

which is used in Sections 4 and 5.

**Annex 4: QNM expansion: derivation of Eqs. (5.4) and (5.5)**

This annex highlights basic aspects, such as the role of dispersion on QNM expansion formalisms, and documents the derivation of the expansion coefficients using orthogonality relations. The derivation does not rely on advanced mathematics and is provided in its entirety. It strictly follows [Sau13,Yan17], the only difference is that a scattered field formulation (Annex 2) is used here for the sake of consistency.

**Resonator with frequency-dependent permittivities.** To derive a closed-form expression for the $\alpha_m$, we expand the scattered field at frequency $\omega$ into the QNM basis $\begin{bmatrix} E_S(r,\omega) \\ H_S(r,\omega) \end{bmatrix} = \sum_p \alpha_p(\omega) \begin{bmatrix} \widetilde{E}_p \\ \widetilde{H}_p \end{bmatrix}$. Simply inserting the expansion into the equation relation linking the scattered field and every individual QNMs, we obtain the set of equations

$$\sum_p B_{mp}(\omega) \alpha_p(\omega) = \omega \iiint \Delta \varepsilon(r,\omega) E_b(r,\omega) \cdot \widetilde{E}_m d^3r, \tag{A4.1}$$

with $B_{mp}(\omega) = \iiint_\Omega [\widetilde{E}_p \cdot (\omega \varepsilon(\omega) - \widetilde{\omega}_m \varepsilon(\widetilde{\omega}_m)) \widetilde{E}_m - \widetilde{H}_p \cdot (\omega \mu(\omega) - \widetilde{\omega}_m \mu(\widetilde{\omega}_m)) \widetilde{H}_m] d^3r$. Since $B_{mp}(\widetilde{\omega}_p) = 0$, we may note $B_{mp}(\omega) = (\omega - \widetilde{\omega}_p) A_{mp}(\omega)$ and the previous set of equations can be recast into the following linear system

$$\begin{bmatrix} A_{11} & & A_{1N} \\ & & \\ A_{N1} & & A_{NN} \end{bmatrix} \begin{bmatrix} (\omega - \widetilde{\omega}_1)\alpha_1 \\ \ldots \\ (\omega - \widetilde{\omega}_N)\alpha_N \end{bmatrix} = \omega \iiint_{V_r} \Delta \varepsilon(r,\omega) E_b(r,\omega) \cdot \begin{bmatrix} \widetilde{E}_1 \\ \ldots \\ \widetilde{E}_N \end{bmatrix} d^3r, \tag{A4.2}$$

with

$$A_{mn}(\omega) = (\omega - \widetilde{\omega}_n)^{-1} \iiint_\Omega \{\widetilde{E}_n(\omega\varepsilon - \widetilde{\omega}_m \varepsilon(\widetilde{\omega}_m))\widetilde{E}_m - \widetilde{H}_n(\omega\mu - \widetilde{\omega}_m \mu(\widetilde{\omega}_m))\widetilde{H}_m\} d^3r. \tag{A4.3}$$

The matrix $A$ depends on the driving frequency $\omega$ and can be safely inverted with standard inversion routines since the zeros have been removed. Its computation requires performing pairwise overlaps between the QNMs. The constant term on the right of Eq. (A4.2) also depends on $\omega$ and requires computing several overlap integrals between the background field and the QNMs. Equation (A4.2) is accurate if enough QNMs are retained in the computations.

**Resonator with frequency-independent permittivities.** If the permittivity and permeability $\varepsilon$ and $\mu$ are frequency-independent, Eq. (A4.3) can be simply rewritten $A_{mn}(\omega) = \frac{\omega-\widetilde{\omega}_m}{\omega-\widetilde{\omega}_n} \iiint_\Omega \{\widetilde{E}_n \cdot \varepsilon \widetilde{E}_m - \widetilde{H}_n \cdot \mu \widetilde{H}_m\} d^3r$ and, due to the orthogonality condition for nondispersive media, we find that $A_{mn}(\omega) = \delta_{mn}$ for normalized modes, with $\delta_{mn} = 0$ if $m \neq n$ and 1 otherwise. Since all the off-diagonal terms in matrix $A$ are null, solving the linear system of equations (A4.2) becomes trivial and we get

$$\alpha_m(\omega) = \frac{\omega}{\omega-\widetilde{\omega}_m} \iiint_{V_r} \Delta\varepsilon(r) E_b(r,\omega) \cdot \widetilde{E}_m d^3r. \tag{A4.4}$$

Equation (A4.4) is exact and valid for any driving field, a far-field illumination or a localized near-field source. Indeed, the oscillations of a resonant structure submitted to harmonic excitation reaches its maximum amplitude at the resonance frequency $Re(\widetilde{\omega}_m)$. At low driving frequencies, $\omega < Re(\widetilde{\omega}_m)$, its response is in phase with the forcing but becomes out of phase just beyond for $\omega > Re(\widetilde{\omega}_m)$.

### Annex 5: Green's matrix approach

This annex presents a method, known in the literature as the "Green's matrix approach" to compute the resonant modes of an ensemble of coupled oscillators. The formalism is presented here for 3D ensembles and was used for 2D ensembles in Section 9.3 to generate Fig. 18.

In a multiple scattering problem, the field $E_{exc}$ exciting the individual scatterers depends on the field scattered by all the other scatterers, and is expressed using a self-consistent expression

$$E_{exc}(r_j;\omega) = E_{inc}(r_j;\omega) + \frac{\omega^2}{c^2}\sum_{i\neq j} G_0(r_j,r_i;\omega)\alpha_i(\omega)E_{exc}(r_i;\omega), \tag{A5.1}$$

with the $i$'s and $j$'s labelling the scatterers. $G_0$ is the 3D electromagnetic dyadic Green's function in a uniform and isotropic medium of refractive index $n$, defined as

$$G_0(r_j,r_i;\omega) = \frac{\exp[in\omega R_{ij}/c]}{4\pi R_{ij}}\left[I - \widehat{R}_{ij}\widehat{R}_{ij} - \left(\frac{c}{in\omega R_{ij}} + \frac{c^2}{(n\omega R_{ij})^2}\right)(I - 3\widehat{R}_{ij}\widehat{R}_{ij})\right], \tag{A5.2}$$

with $R_{ij} = |R_{ij}| = |r_i - r_j|$ and $\widehat{R}_{ij} = R_{ij}/R_{ij}$. The resonant modes are found by setting the driving field $E_{inc} = 0$. For the sake of the illustration, we hereafter assume that the oscillators are all identical with a diagonal and isotropic polarizability-tensor, i.e., $\alpha_i(\omega) = \alpha(\omega)I$, with

$$\alpha(\omega) = \frac{3\pi c^3}{\omega^3}\frac{\gamma_e/n}{\omega_e-\omega-i\gamma_e/2}, \tag{A5.3}$$

where $\omega_e$ is the transition frequency and $\gamma_e$ is the decay rate of the excited state in the medium of refractive index $n$, see Eq. (8.1). Equation (A5.1) can then be rewritten as

$$\left[(1-\delta_{ij})\frac{6\pi c}{\omega}G_0(r_j,r_i;\omega) - \delta_{ij}\frac{6\pi c^3}{\omega^3}\alpha^{-1}(\omega)I\right]E_{exc}(r_i;\omega) = 0. \tag{A5.4}$$

The prefactor $\frac{6\pi c}{\omega}$ has been added for convenience, as shown below. The roots of Eq. (A5.4), which can be found numerically by pole-searching methods, give the QNMs of the coupled dipole ensemble. All QNMs, however, cannot be found at once. Indeed, Eq. (A5.4) readily takes the form of a nonlinear eigenvalue problem $T(\Lambda)\widetilde{E}_{exc} = 0$, where $\Lambda$ is the eigenvalue. The nonlinearity here is clearly due to the frequency-dependence of the dyadic Green's function. This difficulty however can be circumvented for ensembles of high-$Q$ resonators. In this case, the dyadic Green's function is expected to vary slowly spectrally compared to the polarizability and can thus be evaluated at the resonance frequency. In the limits where $\gamma_e \ll \omega_e$, that is for spectrally sharp scatterer resonances, and $\max(R_{ij})/c \ll 1/\gamma_e$, that is when the time needed to cross the system without scattering is negligible compared to the resonance lifetime of the scatterer, we have $\frac{\omega^2}{c^2} G_0(r_j, r_i; \omega)\alpha(\omega) \approx \frac{\omega_e^2}{c^2} G_0(r_j, r_i; \omega_e)\alpha'(\omega)$ with a slightly modified expression of the polarizability

$$\alpha'(\omega) = \frac{3\pi c^3}{\omega_e^3} \frac{\gamma_e/n}{\omega_e - \omega - i\gamma_e/2}. \tag{A5.5}$$

Repeating the steps above, we then obtain

$$\left[ (1-\delta_{ij}) \frac{6\pi c}{\omega_e} G_0(r_j, r_i; \omega_e) - \delta_{ij} \frac{6\pi c^3}{\omega_e^3} \alpha'^{-1}(\omega) I \right] E_{exc}(r_i; \omega) = 0. \tag{A5.6}$$

The equation now takes the form of a linear eigenproblem with $T = A - \Lambda I$ and $A$ is the so-called *Green's matrix*. The eigenvalues $\Lambda$ of $A$ directly provide the complex frequencies associated to the modes of the system, $\Lambda_m = \frac{6\pi c^3}{\omega_e^3} \alpha'^{-1}(\widetilde{\omega}_m) = \frac{2n}{\gamma_e}\left(\omega_e - \widetilde{\omega}_m - i\frac{\gamma_e}{2}\right)$, yielding

$$\widetilde{\omega}_m = \left(\omega_e - i\frac{\gamma_e}{2}\right) - \frac{\gamma_e}{2n}\Lambda_m. \tag{A5.7}$$

In sum, the numerical computation of the modes of a coupled dipole system reduces simply to computing the eigenvalues and eigenproblem of a $3N \times 3N$ matrix ($N$ being the number of dipole scatterers). The corresponding eigenvectors, $\widetilde{E}_{exc,m}(r_i)$, provide the excitation fields of the individual modes at the scatterer positions and the QNM field distributions at a point $r$ can finally be computed as

$$\widetilde{E}_m(r) = \frac{\widetilde{\omega}_m^2}{c^2} \alpha'(\widetilde{\omega}_m) \sum_{i=1}^{N} G_0(r, r_i; \widetilde{\omega}_m) \widetilde{E}_{exc,m}(r_i). \tag{A4.8}$$